\documentclass[10pt,aps,floats,floatfix,amssymb,amsmath,prb,nofootinbib,twocolumn,longbibliography]{revtex4-2}

\usepackage[caption=false]{subfig}
\usepackage{ragged2e}
\usepackage{physics}
\usepackage{psfrag}
\usepackage{graphicx}
\usepackage{color}
\usepackage{esint} 
\usepackage[dvipsnames]{xcolor}
\usepackage[unicode=true,pdfusetitle,
 bookmarks=true,bookmarksnumbered=false,bookmarksopen=false,
 breaklinks=false,pdfborder={0 0 0},backref=false,colorlinks=true,
 allcolors=blue]
 {hyperref}
\usepackage{geometry}
\usepackage{subfig}
\newcommand{\poc}{\hspace*{8pt}}
\newcommand{\pas}{\newline \hspace*{8pt}}
\newcommand{\latsize}{N}
\newcommand{\bose}{n_{\mathrm{ph}}}

\newcommand{\complexen}{\tilde{E}_{\bf k}}
\renewcommand{\appendixname}{APPENDIX}

\newcommand{\phset}{\{ n_p \}}
\newcommand{\phstate}{\tilde{n}_p}

\definecolor{crvena}{rgb}{1.0, 0.0, 0.00} 

\usepackage[normalem]{ulem}
\newcommand\redsout{\bgroup\markoverwith{\textcolor{red}{\rule[0.5ex]{2pt}{1.0pt}}}\ULon}

\begin{document}
\title{Cumulant expansion in the Holstein model: Spectral functions and mobility}

\author{Petar Mitri\'c}
\affiliation{Institute of Physics Belgrade, University of Belgrade, 
Pregrevica 118, 11080 Belgrade, Serbia}
\author{Veljko Jankovi\'c}
\affiliation{Institute of Physics Belgrade, University of Belgrade, 
Pregrevica 118, 11080 Belgrade, Serbia}
\author{Nenad Vukmirovi\'c}
\affiliation{Institute of Physics Belgrade, University of Belgrade, 
Pregrevica 118, 11080 Belgrade, Serbia}
\author{Darko Tanaskovi\'c}
\affiliation{Institute of Physics Belgrade, University of Belgrade, 
Pregrevica 118, 11080 Belgrade, Serbia}

\begin{abstract}
We examine the range of validity of the second-order cumulant expansion (CE) for the calculation of spectral functions, quasiparticle properties, and mobility of the Holstein polaron. We devise an efficient numerical implementation that allows us to make comparisons in a broad interval of temperature, electron-phonon coupling, and phonon frequency. For a benchmark, we use the dynamical mean-field theory (DMFT) which gives, as we have recently shown, rather accurate spectral functions in the whole parameter space, even in low dimensions. We find that in one dimension the CE resolves well both the quasiparticle and the first satellite peak in a regime of intermediate coupling. At high temperatures, the charge mobility assumes a power law $\mu\propto T^{-2}$ in the limit of weak coupling and $\mu\propto T^{-3/2}$  for stronger coupling. We find that, for stronger coupling, the CE gives slightly better results than the self-consistent Migdal approximation (SCMA), while the one-shot Migdal approximation is appropriate only for a very weak electron-phonon interaction. We also analyze the atomic limit and the spectral sum rules. We derive an  analytical expression for the moments in CE and find that they are exact up to the fourth order, as opposed to the SCMA where they are exact to the third order. Finally, we analyze the results in higher dimensions.
\end{abstract}

\maketitle

\section{Introduction} \label{Sec:Introduction}
\poc
The cumulant expansion (CE) method presents an alternative to the usual Dyson equation approach in the calculation of spectral functions of interacting quantum many-particle systems \cite{2000_Mahan}. In this method, we express the Green's function in real time as an exponential function of an auxiliary quantity $C(t)$, called the cumulant,
which can be calculated perturbatively \cite{1962_Kubo}.  In the late 1960s, it was established that the lowest order CE gives the exact solution of the problem of a core hole coupled to bosonic excitations (plasmons or phonons) \cite{1969_Lundqvist, 1970_Langreth}. While there were early papers, that emphasized the potential role of CE as an approximate method to treat the electronic correlations in metals beyond the {\it GW} approximation \cite{1980_Hedin,1999_Hedin, 1996_Aryasetiawan, 1997_Holm} and the electron-phonon interaction in semiconductors and narrow band metals beyond the Migdal approximation \cite{1966_Mahan, 1975_Dunn, 1994_Gunnarsson}, a surge of studies of CE has appeared only recently.
\pas
Renewed interest has emerged due to the possibility of combining CE with {\it ab initio} band structure calculations. The CE for the electron-phonon interaction was used to obtain the spectral functions of several doped transition-metal oxides  \cite{2017_Verdi,2020_Antonius}, showing a favorable comparison with angle-resolved photoemission spectroscopy (ARPES)  \cite{2013_Moser}. A particularly appealing feature of the CE approach is that it describes the quasiparticle part of the spectrum as well as the satellite structure (sidebands). Combining the CE with the Kubo formula for charge transport gives an attractive route to calculate mobility in semiconductors, beyond the Boltzmann approach which is applicable only for weak electron-phonon coupling \cite{2017_Giustino}. This was very recently demonstrated for $\mathrm{SrTiO_3}$ \cite{2019_Zhu} and naphthalene~\cite{2022_Chang}. CE was also applied to elemental metals where a correction to the standard Migdal approximation is discussed 
\cite{2014_Story}.  Similarly, the CE is successfully used to treat the electronic correlations beyond the {\it GW} approximation \cite{2014_Kas, 2014_Lischner, 2015_Caruso_Giustino, 2015_Zhou, 2016_Gumhalter, 2016_Vigil-Fowler, 2018_Zhou_Gatti}.  Furthermore, CE was used to study absorption spectra in molecular aggregates representative of photosynthetic pigment-protein complexes \cite{2015_Ma,2020_Cupellini, 2022_Nothling}.
\pas
{ Despite the wide use of the lowest order CE, there seems to be a lack of studies establishing its range of validity, which represents the central motivation for this paper.} To achieve this, we turn to simplified models of the electron-phonon interaction. CE for the
Fr{\"o}hlich model \cite{2018_Nery,2022_Kandolf} gives the ground-state energy and the
effective mass similar to the exact QMC calculations for moderate interaction \cite{2000_Mishchenko}. This is in contrast to the Dyson-Migdal approach which severely underestimates mass renormalization. A comparison of the corresponding spectral functions is, however, missing since reliable QMC results are not available due to the well-known problems with analytical continuation. The Holstein polaron model gives a unique opportunity to explore the applicability of the CE since various numerically exact methods are developed and applied to this model covering different parameter regimes \cite{1959_Holstein, 1962_Lang_Firsov, 2007_Alexandrov,1998_Jeckelmann,1998_Kornilovitch,1998_Romero,2003_Fratini, 2006_Fratini_Ciuchi,1999_Zhang,2006_Goodvin, 2006_Berciu,2008_Ciuchi,2019_Bonca, 2019_Prodanovic,2020_Jansen,2022_Jankovic,2022_Bonca, 2014_Mishchenko}. This was the approach of a very recent work by Reichman and collaborators \cite{2022_Robinson_1,2022_Robinson_2}. Still, there are several questions that remained unresolved. Most importantly, a comparison of spectral functions was made just for a small set of parameters on a finite-size lattice,  where the benchmark  spectral functions were available from the finite-temperature Lanczos results, while the charge transport was not examined.
\pas
In our recent work \cite{2022_Mitric} we have established that the dynamical mean-field theory (DMFT) \cite{1997_Ciuchi} gives close to exact spectral functions of the Holstein polaron for different phonon frequencies, electron-phonon couplings, and temperatures even in low dimension, covering practically the whole parameter space. This method is computationally very fast and precise which makes us ideally positioned to perform comprehensive comparisons with the CE method, which is the goal of this paper. Within the CE, we calculate the spectral functions and charge mobility for a broad set of parameters and make detailed comparisons with DMFT and (self-consistent) Migdal approximation. We find that the one-shot Migdal approximation is appropriate only for very weak electron-phonon coupling. The validity of the CE and self-consistent Migdal approximation (SCMA) is much broader and for intermediate interaction CE even outperforms SCMA. We also derive analytical CE expressions for the ground-state energy, renormalized mass, and scattering rate, as well as the spectral sum rules, and make comparisons between the methods. We establish a power law behavior for the charge mobility at high temperatures. We also compare the performance of different methods as the bandwidth is reduced toward the atomic limit.
\pas
The remaining part of the paper is organized as follows. In Sec.~\ref{Sec:Model_Methods}, we introduce the CE method and present details of its implementation on the Holstein model.
DMFT and SCMA are here introduced as benchmark methods. Representative spectral functions are shown in Sec.~\ref{Sec:SpecF} from weak toward the strong coupling. The high-temperature and atomic limits are analyzed in detail, as well as the spectral sum rules.  In Sec.~\ref{Sec:QP_prop}, we present the results for the effective mass and ground-state energy. The temperature dependence of the electron mobility is analyzed in Sec.~\ref{Sec:Mobility_diff_methods}, and Sec.~\ref{Sec:discussion} contains our conclusions. Some details concerning numerical implementations and additional figures for various parameters are shown in the Appendix and in the Supplemental
Material (SM) \cite{SuppMat}.
\section{MODEL AND METHODS} \label{Sec:Model_Methods}
\poc
The Holstein model is the simplest model of the lattice electrons interacting with the phonons. It assumes a local electron-phonon interaction and dispersionless phonons. The Hamiltonian is given by
\begin{align} \label{Eq:Holstein_Hamiltonian}
    H =& -t_0 \sum_{\langle ij \rangle} 
    \left( c_i^\dagger c_j + \mathrm{H.c}. \right) \nonumber \\
    &-g \sum_i n_i \left( a_i^\dagger + a_i \right)
    + \omega_0 \sum_i a_i^\dagger a_i.
\end{align}
Here, $t_0$ is the hopping parameter between the nearest neighbors and $\omega_0$ is the phonon frequency. $c_i$ and $a_i$ are the electron and the phonon annihilation operators, $n_i = c_i^\dagger c_i$ and $g$ denotes the electron-phonon coupling strength. We set $\hbar, k_B$, elementary charge $e$, and lattice constant to $1$. We also often use a parameter $\alpha = g/\omega_0$. We study the model in the thermodynamic limit (number of sites $N\to\infty$). Furthermore, we consider a dynamics of a single electron in the conduction band and treat the electrons as spinless, since we are interested only in weakly doped semiconductors. This is equivalent to setting the chemical potential far below the conduction band, i.e., considering the limit $\tilde{\mu}\to -\infty$. This case is often referred to as the Holstein polaron problem. We mostly focus on the one-dimensional (1D) system, but we also consider the system in 2D  and 3D.

\vspace*{-0.25cm}
\subsection{Cumulant expansion} \label{SubSec:CE}
\vspace*{-0.25cm}
\subsubsection{General theory} \label{SubSubSec:General_theory}
\vspace*{-0.25cm}
\poc
The central quantity of this paper is the electron spectral function $A_{\bf k}(\omega) = (-1/\pi) \mathrm{Im} G_{\bf k}(\omega)$, where ${\bf k}$ is the momentum and  $G_{\bf k}(\omega)$ is the retarded Green's function in frequency domain \cite{2000_Mahan}. Its exact evaluation is often a formidable task, which is why approximate techniques are usually employed. One needs to be careful with such approaches in order not to violate some analytic properties, such as the pole structure of the Green's function, the positivity of the spectral function, or the spectral sum rules. At least some of these properties can be easily satisfied if the Green's function is not calculated directly, but instead through some auxiliary quantity, such as the  self-energy $\Sigma_{\bf k}(\omega)$. In the latter case, the connection with the Green's function is  established  via the Dyson equation  
\begin{equation} \label{Eq:Dyson}
    G_{\bf k}(\omega) = \frac{1}{G_{{\bf k},0} (\omega)^{-1} - \Sigma_{\bf k}(\omega)}
    =\frac{1}{\omega - \varepsilon_{\bf k} - \Sigma_{\bf k}(\omega)},
\end{equation}
where $G_{{\bf k},0} (\omega)$ is the noninteracting Green's function and $\varepsilon_{\bf k}$ is the noninteracting dispersion relation.
\pas
An alternative to the Dyson equation based approaches is the so-called cumulant expansion method \cite{2014_Kas}, in which the exponential ansatz is chosen for the Green's function in the time-domain:
\begin{equation} \label{Eq:Cumulant_def}
G_{\bf k} (t) = G_{{\bf k},0} (t) e^{C_{\bf k} (t)} = 
-i \theta (t) e^{-i\varepsilon_{\bf k} t} e^{C_{\bf k} (t)}.
\end{equation}
Here, $\theta(t)$ is the Heaviside step function and $C_{\bf k} (t)$ plays the role of an auxiliary quantity which is called the cumulant. Both Eqs.~\eqref{Eq:Dyson} and~\eqref{Eq:Cumulant_def} would correspond to the same Green's function in frequency and time domain if the cumulant $C_{\bf k} (t)$ and the self-energy $\Sigma_{\bf k}(\omega)$ could be evaluated exactly \cite{2000_Mahan}. In practice, however,  one of these approaches is expected to perform better. 
\pas
The spectral function within the CE can be obtained as follows
\begin{equation} \label{Eq:CE_specF_def}
A_{\bf k}(\omega + \varepsilon_{\bf k}) = \frac{1}{\pi} \mathrm{Re} \int_0^\infty dt
e^{i\omega t} e^{C_{\bf k} (t)}.
\end{equation}
Equation~\eqref{Eq:CE_specF_def} circumvents the Fourier transform of the whole Green's function $A_{\bf k}(\omega) = -\frac{1}{\pi} \mathrm{Im} G_{\bf k}(\omega)$, which is useful in practice, as the free electron part $e^{-i \varepsilon_{\bf k} t}$ typically oscillates much more quickly than $e^{C_{\bf k}(t)}$. 
\pas
The expression for $C_{\bf k}(t)$ in the lowest order perturbation expansion can be obtained by taking the leading terms in the Taylor expansion of the Dyson equation $G_{\bf k}(\omega) = (G_{{\bf k},0}(\omega)^{-1} - \Sigma_{\bf k}(\omega))^{-1} \approx G_{{\bf k},0}(\omega) + G_{{\bf k},0}(\omega) \Sigma_{\bf k}(\omega) G_{{\bf k},0}(\omega)$, taking its inverse Fourier transform and equating it to Eq.~\eqref{Eq:Cumulant_def}, where the cumulant in the exponent is replaced with its linear approximation $e^{C_{\bf k}(t)} \approx 1 + C_{\bf k}(t)$:
\begin{equation} \label{Eq:Cum_1}
    C_{\bf k}(t) = ie^{i\varepsilon_{\bf k} t} \int_{-\infty}^\infty \frac{d\omega }{2\pi}
     \frac{e^{-i\omega t} \Sigma_{\bf k}(\omega)}{(\omega - \varepsilon_{\bf k} + i0^+)^2}.
\end{equation}
Using the spectral representation of the self-energy
\begin{equation}
    \Sigma_{\bf k}(\omega) = \int \frac{d\nu}{\pi} \frac{|\mathrm{Im} \Sigma_{\bf k}(\nu)|}{\omega - \nu + i0^+},
\end{equation}
and the contour integration over $\omega$, Eq.~\eqref{Eq:Cum_1} simplifies to~\cite{2014_Kas}
\begin{equation} \label{Eq:Cumulant_expression}
C_{\bf k}(t) = \frac{1}{\pi} \int_{-\infty}^\infty
d\omega \frac{|\mathrm{Im}\Sigma_{\bf k}(\omega + \varepsilon_{\bf k})|}
{\omega^2} (e^{-i\omega t} + i\omega t -1).
\end{equation}
The corresponding spectral function satisfies the first two sum rules, irrespective of $\Sigma_{\bf k}(\omega)$. This is a consequence of the behavior of $C_{\bf k}(t)$ for small $t$; see Sec.~\ref{Sec:High_temp}. In general, $C_{\bf k}(t=0) = 0$ is sufficient for the first spectral sum rule $\int A_{\bf k} (\omega) d\omega = 1$ to be satisfied. The second sum rule $\int A_{\bf k}(\omega) \omega d\omega = \varepsilon_{\bf k}$ can also be satisfied if we additionally impose that the cumulant's first derivative at $t=0$ is vanishing, $\frac{dC_{\bf k}}{dt}(0)=0$. Both of these conditions are satisfied by the cumulant function in Eq.~\eqref{Eq:Cumulant_expression}, as it is a quadratic function of time for small arguments $e^{-i\omega t} + i\omega t -1 \approx -\omega^2 t^2 /2$ for $t\to 0$.
\pas
The application of  Eq.~\eqref{Eq:Cumulant_expression} is facilitated by the fact that it does not contain any iterative self-consistent calculations. However, one needs to overcome the numerical challenges caused by  the removable singularity at $\omega = 0$ and by the rapidly oscillating trigonometric factor $e^{-i\omega t}$ for large $t$. The latter is important for the weak electron-phonon couplings, where it is necessary to propagate $C_{\bf k}(t)$ up to long times until the Green's function is sufficiently damped out.  The same problem occurs  in other regimes as well (e.g., close to the atomic limit), where the Green's function does not attenuate at all; see Sec.~\ref{Sec:lifetime}.  
\pas
The numerical singularity at $\omega=0$ can be completely avoided if we consider the cumulant's second derivative 
\begin{equation} \label{Eq:Cumulant_sec_deriv_general}
\frac{d^2 C_{\bf k} (t)}{dt^2} =  \int_{-\infty}^\infty
\frac{d\omega}{\pi} \, \mathrm{Im} \Sigma_{\bf k}(\omega + \varepsilon_{\bf k}) \,
e^{-i \omega t} \equiv 2 e^{i \varepsilon_{\bf k}t} \tilde{\sigma}_{\bf k}(t),
\end{equation}
where we used  $\mathrm{Im}\Sigma_{\bf k} (\omega) < 0$ and introduced  $\tilde{\sigma}_{\bf k}(t) \equiv \int_{-\infty}^\infty \mathrm{Im} \Sigma_{\bf k}(\omega)e^{-i\omega t} \frac{d\omega}{2\pi} $. Then, $C_{\bf k}(t)$ is obtained as a  double integral over time of Eq.~\eqref{Eq:Cumulant_sec_deriv_general}
\begin{equation}
    C_{\bf k}(t) = 2\int_0^t dt' \int_0^{t'} dt'' 
    e^{i \varepsilon_{\bf k} t''}  
    \tilde{\sigma}_{\bf k}(t''),
\end{equation}
where the lower boundaries of both integrals have to be zero, as guaranteed by the initial conditions $C_{\bf k}(0) = \frac{d C_{\bf k}}{dt}(0) = 0$. Using the Cauchy formula for repeated integration, this can also be written as a single integral:
\begin{equation} \label{Eq:Cumulant_general_Cauchy}
     C_{\bf k}(t) = 2 \int_0^t (t-x) e^{i\varepsilon_{\bf k}x}
    \tilde{ \sigma}_{\bf k}(x) dx.
\end{equation}
This completely removed the problem of numerical singularities. Still, the problem of rapid oscillations of the subintegral function remains due to the presence of $e^{i \varepsilon_{\bf k} x}$ term. In Sec.~\ref{Sec:Cumulant_for_Holstein} we provide an elegant solution for this issue, focusing on the case of the Holstein model.
\subsubsection{Asymptotic expansion for cumulant when \texorpdfstring{$t\to\infty$}{t → ∞}} \label{Sec:Asyptotic_CE}
\poc
The asymptotic expansion of $C_{\bf k}(t)$ for large times, as we now demonstrate, completely determines the quasiparticle properties within this method. This is one of the main motivations for studying the $t\to\infty$ limit.  
\pas
From Eq.~\eqref{Eq:Cumulant_sec_deriv_general}, we see that
\begin{align} \label{Eq:Cum_first_der_deriv}
    i \frac{dC_{\bf k}}{dt}(t\to\infty) &= 
    i\int_{0}^{\infty} \frac{d^2 C_{\bf k}(t)}{dt^2} dt  \nonumber \\
    &= 
    -\frac{i}{\pi} \int_{-\infty}^\infty
      d\omega |\mathrm{Im} \Sigma_{\bf k}(\omega + \varepsilon_{\bf k})|
      \int_0^{\infty} dt e^{-i\omega t} \nonumber \\
      &= \Sigma_{\bf k}(\varepsilon_{\bf k}),
\end{align}
where we used the identity $\int_0^{\infty} dt e^{-i\omega t} = \pi \delta(\omega) - i \mathcal{P} \frac{1}{\omega}$ and the Kramers-Kronig relations for the self-energy.  Hence, the cumulant function $C_{\bf k}(t)$, and also the whole exponent in Eq.~\eqref{Eq:Cumulant_def} is  a linear function of time $C_{\bf k}(t)-i\varepsilon_{\bf k}t \approx -i \tilde{E}_{\bf k} t + \mathrm{const}$ for $t\to\infty$, where
\begin{equation} \label{Eq:QP_perturbation}
    \complexen = \varepsilon_{\bf k} + \Sigma_{\bf k}(\varepsilon_{\bf k}).
\end{equation}
As a consequence, the Green's function in Fourier space has a simple pole situated at $\complexen$, as seen from the following expression 
\begin{equation}
    G_{\bf k}(\omega) = -i \int_0^\infty e^{it \left( \omega - \varepsilon_{\bf k} - \frac{iC_{\bf k}(t)}{t} \right)} dt.
\end{equation}
Therefore, quasiparticle properties are encoded in  $\complexen$: its real and imaginary parts correspond to the quasiparticle energy and scattering rate, respectively. We note that, in our present analysis, we implicitly assumed that ${\frac{dC_{\bf k}}{dt} (t\to\infty)}$ exists and is finite. Although this is generally true, there are a few exceptions. In the Holstein model, the first assumption is violated at the atomic limit ($t_0 = 0$; see Eq.~\eqref{Eq:CF_in_atomic}), while the second assumption is violated at the adiabatic limit ($\omega_0 = 0$) for $k=0$ or $k=\pm \pi$; see Eqs.~\eqref{Eq:lifetime_CE}~or~\eqref{Eq:Re_Migdal_selfen}.
\pas
The knowledge that we gained about the analytic properties of the $C_{\bf k}(t)$ provides us with an intuitive understanding  of how the shape of the cumulant determines the shape of the spectral function. The asymptotic limits $t\to\infty$ (where $C_{\bf k}(t)$ is linear) and $t\to 0$ (where $C_{\bf k}(t)$ is quadratic) by themselves, to a large extent, describe only the simple one-peak spectral functions, while the crossover between these limits is responsible for the emergence of satellite peaks. This can be explained as follows: If the cumulant was quadratic over the whole $t$ domain $C_{\bf k}(t) = c t^2$, the spectral function would have a simple Gaussian shape. Similarly, the Lorentzian shape would be obtained from the linear cumulant $C_{\bf k}(t) = c t$. This suggests that the simple crossover between quadratic (at small $t$) and linear (at large $t$) behaviors would also give a simple one-peak shape of the spectral function. The information about  phonon satellites is thus completely encoded in the $C_{\bf k}(t)$ for intermediate times $t$, which depends on the system and approximation in which the cumulant function is calculated. 

\vspace*{-0.2cm}
\subsubsection{Second-order cumulant expansion for the Holstein model} \label{Sec:Cumulant_for_Holstein}
\vspace*{-0.2cm}
\poc
Let us now concentrate on a specific example, the Holstein model on a hypercubic lattice in $n$ dimensions. The second-order cumulant is given by Eq.~\eqref{Eq:Cumulant_expression}, where the self-energy is taken to be in the Migdal approximation $\Sigma_{\bf k}(\omega)=\Sigma^{\mathrm{MA}}(\omega)$, i.e., of the second (lowest) order with respect to the electron-phonon coupling $g$. This is in accordance with the derivation from Sec.~\ref{SubSubSec:General_theory}, since we restricted ourselves to the lowest order terms in the Taylor expansion of the Dyson equation and of $e^{C_{\bf k}(t)}$. An alternative derivation of this expression is given in Sec.~I of the SM \cite{SuppMat}. 
\pas
\begin{figure}[!t]
  \includegraphics[width=3.4in,trim=0cm 0cm 0cm 0cm]{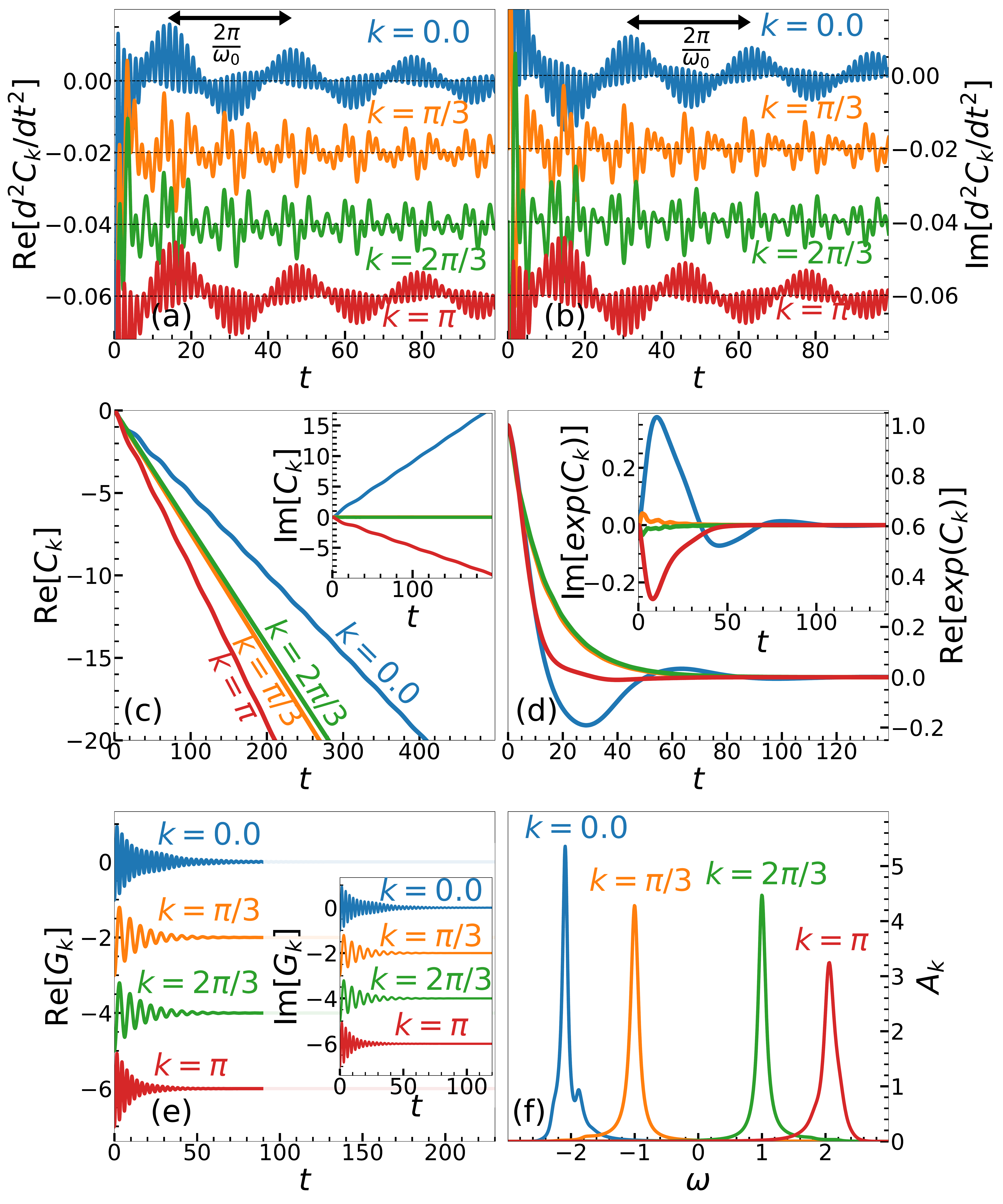} 
 \caption{(a)--(f) The cumulant, Green's, and spectral function on the example of the one-dimensional  Holstein model with the following values of the model parameters: $\omega_0=0.2$, $g=0.2$, $T=0.3$, and $t_0=1$.
 }
 \label{Fig:RC_subint}
\end{figure}
Migdal approximation is briefly discussed in Sec.~\ref{Subsec:Migdal_approximation}. For our present purpose, we only need the expression for the imaginary part of the self-energy
\begin{equation} \label{Eq:Migdal_ImSigma}
    \mathrm{Im} \Sigma^{\mathrm{MA}}(\omega) = 
    - \pi g^2  \left[
	(\bose+1) \rho (\omega - \omega_0) + \bose \rho (\omega + \omega_0)
    \right],
\end{equation}
where $\bose=1/\left(e^{\omega_0/T} -1\right)$ is the Bose factor, $\rho (\omega)=\frac{1}{\latsize}\sum_{\bf k} \delta(\omega-\varepsilon_{\bf k})$ is the density of electron states for the system of size $\latsize$, which we take in the thermodynamic limit $\latsize\to\infty$, and  $\varepsilon_{\bf k} = -2t_0 \sum_{j=1}^n \cos k_j$ is the noninteracting dispersion relation.
\pas
The expression for the cumulant function, as seen from Eq.~\eqref{Eq:Cumulant_general_Cauchy}, is related to the inverse Fourier transform of $\mathrm{Im}\Sigma^{\mathrm{MA}}(\omega)$, which in turn is completely determined by the inverse Fourier transform of the density of states $\tilde{\rho}(t)$. The latter admits a closed-form solution 
\begin{align} \label{Eq:DOS_Fourier}
	\tilde{\rho}(t) &= \int_{-\infty}^\infty  \frac{d\omega e^{-i\omega t}}{(2\pi)^{n+1}} 
	\int_{[0,2\pi)^n} d^n {\bf k}\; \delta \left(\omega + 2t_0 \sum_{j=1}^n \cos k_j \right)  \nonumber \\
	&= \frac{1}{2\pi}\left( \frac{1}{2\pi} \int_0^{2\pi} dk e^{2 i t_0 t \cos k} \right)^n
	= \frac{J_0 (2 t_0 t)^n}{2\pi},
\end{align}
where $J_0$ is the Bessel function of the first kind of order zero. Hence, Eqs.~\eqref{Eq:Cumulant_general_Cauchy},~\eqref{Eq:Migdal_ImSigma}~and~\eqref{Eq:DOS_Fourier} imply that the cumulant function can be written as
\begin{equation} \label{Eq:Cumualnt_wrt_Bessel}
C_{\bf k}(t) = -g^2  \int_0^t dx (t-x) 
iD(x)e^{i x \varepsilon_{\bf k}} J_0(2t_0 x)^n,
\end{equation}
where $iD(t) = (\bose+1) e^{-i\omega_0 t}+ \bose e^{i\omega_0 t}$ is the phonon propagator in real time (for $t>0$). 
\pas
In Fig.~\ref{Fig:RC_subint}, we illustrate the cumulant function, as well as the corresponding Green's function and spectral function. Figures~\ref{Fig:RC_subint}(a)~and~\ref{Fig:RC_subint}(b) show the second derivative of the cumulant
\begin{equation} \label{Eq:Cumulant_dos_derivative}
    \frac{d^2 C_{\bf k}(t)}{dt^2} = -g^2 iD(t) e^{it\varepsilon_{\bf k}} J_0(2t_0 t)^n,
\end{equation}
in order to demonstrate the rapid oscillations that are also present in the cumulant itself. These are not easily observed by inspecting $C_{\bf k}(t)$ directly, as the linear behavior dominates for large times.  We observe that the $k=0$ and $k=\pi$ results possess an oscillating envelope with period $2\pi / \omega_0$, while intermediate momenta have a much less regular structure. This can have direct consequences on the spectral functions, as the satellite peaks are expected to be at a distance $\omega_0$ from each other. To be more explicit, oscillating envelopes suggest that there is a much higher chance for the occurrence of satellite peaks near the bottom ($k\approx 0$) and the top ($k\approx \pi$) of the band, than otherwise. However, that does not guarantee that the satellite peaks will in fact occur. Figure~\ref{Fig:RC_subint}(c) shows that $\mathrm{Re}  C_{\bf k}(t)$ is declining faster for $k>0$ than for $k=0$. As a consequence, $e^{C_{\bf k}(t)}$  in Fig.~\ref{Fig:RC_subint}(d) attenuates slower for $k=0$, having enough time to complete a full period, while $k=\pi$ results are reminiscent of an overdamped oscillator. A similar, although much less evident,  effect can be seen in the Green's function itself; see 
Fig.~\ref{Fig:RC_subint}(e). This is why the $k=\pi$ spectral function in Fig.~\ref{Fig:RC_subint}(f) has a simple one-peak shape, while only the $k=0$ result captures one small satellite peak. 
\pas
From a numerical point of view, Eq.~\eqref{Eq:Cumualnt_wrt_Bessel} is treated using Levin's collocation method \cite{1996_Levin}, which is reviewed in Appendix~\ref{Apendix:Numerical_RC}. It provides a controlled, accurate, and numerically efficient way to integrate the product of trigonometric, Bessel, and some slowly varying function. This approach avoids using a dense $t$ grid, which would otherwise be required, as the subintegral function in Eq.~\eqref{Eq:Cumualnt_wrt_Bessel} has the same type of rapid oscillations  present in $d^2C_{\bf k}(t) /dt^2 $.

%\vspace*{-0.7cm}
\subsubsection{Lifetime} \label{Sec:lifetime}
%\vspace*{-0.3cm}
%
\begin{figure}[!t]
  \includegraphics[width=3.4in,trim=0cm 0cm 0cm 0cm]{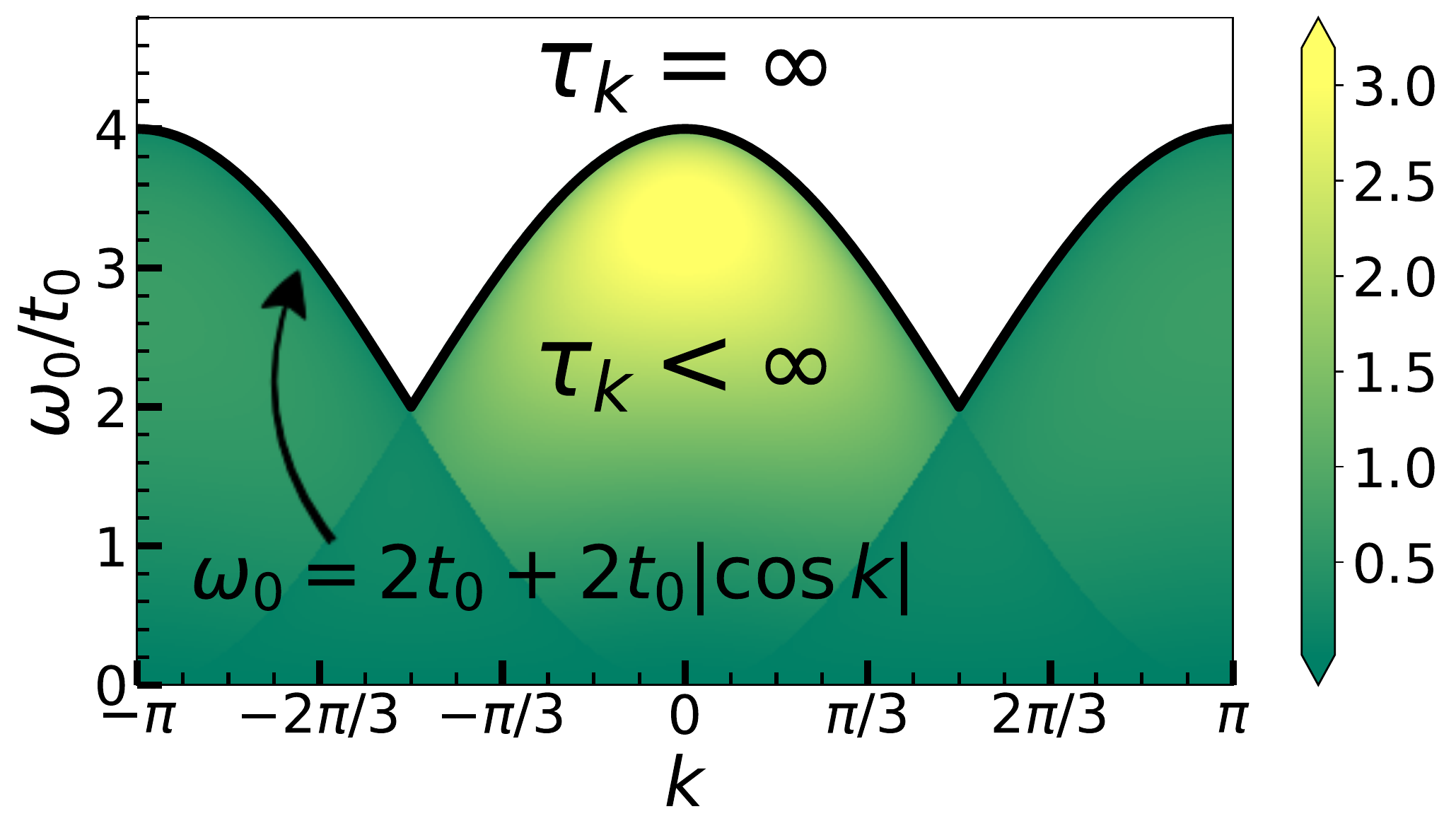} 
 \caption{Quasiparticle lifetime $\tau_{\bf k}$ in the CE method for $T/t_0=2$ and $g/t_0 = 1$. 
 }
 \label{Fig:RC_lifetime}
\end{figure}
\poc
Another question of practical importance is how long should we propagate the cumulant function in real time until the corresponding Green's function attenuates. A rough estimate of such quantity is given by the quasiparticle lifetime $\tau_{\bf k}$. The lifetime is given by $\tau_{\bf k} =1/(2|\mathrm{Im} \complexen|)$, where $\complexen$ is given by Eq.~\eqref{Eq:QP_perturbation}, and the self-energy is taken in the Migdal approximation (see Eq.~\eqref{Eq:Migdal_ImSigma}):
\begin{align} \label{Eq:lifetime_CE}
    \tau_{\bf k}^{-1} =2|\mathrm{Im} \complexen| =& 2g^2 \frac{\theta (4t_0^2-(\varepsilon_{\bf k} - \omega_0)^2)}{\sqrt{4t_0^2-(\varepsilon_{\bf k} - \omega_0)^2}} (\bose+1) \nonumber \\
    +&
    2g^2 \frac{\theta (4t_0^2-(\varepsilon_{\bf k} + \omega_0)^2)}{\sqrt{4t_0^2-(\varepsilon_{\bf k} + \omega_0)^2}} \bose.
\end{align}
This is illustrated in Fig.~\ref{Fig:RC_lifetime}. We observe that there is a considerable part of the parameter space where the lifetime is infinite, which means that the corresponding Green's function never attenuates. This occurs for $\omega_0 > 2t_0 + 2t_0|\cos k|$ in the case of finite temperatures, and for $\omega_0 > 4t_0 \sin^2 k/2$ in the $T=0$ case. In these regimes, one could presume that this is reflected in the spectral functions through the appearance of Dirac delta peaks, which is not expected at finite temperatures.  This illustrates one of the limitations of this method.

\subsection{Benchmark methods} \label{Sec:Benchmark_methods}
\begin{figure}[!t]
  \includegraphics[width=3.4in,trim=0cm 0cm 0cm 0cm]{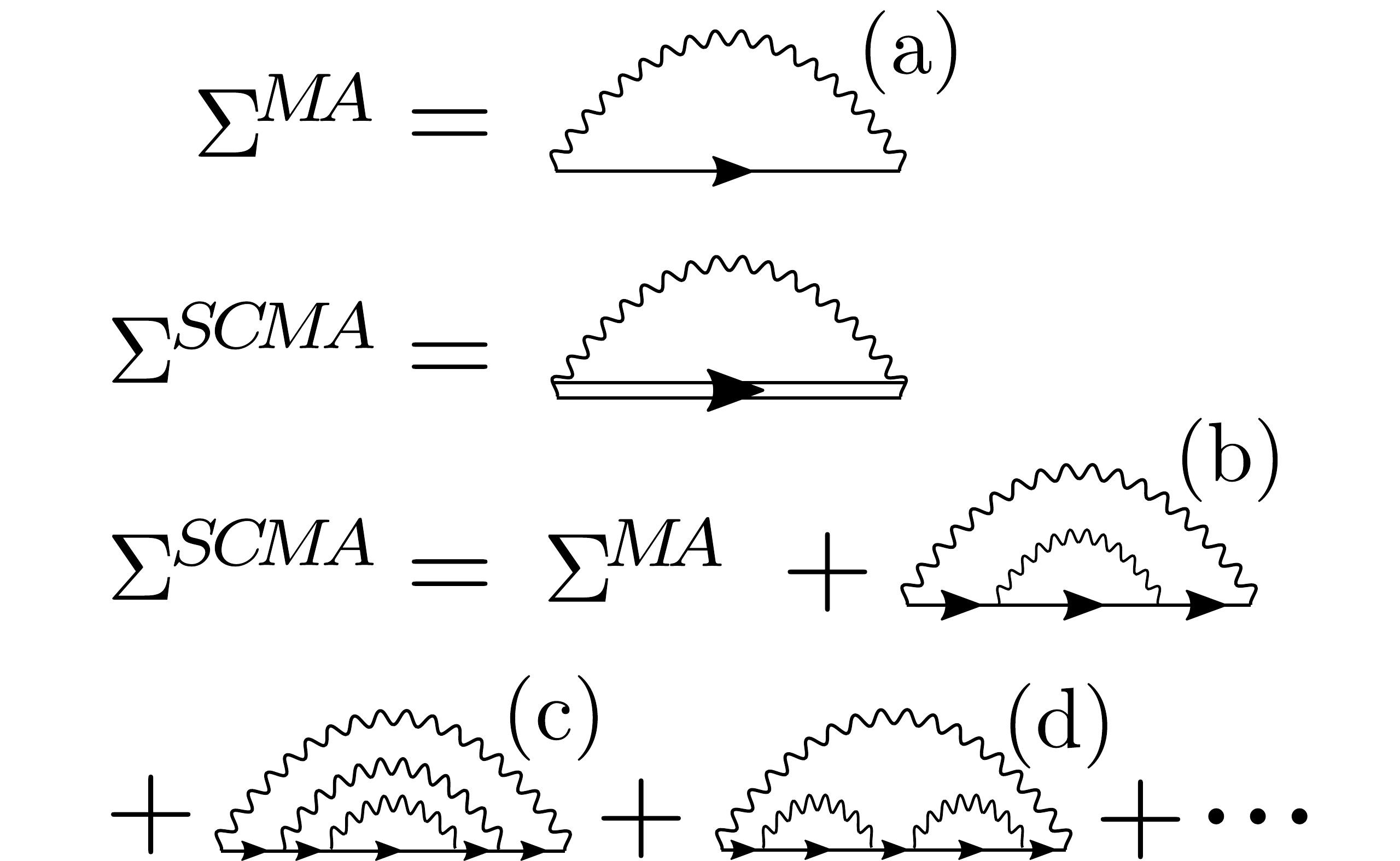} 
 \caption{Feynman diagrams in the Migdal approximation and the  self-consistent Migdal approximation.
 }
 \label{Fig:Feynman_diagrams}
\end{figure}
\subsubsection{Migdal and self-consistent Migdal approximation} 
\label{Subsec:Migdal_approximation}
\poc
The Migdal approximation (MA) \cite{1958_Migdal} is the simplest perturbation approach, whose  self-energy is represented with a single, lowest order Feynman diagram, as shown in Fig.~\ref{Fig:Feynman_diagrams}(a). The imaginary part of the self-energy is given by Eq.~\eqref{Eq:Migdal_ImSigma} in the case when there is just a single electron in the  band,   regardless of the dispersion relation or the number of dimensions of the system. The corresponding real part is obtained using the Kramers-Kronig relations, and in 1D reads as
\begin{align} \label{Eq:Re_Migdal_selfen}
\mathrm{Re} \Sigma^{\mathrm{MA}}(\omega) =&
g^2 (\bose+1) \frac{\theta \left( (\omega-\omega_0)^2 - 4t_0^2 \right) \mathrm{sgn}  (\omega-\omega_0)}{\sqrt{(\omega-\omega_0)^2 - 4t_0^2}} \nonumber \\
+& g^2 \bose\frac{\theta \left( (\omega+\omega_0)^2 - 4t_0^2 \right) \mathrm{sgn}  (\omega+\omega_0)}{\sqrt{(\omega+\omega_0)^2 - 4t_0^2}}.
\end{align}
The range of validity of the Migdal approximation can be extended if we substitute the noninteracting electron propagator in Fig.~\ref{Fig:Feynman_diagrams}(a) with an interacting one. At the same time,  the interacting propagator itself is expressed through the self-energy via the Dyson equation. These relations constitute the self-consistent Migdal approximation.  Figure~\ref{Fig:Feynman_diagrams} illustrates that the SCMA self-energy consists of a series of noncrossing diagrams, whose lowest order coincides with the Migdal approximation. Figure~\ref{Fig:Feynman_diagrams}(b) shows the second-order contribution, while 
the third-order contributions are shown in Figs.~\ref{Fig:Feynman_diagrams}(c)~and~\ref{Fig:Feynman_diagrams}(d). 
\pas
Mathematically, the self-consistency relations are straightforwardly derived and, in our case, read as 
\begin{subequations} \label{Eq:SCMA}
\begin{align}  
\Sigma^{\mathrm{SCMA}}(\omega) &= g^2 (\bose+1) G(\omega - \omega_0)
+ g^2 \bose G(\omega + \omega_0), \label{Eq:SCMA_selfen} \\
G(\omega) &= 
\frac{1}{(2\pi)^n}\int_{-\pi}^{\pi} d^n{\bf k} \frac{1}{\omega - \varepsilon_{\bf k} - 
\Sigma^{\mathrm{SCMA}}(\omega)}, \label{Eq:SCMA_local_Green}
\end{align}
\end{subequations}
where $G(\omega)$ is the local Green's function. We see that in the case of the Holstein model, the SCMA self-energy is ${\bf k}$ independent.

\subsubsection{Dynamical mean-field theory} \label{SubSec:dmft}
\poc
Dynamical mean-field theory (DMFT) is a nonperturbative approximate method, that represents a natural generalization of the traditional mean-field theory \cite{1996_Georges}. It simplifies the original lattice problem by mapping it to a single site impurity problem, embedded into an external bath that is described with a frequency-dependent (i.e., dynamical) field $G_0(\omega)$, which needs to be determined self-consistently. This simplification is reflected on the self-energy, which is assumed to be ${\bf k}$ independent $\Sigma_{\bf k}(\omega) = \Sigma(\omega)$. The DMFT becomes exact in the limit of infinite dimensions or, equivalently, infinite coordination number.
\pas
In practice, $G_0(\omega)$ and $\Sigma(\omega)$ are determined self-consistently, by imposing that the local Green's function of the lattice problem 
\begin{equation} \label{Eq:DMFT_local_green_int_dos}
    G(\omega) = \int_{-\infty}^\infty \frac{\rho(\epsilon)d\epsilon}{\omega - \Sigma(\omega) - \epsilon},
\end{equation}
and the self-energy $\Sigma(\omega)$ coincide with the corresponding quantities of the impurity problem. Here, $\rho(\epsilon)$ is the noninteracting density of states. The self-consistent loop is closed using the Dyson equation $G_0(\omega) = (G^{-1}(\omega)+\Sigma(\omega))^{-1}$. 
\pas
In the case of the Holstein model, the (polaron) impurity problem can be solved exactly,  directly on the real-frequency axis, in terms of the continued fraction expansion  \cite{1997_Ciuchi}. Furthermore, in the one-dimensional case Eq.~\eqref{Eq:DMFT_local_green_int_dos} assumes a closed-form solution and reads as 
\begin{equation} \label{Eq:Analytic_G}
    G(\omega) = \mathrm{Re} \frac{1}{2t_0B(\omega) \sqrt{1-\frac{1}{B(\omega)^2}}} + i \mathrm{Im}
    \frac{-i}{2t_0\sqrt{1-B(\omega)^2}},
\end{equation}
where $B(\omega) = (\omega - \Sigma(\omega))/(2t_0)$; see  Supplemental Material of Ref. \cite{2022_Mitric}. We note that Eq.~\eqref{Eq:Analytic_G} can also be used for the SCMA in Eq.~\eqref{Eq:SCMA}.
\pas
We have very recently shown \cite{2022_Mitric}, by using extensive comparisons with several numerically exact methods covering various parameter regimes, that the DMFT can provide a rather accurate solution for the Holstein polaron even in low dimension. Hence, the DMFT has
emerged as a unique numerical method that gives close to exact spectral functions in practically the whole space of parameters, irrespective of the number of dimensions. This makes the DMFT an ideal benchmark method for comparisons with the CE results for the Holstein model.
\vspace*{-0.4cm}
\section{Spectral functions} \label{Sec:SpecF}
\poc
In this section, we present the  CE spectral functions of the 1D Holstein model. The DMFT is used as a benchmark, while  MA and SCMA represent the main competitors and alternatives to the CE method. Section~\ref{Sec:SpecF_k=0_res} shows the results for $k=0$, whereas heat plots and the $k=\pi$ results are shown in Sec.~\ref{Sec:SpecF_k=pi_res}. High-temperature spectral functions and spectral sum rules are presented in Sec.~\ref{Sec:High_temp}. The behavior near the atomic limit is discussed in Sec.~\ref{SubSec:atomic_limit}. We present only the results for $\omega_0=0.5$, while the results for other phonon frequencies and various momenta are shown in Sec.~II of SM \cite{SuppMat}. The 2D spectral functions are presented in Appendix~\ref{AppSec:2DspecF}.
\vspace*{-0.5cm}
\subsection{Low and intermediate temperatures for \texorpdfstring{$k=0$}{k=0}}
\label{Sec:SpecF_k=0_res}
\begin{figure}[!t]
  \includegraphics[width=3.2in,trim=0cm 0cm 0cm 0cm]{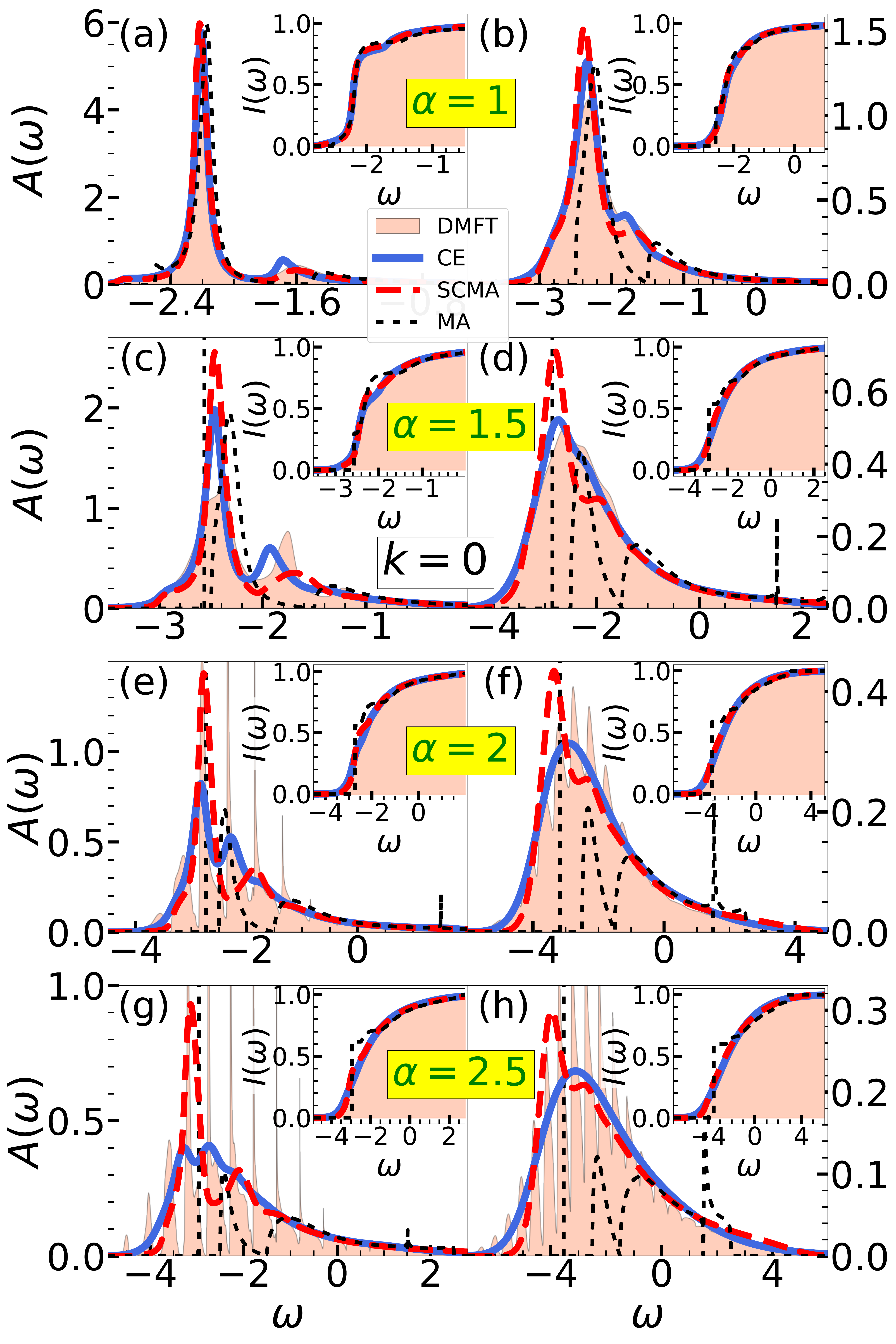} 
 \caption{
 (a)--(h) Spectral functions for $t_0=1$, $\omega_0=0.5$ and $k=0$. In the left panels 
$T = 0.3$, while $T = 0.7$ in the right 
panels. Insets show the integrated spectral weights $I_{\bf k}(\omega) = \int_{-\infty}^\omega  A_{\bf k}(\nu) d\nu$.
 }
 \label{SubFig:SpecF_w_0.5_k_0}
\end{figure}
\poc
In the weak-coupling limit $\alpha \to 0$, all these approximate methods (DMFT, CE, SCMA, MA) provide accurate results. In Fig.~\ref{SubFig:SpecF_w_0.5_k_0}, we investigate  how far from this strict limit each of our methods continues to give reasonably accurate spectral functions. In Fig.~\ref{SubFig:SpecF_w_0.5_k_0}(a), we see that for $\alpha=1$ all methods correctly capture the QP peak, which dominates in the structure of the spectrum. The MA satellite peak is slightly shifted towards higher frequencies, which becomes significantly more pronounced at higher temperatures; see Fig.~\ref{SubFig:SpecF_w_0.5_k_0}(b). The limitations of the MA  become more obvious for stronger couplings, where even the position and weight of the QP peak are inaccurate; see Figs.~\ref{SubFig:SpecF_w_0.5_k_0}(c)--~\ref{SubFig:SpecF_w_0.5_k_0}(h). 
\pas
While the QP properties of the CE and SCMA seem to be quite similar if $\alpha$ is not too large, some difference in satellite peaks is already visible in Figs.~\ref{SubFig:SpecF_w_0.5_k_0}(b)~and~\ref{SubFig:SpecF_w_0.5_k_0}(c). Figure~\ref{SubFig:SpecF_w_0.5_k_0}(c) shows that SCMA gives broader satellites than the DMFT benchmark, whereas CE slightly underestimates the position of the satellite. Neither CE nor SCMA can be characterized as distinctly better in this regime. On the other hand, Figs.~\ref{SubFig:SpecF_w_0.5_k_0}(e)~and~\ref{SubFig:SpecF_w_0.5_k_0}(g) display a clear advantage of the CE. We see that it captures rather well the most distinctive features of the solutions, which are the first few satellites. This is not the case for SCMA. 
\pas
Figures~\ref{SubFig:SpecF_w_0.5_k_0}(f)~and~\ref{SubFig:SpecF_w_0.5_k_0}(h) demonstrate that the CE gives a rather quick crossover toward the high-temperature limit, as it predicts a simple broad one-peak structure for the spectral function already for $T=0.7$. This large difference between the spectral functions for $T_1 = 0.3$ and $T_2 = 0.7$ can be understood by examining the ratio of their corresponding lifetimes $\tau(T_1)/\tau(T_2) = \bose (T_2) / \bose (T_1) \approx 8.5$.  This implies that $\mathrm{Re} C_{\bf k}(t)$ for $T=0.7$ has a much steeper slope as a function time, which suppresses the appearance of satellites, as explained in Sec.~\ref{Sec:Cumulant_for_Holstein}.
\vspace*{-0.2cm}
\subsection{Low and intermediate temperatures for  \texorpdfstring{$k \neq 0$}{k ≠ 0}} \label{Sec:SpecF_k=pi_res}
\begin{figure}[!t]
  \includegraphics[width=3.2in,trim=0cm 0cm 0cm 0cm]{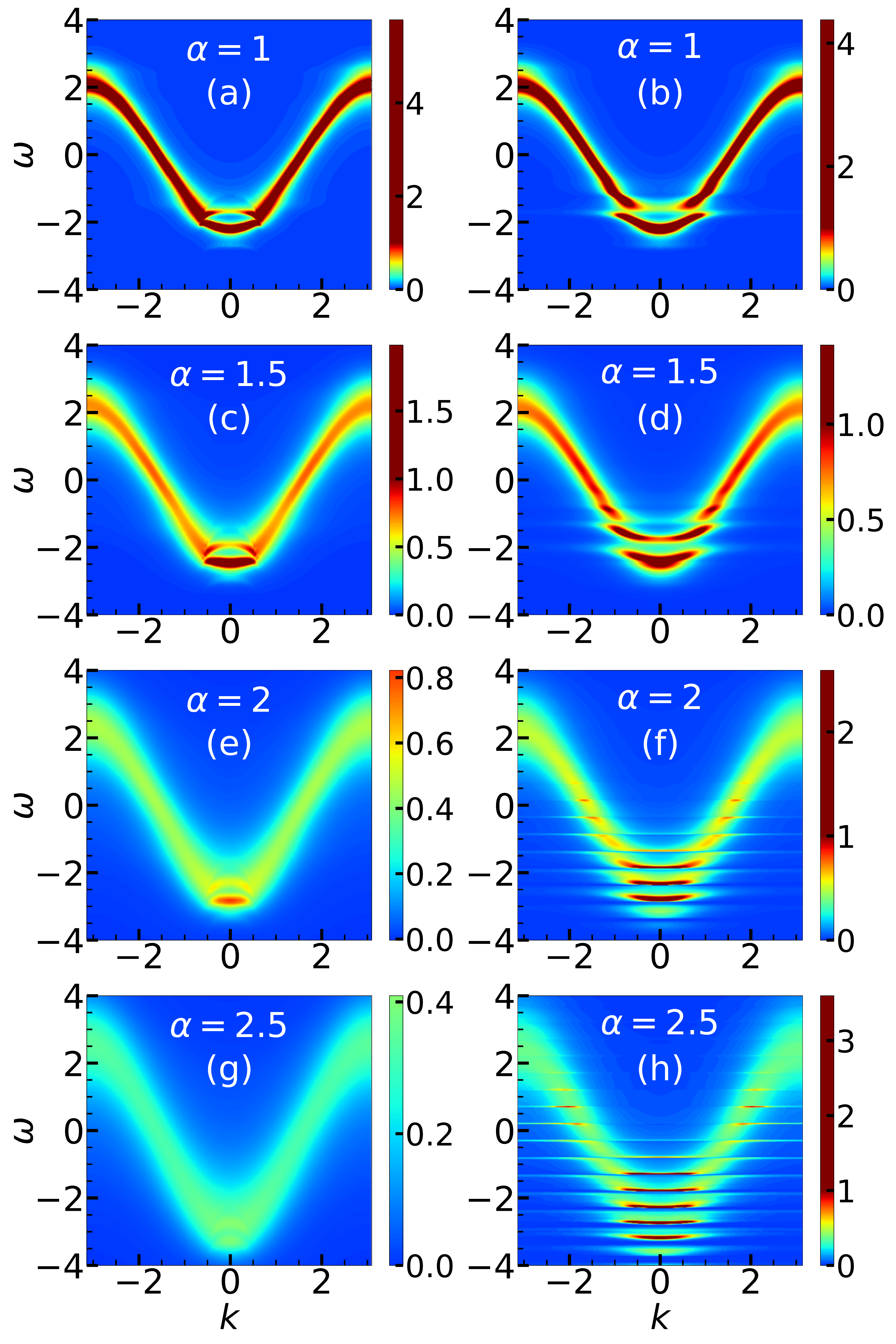} 
 \caption{
 (a)--(h) Heat maps of $A_{\bf k}(\omega)$ for $t_0=1$, $\omega_0=0.5$ and $T=0.3$. In the left panels, we present CE results, while the DMFT benchmark is presented in the right panel. All plots use the same color coding.
 }
 \label{SubFig:HeatMap_w_0.5}
\end{figure}
\begin{figure}[!t]
  \includegraphics[width=3.2in,trim=0cm 0cm 0cm 0cm]{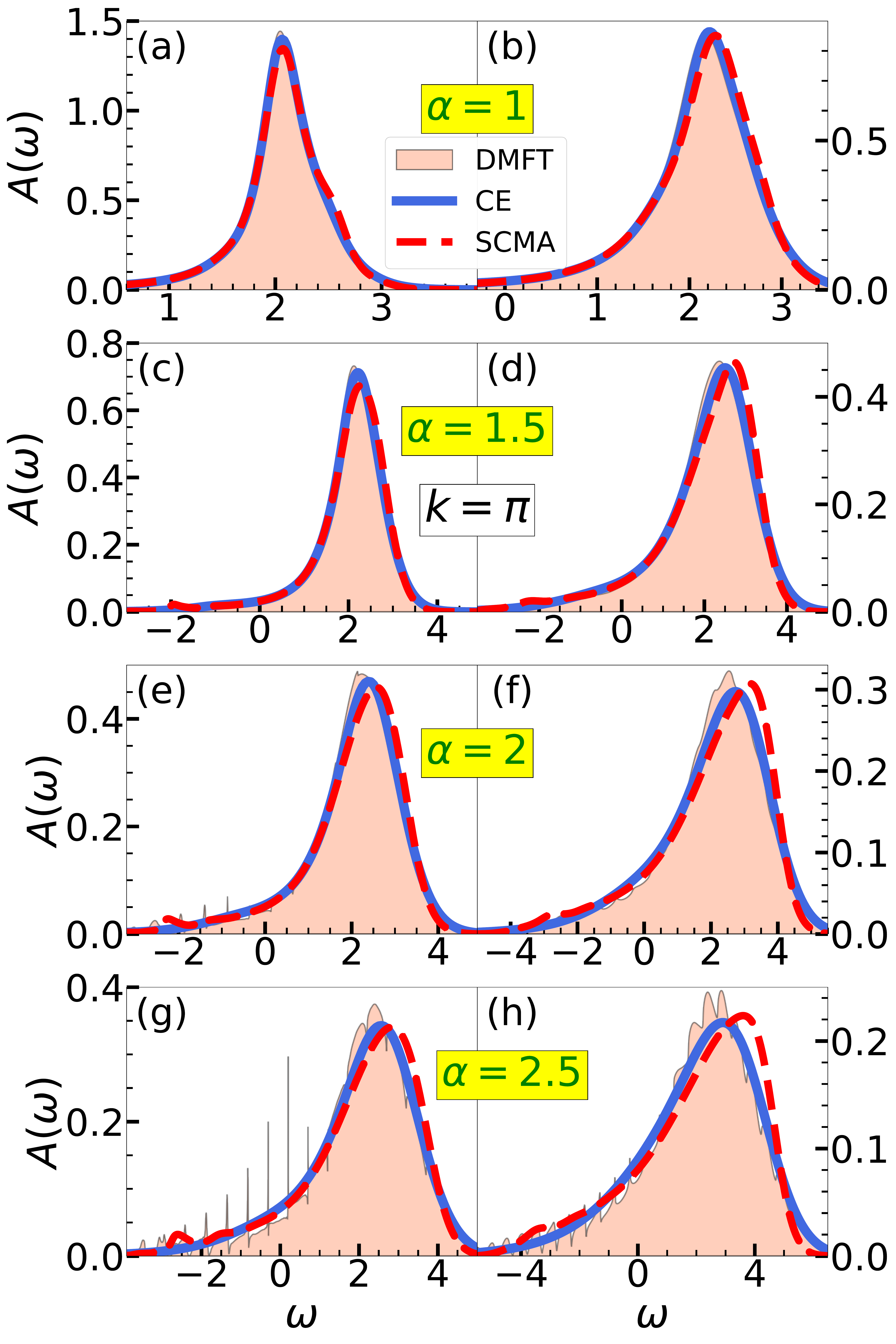} 
 \caption{(a)--(h) Spectral functions for $t_0=1$, $\omega_0=0.5$ and $k=\pi$. In the left panels $T=0.3$, while $T=0.7$ in the right panels.
 }
 \label{Fig:Aw_1D_w=0-5_k=pi}
\end{figure}
\poc
To proceed with the analysis of the CE, we want to answer: i) Whether the conclusions that we reached for $k=0$ can be carried over to other momenta as well?; ii) Does CE continue to be better than SCMA at much higher temperatures? 
\pas
The first question is answered in Fig.~\ref{SubFig:HeatMap_w_0.5}, where we compare CE and DMFT heat plots. Figures~\ref{SubFig:HeatMap_w_0.5}(a)~and~\ref{SubFig:HeatMap_w_0.5}(b) demonstrate that CE results are quite reminiscent of the DMFT results for $\alpha=1$, even at non-zero momenta. The same conclusion holds for weaker couplings as well. On the other hand, there are differences between the results for somewhat stronger coupling $\alpha=1.5$, as shown in Figs.~\ref{SubFig:HeatMap_w_0.5}(c)~and~\ref{SubFig:HeatMap_w_0.5}(d). While the polaron bands in both of these figures are convex, the CE predicts the first satellite to be concave, unlike the DMFT. In other words, CE predicts that the distance between the polaron peak and the satellites decreases, as we increase the momentum. This is counterintuitive, as the satellites are perceived as the QP that absorbed or emitted a phonon, which should consequently be just at energy distance  $\omega_0$ apart. These limitations of the CE are much more pronounced for stronger electron-phonon couplings. While the DMFT solution in Figs.~\ref{SubFig:HeatMap_w_0.5}(f)~and~\ref{SubFig:HeatMap_w_0.5}(h) exhibits a series of distinct bands, Figs.~\ref{SubFig:HeatMap_w_0.5}(e)~and~\ref{SubFig:HeatMap_w_0.5}(g) demonstrate that  the polaron and  satellite bands of the CE  merge into a single band at higher momenta. However, the most noticeable feature here is the fact that the CE is too smeared, as if the temperature is too high. This is a consequence of the fact that the lifetime in Eq.~\eqref{Eq:lifetime_CE}  scales as $\tau_{\bf k}\sim 1/g^2$.
\pas
\begin{figure*}[!t]
  \includegraphics[width=7in,trim=0cm 0cm 0cm 0cm]{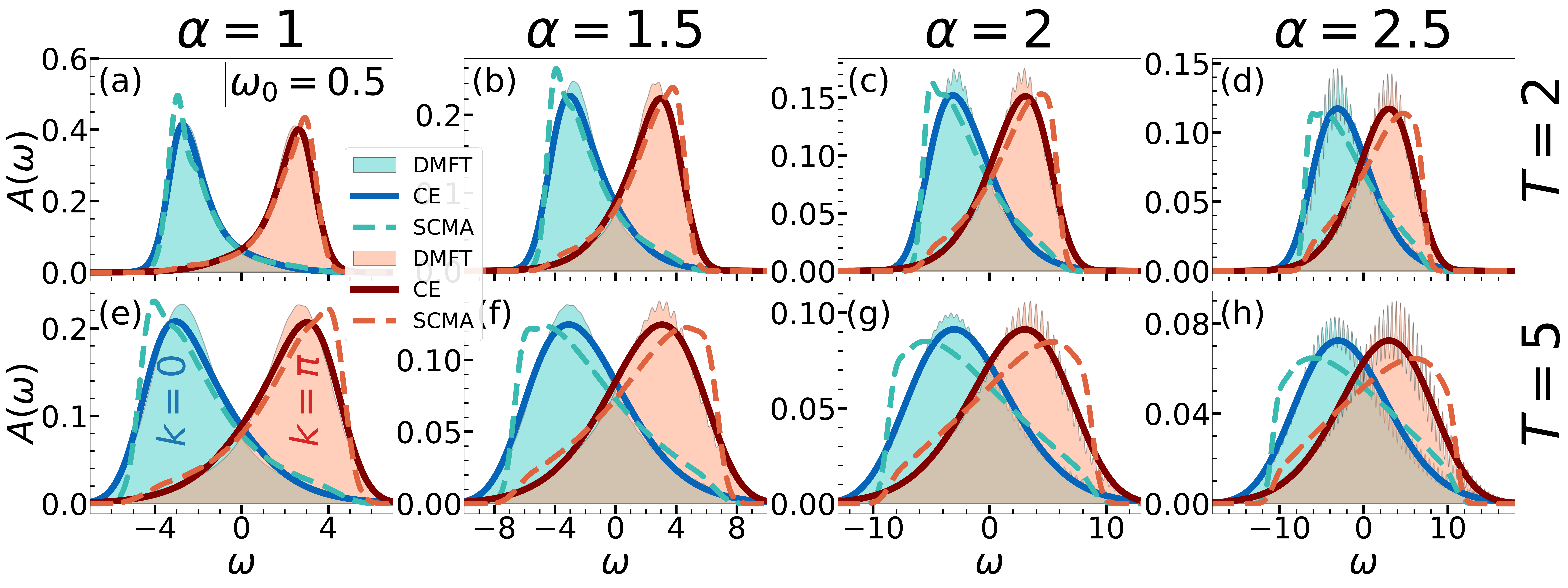} 
 \caption{(a)--(h) CE, DMFT, and SCMA  spectral functions in 1D for $t_0=1$, $\omega_0 = 0.5$, and $k=0,\pi$.
 }
 \label{Fig:highT_w_0.5}
\end{figure*}
While the heat maps reveal noticeable discrepancies between the DMFT and CE for $k \neq 0$, it seems that these differences are much less pronounced around ${k = \pi}$. A more detailed comparison is presented in Fig.~\ref{Fig:Aw_1D_w=0-5_k=pi} that shows the results for the same regimes as in Fig.~\ref{SubFig:SpecF_w_0.5_k_0}. The DMFT solution in Figs.~\ref{Fig:Aw_1D_w=0-5_k=pi}(a)--\ref{Fig:Aw_1D_w=0-5_k=pi}(d) shows that the main feature of the spectral function is a single broad peak  for $\alpha \lesssim 1.5$, which is in agreement with the CE results. This is also the case for the SCMA, although we observe a slight tendency of the main peak to lean toward higher frequencies at higher temperatures. For larger interaction strengths, CE cannot fully reproduce the sharp peaks at lower frequencies of the low-temperature spectral function or the fine structure  of the main peak at higher temperatures; see Figs.~\ref{SubFig:SpecF_w_0.5_k_0}(e)--\ref{SubFig:SpecF_w_0.5_k_0}(h). Similarly, CE  misses the quasiparticle peak as well, situated at low energy, although it is typically tiny and not (clearly) visible in Figs.~\ref{Fig:Aw_1D_w=0-5_k=pi}(a)--\ref{Fig:Aw_1D_w=0-5_k=pi}(h) (see Appendix~\ref{App:Compare_Reichman}). A detailed comparison of the spectral functions for other momenta and phonon frequencies is presented in Sec.~II of the SM~\cite{SuppMat}.
\pas
Overall, we find that the CE gives the most accurate results for $k=0$ and $k=\pi$ and that it is less accurate for other momenta. Although it cannot fully reproduce a tiny quasiparticle peak for $k=\pi$, it describes well a wide single-peak structure, which is the most prominent feature of the spectrum. A much larger discrepancy for $k=\pi$, between the CE and a reliable benchmark,  was reported in Ref.~\cite{2022_Robinson_1}, by examining the system on a finite lattice system with $N=6$. In Appendix~\ref{App:Compare_Reichman}, we examine the same parameter regime as in Ref. \cite{2022_Robinson_1} and show that these discrepancies are significantly reduced in the thermodynamic limit.
\subsection{Spectral functions at high temperatures and spectral sum rules} \label{Sec:High_temp}
\poc
In Fig.~\ref{Fig:highT_w_0.5}, we show CE, SCMA, and DMFT spectral functions at high temperatures, for the same electron-phonon couplings as in Figs.~\ref{SubFig:SpecF_w_0.5_k_0}~and~\ref{Fig:Aw_1D_w=0-5_k=pi}. We see that CE performs very well, both for $k=0$ and $k=\pi$. There are only small discrepancies at stronger interactions (see, e.g., Fig.~\ref{Fig:highT_w_0.5}(c)). In contrast, the SCMA solution gets tilted relative to the DMFT and CE. In addition, it poorly reproduces the low-frequency part of the spectrum.
\pas
It is not obvious whether the CE method is exact in the high-temperature limit $T \to \infty$. As we now demonstrate, this can be answered by examining the spectral sum rules:
\begin{equation} \label{Eq:sum_rule_def}
    \mathcal{M}_n ({\bf k}) = \int_{-\infty}^\infty A_{\bf k}(\omega) \omega^n d\omega.
\end{equation}
These can be calculated both exactly 
\begin{equation}
    \mathcal{M}_n^{\mathrm{exact}} ({\bf k}) = \left\langle 
    \underbrace{\left[\dots \left[\left[ c_{\bf k}, H\right],H\right] \dots ,H\right]}_{n\; \mathrm{times}} c_{\bf k}^\dagger
    \right\rangle_{T},
\end{equation}
and within the CE approximation, where by combining Eqs.~\eqref{Eq:CE_specF_def}~and~\eqref{Eq:sum_rule_def} we find
\begin{multline}
\mathcal{M}_n^{\mathrm{CE}}({\bf k}) = \mathrm{Re} \left[ 
    i^n \left( \frac{d}{dt} \right)^n e^{C_{\bf k}(t)} 
    \right] \bigg\rvert_{ t=0} \\
    -
    \sum_{p=1}^n {\binom{n}{p}} (-\varepsilon_{\bf k})^p \mathcal{M}_{n-p}^{\mathrm{CE}}({\bf k}).
\end{multline}
The difference between these quantities $\mathcal{M}_n^{\mathrm{CE}} ({\bf k}) - \mathcal{M}_n^{\mathrm{exact}} ({\bf k})$ is zero for $n=0$ and $n=1$, as noted in Sec.~\ref{SubSubSec:General_theory}. Higher order sum rules for the CE method are easily calculated, while the evaluation of the exact sum rules quickly becomes cumbersome for increasing $n$. The first five ($0\leq n \leq 4$) sum rules were already calculated by Kornilovitch \cite{2003_Kornilovitch}: 
\begin{subequations} \label{Eq:2_sum_rule}
\begin{align}
    \mathcal{M}_2 ({\bf k}) &=  \varepsilon_{\bf k}^2 + (2\bose + 1) g^2,  \\
    \mathcal{M}_3 ({\bf k}) &=  \varepsilon_{\bf k}^3 + g^2 \omega_0 + 2g^2(2\bose + 1)  \varepsilon_{\bf k}, \\
    \mathcal{M}_4 ({\bf k}) &=  \varepsilon_{\bf k}^4 + 
    2g^2 \varepsilon_{\bf k} \omega_0 + g^2(2\bose+1)  (2t_0^2+3\varepsilon_{\bf k}^2 + \omega_0^2) \nonumber \\
    &+ 3g^4(2\bose+1)^2.
\end{align}
\end{subequations}
All of these are correctly predicted by the CE. However, the disagreement between $\mathcal{M}_n^{\mathrm{exact}}$ and $\mathcal{M}_n^{\mathrm{CE}}$ appears for $n=5$, where we find
\begin{subequations} 
\begin{align}
    \mathcal{M}_5^{\mathrm{exact}} ({\bf k}) &=  \varepsilon_{\bf k}^5 + 3g^2 \omega_0 (2t_0^2+\varepsilon_{\bf k}^2) + g^2 \omega_0^3  \nonumber \\
    &+ 2g^2 \left(
    2 \varepsilon_{\bf k}^3 \! + \! 5g^2 \omega_0 \! +  \! \varepsilon_{\bf k}\omega_0^2 \! + \! 2t_0^2 \varepsilon_{\bf k}
    \right) (2\bose \! + \! 1) \nonumber \\
    & + 7g^4 \varepsilon_{\bf k}(2\bose+1)^2, \\
     \mathcal{M}_5^{\mathrm{CE}} ({\bf k}) &= 
     \mathcal{M}_5^{\mathrm{exact}} ({\bf k}) 
     -2g^4 \varepsilon_{\bf k}(2\bose+1)^2. \label{Eq:sum_rules_CE_minus_exact}
\end{align}
\end{subequations}
Hence, CE cannot be exact in the limit $T \to\infty$. However, we see that there are two limits where CE can potentially be exact: the weak-coupling limit $g\to 0$ and the atomic limit $\varepsilon_{\bf k} \to 0$. It turns out that CE is actually exact in both of these limits, as seen from  Eqs.~\eqref{Eq:Cumulant_def},~\eqref{Eq:Cumulant_expression},~and~\eqref{Eq:Migdal_ImSigma} for the weak-coupling and  Sec.~\ref{SubSec:atomic_limit} for the atomic limit. We note that the SCMA gives correct sum rules only for $n \leq 3$ \cite{2006_Berciu}. This is a consequence of the fact that SCMA ignores one of the fourth-order diagrams ($\sim g^4$) since it includes only the non-crossing diagrams. Also, we numerically checked that the DMFT results are in agreement with all of the sum rules that we listed above.
\pas
\subsection{Atomic limit} \label{SubSec:atomic_limit}
\begin{figure}[!t]
  \includegraphics[width=3.2in,trim=0cm 0cm 0cm 0cm]{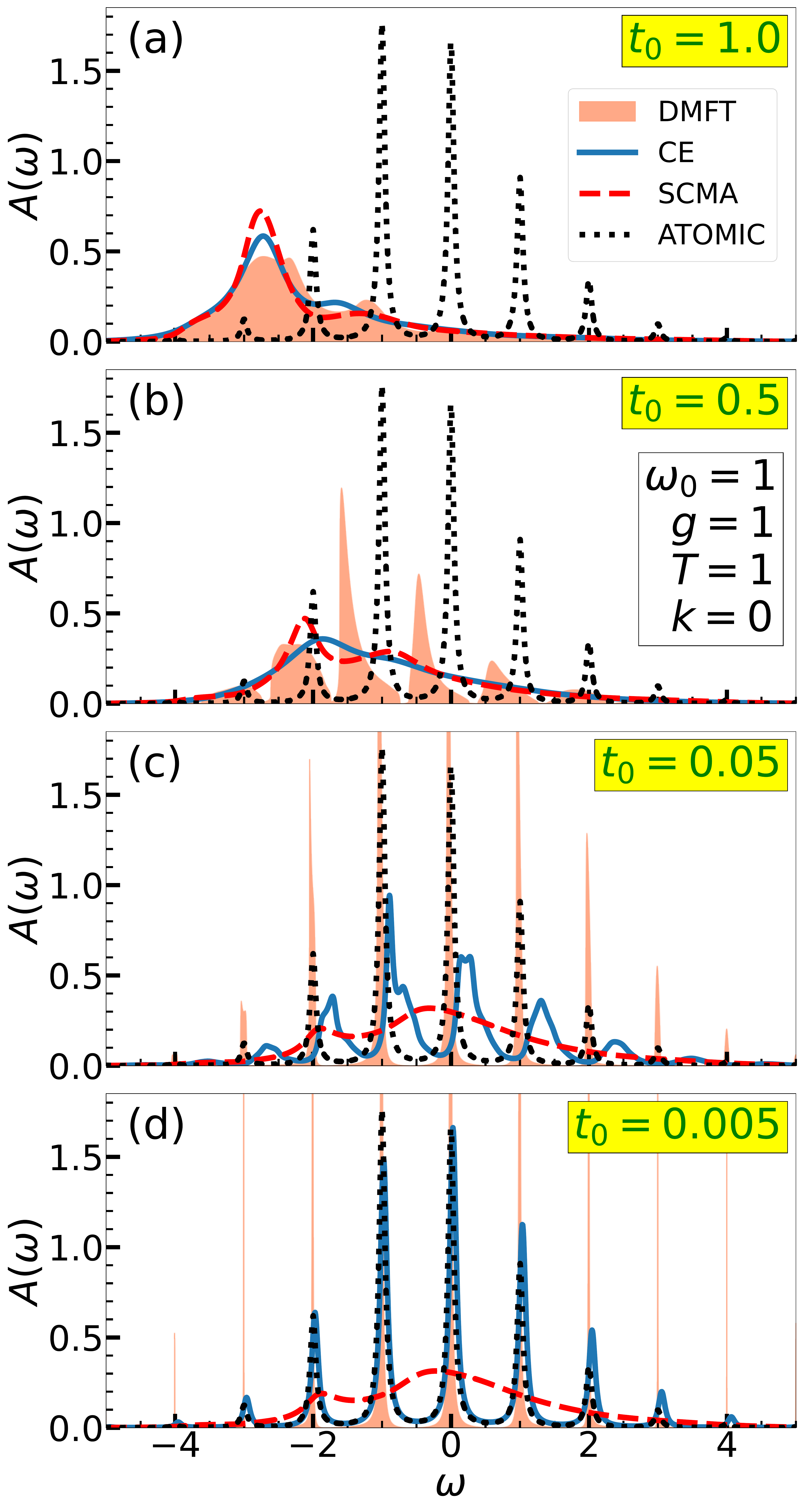} 
 \caption{(a)--(f) CE, DMFT, and SCMA  spectral functions close to the atomic limit. Here, we use artificial Lorentzian broadening with half width set to $\eta=0.05$.
 }
 \label{Fig:atomicSpecF}
% \vspace*{0.5cm}
\end{figure}

\poc
In the atomic limit ($t_0=0$), the cumulant function can be evaluated exactly
\begin{equation} \label{Eq:CF_in_atomic}
C(t) = \alpha^2 (-2\bose-1 +it\omega_0 + iD(t)).
\end{equation}
This follows from Eq.~\eqref{Eq:Cumualnt_wrt_Bessel}, using $J_0(0)=1$. If we express the phonon propagator as $iD(t) = 2\sqrt{\bose(\bose+1)}  \cos \left[ \omega_0 \left(t + \frac{i}{2T} \right) \right]$ and use the modified Jacobi-Anger identity
\begin{multline}
     e^{2\alpha^2 \sqrt{\bose(\bose+1)}
    \cos \left[ \omega_0 \left(t + \frac{i}{2T} \right) \right]} \\=
    \sum_{l=-\infty}^\infty  I_l \left( 2\alpha^2 \sqrt{\bose(\bose+1)} \right)
    e^{-i l \omega_0 t} e^{\frac{l\omega_0}{2T}},
\end{multline}
where $I_l$ are the modified Bessel function of the first kind,  the spectral function (see Eqs.~\eqref{Eq:Cumulant_def}~and~\eqref{Eq:CE_specF_def}) can be calculated analytically and reads as
\begin{multline} \label{Eq:CE_atomit_finite_T}
    A(\omega) = e^{-\alpha^2 (2\bose+1)} \\
    \times \sum_{l=-\infty}^{\infty}
    I_l \left(2\alpha^2 \sqrt{\bose(\bose+1)}  \right) e^{\frac{l\omega_0}{2T}}
    \delta(\omega + \alpha^2 \omega_0 - l \omega_0).
\end{multline}
In the limit $T \to 0$, the previous expression reduces to 
\begin{equation}  \label{Eq:CE_atomit_zero_T}
A(\omega) = e^{-\alpha^2} \sum_{l=0}^\infty \frac{\alpha^{2l}}{l!} 
\delta (\omega + \omega_0 (\alpha^2 - l)).
\end{equation}
This proves that CE gives correct results in the atomic limit, as Eqs.~\eqref{Eq:CE_atomit_finite_T}~and~\eqref{Eq:CE_atomit_zero_T} coincide with the known exact results \cite{2000_Mahan,2019_Bonca}.
\newpage
\poc
In contrast, the SCMA (let alone the MA) does not share this property, which is easy to show at zero temperature. In this case, Eq.~\eqref{Eq:SCMA_selfen} and the Dyson equation imply that 
\begin{equation}
G(\omega) = \frac{1}{\omega - g^2 G(\omega-\omega_0)}.
\end{equation}
The previous equation can be solved by the iterative application of itself in terms of the continued fraction 
\begin{equation}
G(\omega) = \frac{1}{\omega - \frac{g^2}{\omega - \omega_0 - 
\frac{g^2}{\omega - 2\omega_0-\frac{g^2}{\omega-3\omega_0 - \dots}}}}.
\end{equation}
This does not coincide with Eq.~(35) from Ref.~\cite{1997_Ciuchi}, which 
represents the exact solution. Thus, SCMA cannot reproduce the correct result in the atomic limit.
\pas
While the CE is exact in the atomic limit ($t_0 = 0$), it is not immediately obvious how far from this limit it continues to give reliable results. This is why we now examine the regimes with small hopping parameter $t_0$.  Since the lifetime is infinitely large in some of these regimes (see Fig.~\ref{Fig:RC_lifetime}), we  introduce artificial attenuation $\eta$ for the Green's function in real time by making a replacement $G(t)\to G(t)e^{-\eta t}$. The results are presented in Fig.~\ref{Fig:atomicSpecF}. Here, the dotted line is the analytic solution in the atomic limit ($t_0=0$), determined by  Eq.~\eqref{Eq:CE_atomit_finite_T}, where the Dirac delta functions have been replaced by Lorentzians of half width  $\eta$. It is used as a measure to see how far the regime we are examining is from the exact atomic limit. In Fig.~\ref{Fig:atomicSpecF}(a), we see that DMFT, SCMA, and CE spectral functions are in agreement. This regime is quite far from the atomic limit, as indicated by the dotted line. Figure~\ref{Fig:atomicSpecF}(b) shows that the DMFT spectral function already consists of a series of peaks for $t_0 = 0.5$, while the CE and SCMA spectral functions are too flattened out. While the CE solution significantly improved in Fig.~\ref{Fig:atomicSpecF}(c), it is still not giving satisfactory results, even though the DMFT suggests that we are already  close to the atomic limit. Only for $t_0 \lesssim 0.005$  does the CE solution give accurate results; see Fig.~\ref{Fig:atomicSpecF}(d). However, this is practically already at the atomic limit. It is interesting to note that while both the DMFT and the CE are exact in the weak-coupling and in the atomic limit, their behavior in other regimes can be quite  different. 

\section{Quasiparticle properties} \label{Sec:QP_prop}
\poc
We now investigate the quasiparticle properties obtained from  the CE method and compare them extensively to the results obtained from  the DMFT and SCMA. We note that the lifetime within the CE was already studied in Sec.~\ref{Sec:lifetime}, so we supplement that study here with the results for the  ground-state energy and the effective mass. Here we show the results in one, two, and three dimensions. Comparison with the MA ground-state energy, in the 1D case, is presented in Sec.~III of the SM \cite{SuppMat}.

%\vspace*{1.4cm}

\subsection{Ground-state energy}
\poc
The polaron band dispersion $E_{p,\bf k}$  within the CE is given by the real part of  Eq.~\eqref{Eq:QP_perturbation}, where the self-energy is taken  in the Migdal approximation:
\begin{equation} \label{Eq:polaron_en_def}
    E_{p,\bf k} = \varepsilon_{\bf k} + \mathrm{Re}\Sigma^{\mathrm{MA}}(\varepsilon_{\bf k}).
\end{equation}
Since we deal with a single electron in the band, the ground-state energy $E_{p}$ is given by $ E_{p,{\bf k}=0}$ evaluated at zero temperature. In the 1D case, $E_p$ is straightforwardly evaluated using Eq.~\eqref{Eq:Re_Migdal_selfen} and reads as follows:
\begin{equation} \label{Eq:1D_ground_state_energy}
    E^{\mathrm{1D}}_p = -2t_0 -\frac{\alpha^2 \omega_0^2}{\sqrt{\omega_0^2 + 4 \omega_0 t_0}}.
\end{equation}
For the expression in higher dimensions, we need to go back to Eq.~\eqref{Eq:Migdal_ImSigma} that holds in any number of dimensions. At $T=0$, it  reads as
\begin{equation}
    \mathrm{Im}\Sigma^{\mathrm{MA}}(\omega) = -\pi \alpha^2 \omega_0^2 \rho (\omega - \omega_0).
\end{equation}
The real part of $\Sigma^{\mathrm{MA}}(\omega)$, which we are interested in, is obtained using the Kramers-Kronig relation
\begin{equation}
    \mathrm{Re}\Sigma^{\mathrm{MA}}(\omega) = \pi \alpha^2 \omega_0^2 \mathcal{H}[\rho](\omega-\omega_0),
\end{equation}
where $\mathcal{H}[\rho](\omega) = \mathcal{P} \int_{-\infty}^\infty \frac{d\nu}{\pi} \frac{\rho(\nu)}{\omega - \nu}$ is the Hilbert transform of the density of states $\rho(\omega)$ and $\mathcal{P}$ is the Cauchy principle value. The evaluation of the Hilbert transform may be reduced to the evaluation of the Fourier transform $\mathcal{F}$, using the following identity:
\begin{equation} \label{Eq:Fourier_identity}
    \mathcal{F}^{-1} \mathcal{H}[\rho](t) = -i \; \mathrm{sgn}(t) \; \mathcal{F}^{-1} [\rho](t).
\end{equation}
The inverse Fourier transform of the density of states on the right-hand side was already calculated in  Eq.~\eqref{Eq:DOS_Fourier} for the case of the hypercubic lattice with the nearest neighbor hopping. Hence, $\mathcal{H}[\rho](\omega)$ is obtained by applying $\mathcal{F}$ on both sides of Eq.~\eqref{Eq:Fourier_identity},
\begin{equation}
    \mathcal{H}[\rho](\omega) = \frac{1}{\pi} \int_0^\infty dx 
    J_0 (2t_0 x)^n \sin (x\omega),
\end{equation}
where $n$ is the number of dimensions. The polaron  band dispersion  then  reads as
\begin{equation} \label{Eq:polaron_en_integral}
    E_{p, {\bf k}} = \varepsilon_{\bf k} + \alpha^2 \omega_0^2  \int_0^\infty dx 
    J_0 (2t_0 x)^n \sin \left(x  (\varepsilon_{\bf k} -\omega_0) \right).
\end{equation}
$E_p $ is thus a linear function with respect to $\alpha^2$, whose intercept is  $\varepsilon_{\bf k}$, while its slope can be calculated accurately using  the numerical scheme described in Appendix~\ref{Apendix:Numerical_RC}.
\begin{figure}[!t]
  \includegraphics[width=3.6in,trim=0cm 0cm 0cm 0cm]{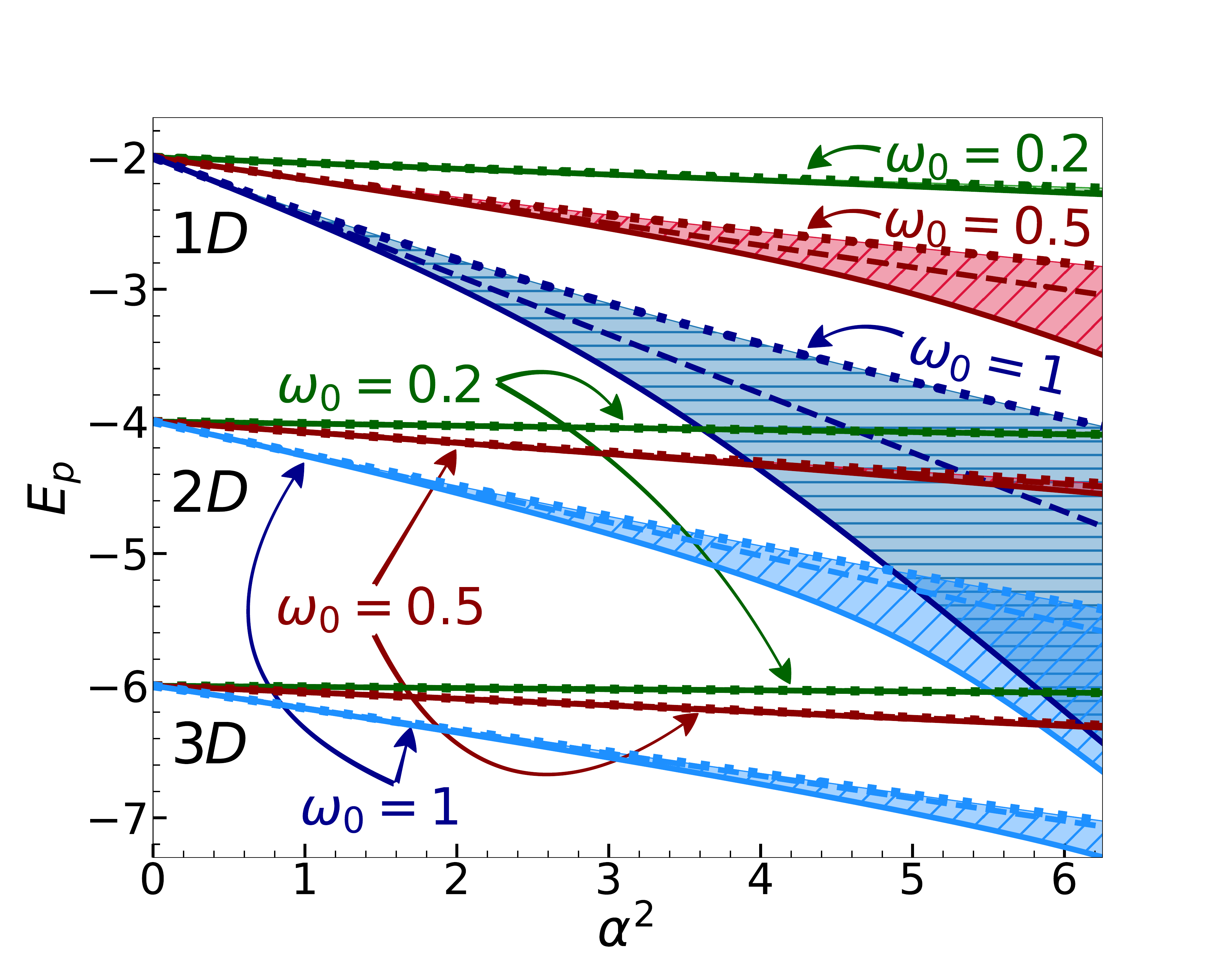} 
 \caption{Ground-state energy within the DMFT (solid line), CE (dashed line), and SCMA (dotted line). Here, $t_0=1$ and $T=0$.
 }
 \label{Fig:GS_en}
\end{figure}
In the 2D case, it  admits an analytical solution 
\begin{equation} \label{Eq:2D_ground_state_energy}
    E^{\mathrm{2D}}_p = -4t_0 - \frac{2\alpha^2 \omega_0^2}{\pi (4t_0+\omega_0)} K \left( \frac{4t_0}{4t_0+\omega_0} \right),
\end{equation}
where $K(k) = \int_0^{\pi/2} d\theta / \sqrt{1-k^2 \sin^2 \theta}$ is the complete elliptic integral of the first kind. In the case  $n=3$, the integral in Eq.~\eqref{Eq:polaron_en_integral} does not admit a closed-form solution and thus requires numerical calculation.
\pas
The polaron band dispersion $E_{p,{\bf k}}$ (and thus the ground-state $E_p$) within the DMFT and SCMA is obtained numerically, as the smallest solution of the following equation:
\begin{equation}
  E_{p,{\bf k}} = \varepsilon_{\bf k} + \mathrm{Re}\Sigma(E_{p,{\bf k}}).
\end{equation}
Results for the 1D, 2D, and 3D case are presented in Fig.~\ref{Fig:GS_en}. The DMFT benchmark, which is known to be very accurate \cite{2022_Mitric}, always gives the lowest ground-state energy predictions in comparison to the CE and SCMA. We see that CE always outperforms the SCMA, despite the fact that its predictions of the energy are always a linear function of $\alpha^2$. In the 1D case, we see that CE results for $\omega_0=0.5$  start to deviate more significantly from the DMFT just around $\alpha = 2.5$. Hence, the range of validity for the CE is similar as for the spectral functions in Fig.~\ref{SubFig:SpecF_w_0.5_k_0}. The analogous conclusions can also be drawn from $\omega_0=1$ data as well. In contrast, all three methods seem to be in agreement for $\omega_0=0.2$ in the whole range of presented values of $\alpha$. This is a consequence of the fact that the ground-state energy correction is small, as seen from  Eqs.~\eqref{Eq:1D_ground_state_energy},\eqref{Eq:polaron_en_integral}~and~\eqref{Eq:2D_ground_state_energy} by fixing $\alpha$ and decreasing $\omega_0$. However, if we fix $g=\omega_0 \alpha$ and then decrease $\omega_0$, the ground-state energy would change substantially (see, e.g., Eq.~\eqref{Eq:1D_ground_state_energy}), and the CE would certainly give poorer results. %In Sec.~III of the SM \cite{SuppMat} we also present the comparison with the MA.
\pas
Similar trends are observed in higher dimensions as well. Seemingly, the range of validity of the CE is increased in higher dimensions. However, one should keep in mind that the hopping parameter is always taken to be unity, which means that the bandwidth of the 2D and 3D systems are, respectively, two and three times larger than their 1D counterpart. Therefore, the correlation is weaker for a given coupling $\alpha$.

\clearpage
\newpage
\subsection{Effective mass}
\poc
Around the bottom ($|{\bf k}| \approx 0$) of the conduction band, the dispersion $E_{p,{\bf k}}$ assumes the following parabolic form:
\begin{equation}
    E_{p, {\bf k}} \approx \mathrm{const} + \frac{{\bf k}^2}{2m^*},
\end{equation}
where $m^*$ is the effective mass, which we now calculate.
\pas
In the 1D case, one obtains the analytical result for the effective mass using Eqs.~\eqref{Eq:Re_Migdal_selfen}~and~\eqref{Eq:polaron_en_def},
\begin{equation} \label{Eq:1D_effective_mass_T=0}
\frac{m^*}{m_0}\bigg\rvert_{\mathrm{1D}, T=0} = \frac{1}{1-\frac{  (2t_0+\omega_0)\alpha^2 \sqrt{\omega_0}}{ (4t_0+\omega_0)^{3/2}}},
\end{equation}
where $m_0 = 1/(2t_0)$ is the band mass which remains the same irrespective of the number of dimensions. Results for the higher number of dimensions are evaluated using Eq.~\eqref{Eq:polaron_en_integral}. As for the ground-state energy, the 2D case  admits an  analytic solution
\begin{equation} \label{Eq:2D_effective_mass_T=0}
\frac{m^*}{m_0}\bigg\rvert_{\mathrm{2D}, T=0} = \frac{1}{1-\frac{2 \alpha^2 \omega_0 }{ \pi (8t_0+\omega_0) } E\left( \frac{4t_0}{4t_0+\omega_0} \right)},
\end{equation}
where $E(k) = \int_0^{\pi/2} d\theta \sqrt{1-k^2 \sin^2 \theta}$ is the complete elliptic integral of the second kind. Results in the  $n$-dimensional case are given by 
\begin{equation}  \label{Eq:nD_effective_mass_T=0}
    \frac{m^*}{m_0}\bigg\rvert_{ T=0} = \frac{1}{1+\pi \alpha^2 \omega_0^2 \frac{d\mathcal{H}[\rho]}{d \omega}\big\rvert_{\omega = -2nt_0-\omega_0} },
\end{equation}
and require numerical calculation in the general case. From Eq.~\eqref{Eq:nD_effective_mass_T=0} we see that  $m_0/m^*$ is a linear function of $\alpha^2$. This linear behavior has to break down at one point, as $m_0 /m^*$ cannot be negative. This happens for strong interaction, where the CE is certainly not expected to be reliable. 
\pas
\begin{figure}[!t]
  \includegraphics[width=3.19in,trim=0cm 0cm 0cm 0cm]{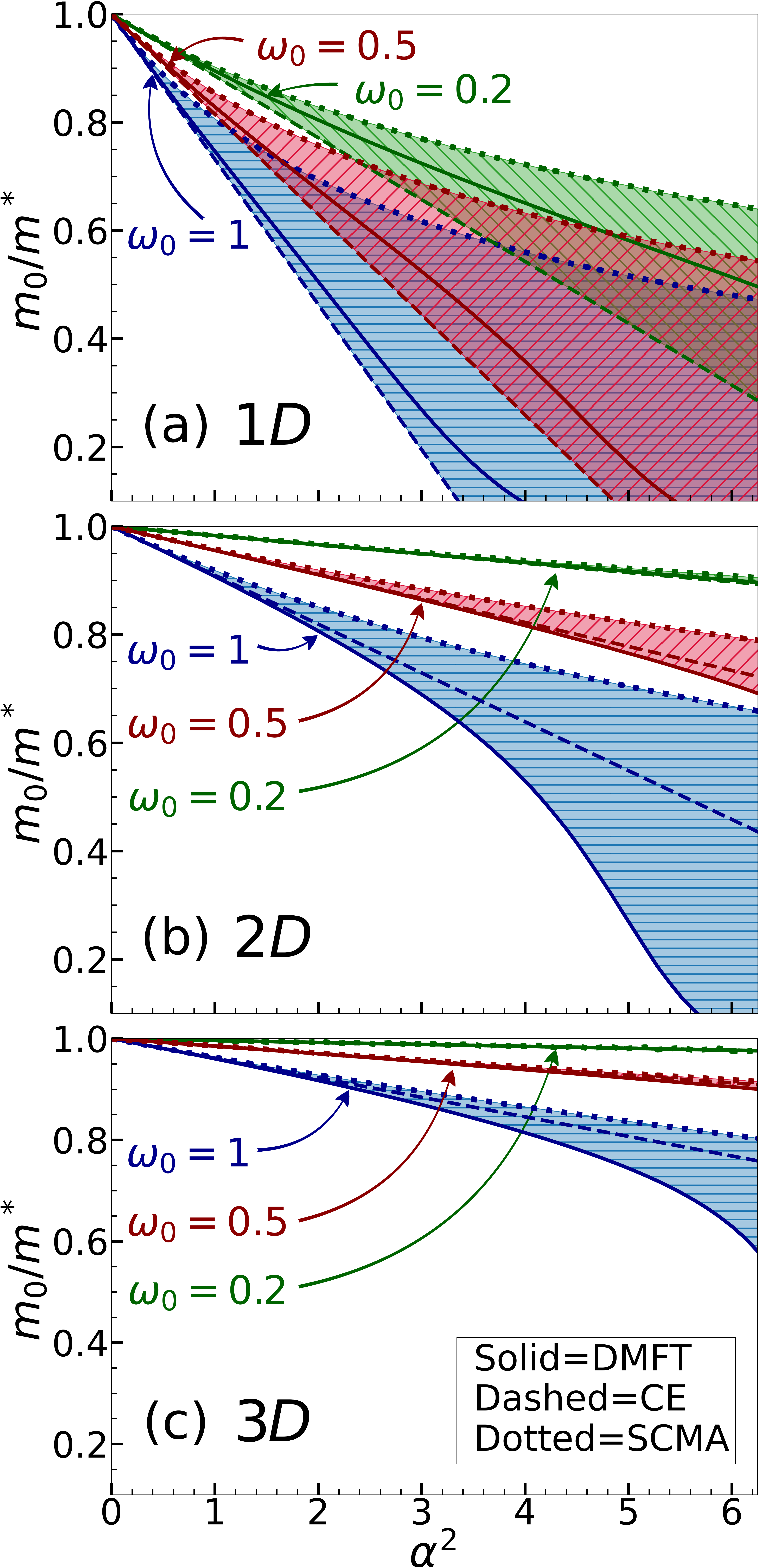} 
 \caption{Effective mass results within the DMFT, CE, and SCMA for $t_0=1$ and $T=0$.
 }
 \label{Fig:mass_renorm}
\end{figure}
The mass renormalization within the DMFT and SCMA is calculated numerically as 
\begin{equation}
     \frac{m^*}{m_0}\bigg\rvert_{ T=0} = 1-\frac{d\Sigma (\omega)}{d\omega} \bigg\rvert_{\omega = E_p}, 
\end{equation}
where $E_p$ is the ground-state energy. Results for the DMFT, CE, and SCMA effective mass, in different parameter regimes and for different number of dimensions, are presented in Fig.~\ref{Fig:mass_renorm}. In the 1D case, we see that the CE always underestimates, while the SCMA overestimates the results from the DMFT benchmark. Still, CE clearly outperforms the SCMA for  $\omega_0=1$ and $\omega_0=0.5$, while the results in the vicinity of the adiabatic limit ($\omega_0=0.2$) seem to be equally well (poor) represented by both methods.
%
% \pas
% %
% The presented results provide valuable insight into the results from Figs.~\ref{Fig:GS_en}~and~\ref{SubFig:SpecF_w_0.5_k_0} that we already examined.  
% For example, while Fig.~\ref{Fig:GS_en} suggested that  the 1D case around $\alpha\approx 2.5$ for $\omega_0=0.2$   is weakly renormalized and thus uninteresting, we now see that its effective mass is actually $2$ times larger than the corresponding band mass. Similar analysis can be conducted for other regimes as well.
%
%\pas
%
\newpage
In the higher-dimensional case, we see that the CE is always a clearly better approximation than the SCMA, while both of them overestimate the DMFT predictions. As for the ground-state energy, we emphasize again that the hopping parameter was set to $1$. As a consequence, the system has a larger bandwidth in the higher-dimensional case and, therefore, the correlations are weaker. 
\vfill

%\clearpage
%\newpage
\section{Mobility} \label{Sec:Mobility_diff_methods}
\poc
The mobility is defined as the DC conductivity, normalized to the concentration of charge carriers $n_e$ (and their unit charge which we set to $e=1$), i.e., ${\mu = \sigma^{\mathrm{DC}} / n_e}$. It can be calculated using the Kubo formalism, which relates $\mu$ to the current-current correlation function \cite{2000_Mahan}. The latter can be written as a sum of the so-called bubble part, which is completely determined by the spectral functions $A_{\bf k}(\omega)$, and the vertex corrections. Within the DMFT, the vertex corrections vanish \cite{1996_Georges,1990_Khurana}, while estimating their  contribution in the general case is beyond the scope of this paper. In the following, we calculate the mobility solely from the bubble part. 
\pas
In the case of a 1D system with a single spinless electron in the band, the mobility in the bubble approximation can be written as \cite{2000_Mahan,2001_Fratini}
\begin{equation} \label{Eq:mobility_from_spectral}
    \mu = \frac{4\pi t_0^2}{T} 
    \frac{\sum_k \int_{-\infty}^\infty d\nu A_k (\nu)^2 e^{-\nu/T} \sin^2 k}{\sum_k \int_{-\infty}^\infty d\nu A_k (\nu) e^{-\nu/T}}.
\end{equation}
The processing time required for the calculation of $\mu$ within the CE method rises linearly with the number of $k$-points we sum over.  This is not the case for the DMFT and SCMA, as their self-energies are $k$ independent, and thus need to be calculated only once for a given parameter set. In every parameter regime the CE was applied to, we checked that $64$ sampling points in the Brillouin zone are enough to be representative of the thermodynamic limit. This was also crosschecked using the DMFT.
 \pas
\begin{figure}[!t]
  \includegraphics[width=3.45in,trim=0cm 0cm 0cm 0cm]{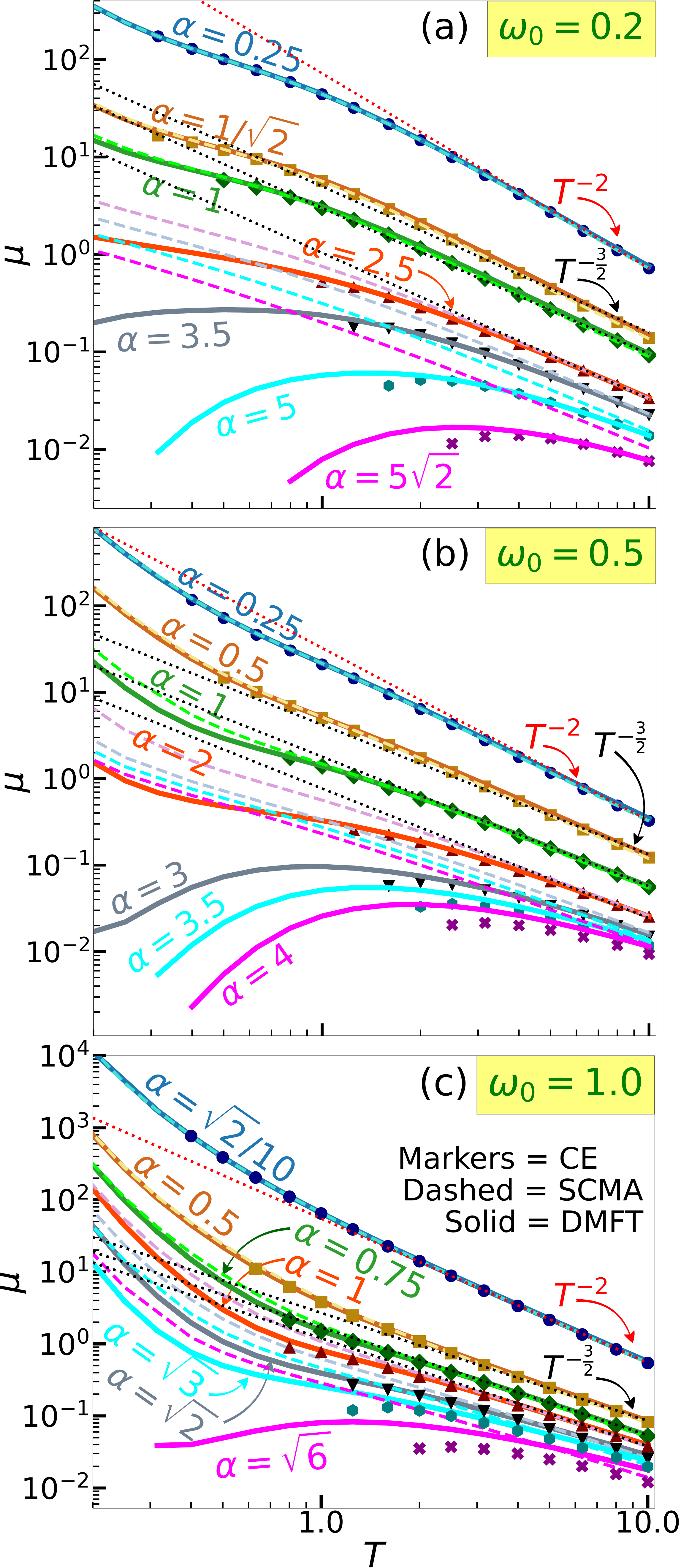} 
 \caption{Temperature dependence of the mobility for the CE, DMFT, and SCMA. The dotted red (black) lines are auxiliary lines with the power law behavior  $\mu  \propto T^{-2}$ ($\mu  \propto T^{-3/2}$). Here $t_0=1$.
 }
 \label{Fig:mobility}
\end{figure}
 The exponential term $e^{-\nu /T}$ in Eq.~\eqref{Eq:mobility_from_spectral} has some important implications. Despite the factor $\sin^2 k$, it implies that the largest contribution to the mobility most commonly comes from the spectral functions around the bottom of the band ($k\approx 0$), as they are typically situated at lower frequencies with respect to their higher momentum counterparts. This is actually helpful, as we have seen that the CE is more reliable for $k\approx 0$ than for $0<k<\pi$. However, $e^{-\nu/T}$ also introduces numerical instabilities, as even a small numerical noise of $A_k (\nu)$ at $\nu \ll -1$ will be inflated and give an enormous overall error in the mobility. This is why the integrals in Eq.~\eqref{Eq:mobility_from_spectral} require introducing some kind of negative frequency cutoff $\int_{-\infty}^\infty \to \int_{-\Lambda}^\infty$. We always check that the mobility results converge with respect to $\Lambda$. This is easily done in both the DMFT and SCMA, due to the high numerical accuracy of our numerical implementations. The convergence with respect to $\Lambda$ is much harder to achieve within the CE, as the Green's functions are initially calculated in the time-domain and require the use of  numerical Fourier transform. We have implemented a well-known interpolation scheme \cite{2007_Press} to increase the precision of the Fourier transform. Still, the numerical noise at low temperatures and strong interactions prevented us from precisely calculating the mobility in these regimes. We show only the data where an accurate calculation was possible. 
\pas
In Fig.~\ref{Fig:mobility} we present numerical results for the temperature dependence of the electron mobility. For weak electron-phonon coupling, all methods are in agreement; see Fig.~\ref{Fig:mobility}(a) for $\alpha \le 1$ and Figs.~\ref{Fig:mobility}(b)~and~\ref{Fig:mobility}(c) for $\alpha \le 0.5$. Electron-phonon scattering is weak in these regimes, which is why the quasiparticle lifetime $\tau_k$ is long, and the linear time dependence dominates in the cumulant function. The spectral function and its square can thus be approximated as  $A_k(\omega) \approx \delta(\omega-E_{p,k})$ and $A_k^2 (\omega) \approx \frac{\tau_k}{\pi} \delta(\omega- E_{p,k})$, where $\delta$ is the Dirac delta function and $E_{p,k}$ is given by Eq.~\eqref{Eq:polaron_en_def}. The  mobility from Eq.~\eqref{Eq:mobility_from_spectral}  thus simplifies to 
\begin{equation} \label{Eq:mobility_highT}
    \mu_{\text{weak}} \approx \frac{4t_0^2}{T} 
    \frac{\sum_k \tau_k e^{-E_{p,k}/T} \sin^2 k}{\sum_k  e^{-E_{p,k}/T}}.
\end{equation}
At high temperatures, Eq.~\eqref{Eq:mobility_highT} further simplifies as $e^{-E_{p,k}/T} \approx 1$. In this case, the lifetime is inversely proportional to the temperature $\tau_k \propto 1/T$, as seen from Eq.~\eqref{Eq:lifetime_CE}, which implies the power-law behavior of the mobility  $\mu_{\text{weak}} \propto 1/T^2$. This conclusion holds only for very weak electron-phonon couplings, where the assumption of weak scattering is still satisfied despite the high temperatures; see Figs.~\ref{Fig:mobility}(a)~and~\ref{Fig:mobility}(b) for $\alpha=0.25$ and Fig.~\ref{Fig:mobility}(c) for $\alpha = \sqrt{2}/10$. This assumption is also violated at extremely high temperatures $T\to\infty$. 
\pas
For stronger couplings, in the limit of  high-temperatures $T\gg t_0,\omega_0$, the Green's function in the time domain is quickly damped, which is why $C_k(t)$ can be approximated with just the lowest order (quadratic) Taylor expansion around $t=0$. Hence, Eqs.~\eqref{Eq:Cumulant_def}~and~\eqref{Eq:Cumulant_dos_derivative} imply that the Green's function can be written as
\begin{equation}
    G_k(t) = -i \theta(t) e^{-i\varepsilon_k t} e^{-\frac{g^2}{2}(2\bose+1)t^2},
\end{equation}
while the corresponding spectral function is given by the Gaussian
\begin{equation}
    A_k (\omega) = \frac{e^{-\frac{(\omega - \varepsilon_k)^2}{2g^2 (2\bose+1)}}}{\sqrt{2\pi g^2 (2\bose+1)}} .
\end{equation}
Plugging this back into Eq.~\eqref{Eq:mobility_from_spectral} and changing the sum over momenta to integral, we obtain
\begin{equation} \label{Eq:CE_highT_mob}
    \mu_{\mathrm{high-}T} =  \frac{t_0}{g}\sqrt{\frac{\pi}{2\bose+1}}
     \exp \left(-\frac{g^2 (2\bose+1)}{4T^2} \right)
    \frac{I_1 (\frac{2t_0}{T})}{I_0 (\frac{2t_0}{T})},
\end{equation}
where $I_0$ and $I_1$ are modified Bessel functions of the first kind, of zeroth and first order, respectively. Equation~\eqref{Eq:CE_highT_mob} can be simplified by using the following approximations $2\bose + 1 \approx 2T/\omega_0$ and $I_1(2t_0/T) / I_0(2t_0 / T) \approx t_0/T$, that are valid for large $T$. Such a simplified formula coincides with the mobility obtained by combining the Einstein relation, between the mobility and diffusion coefficient, with the Marcus formula \cite{2019_Prodanovic, 2016_Fratini}. Furthermore,  Eq.~\eqref{Eq:CE_highT_mob} implies the power law behavior for the mobility  $\mu_{\mathrm{high-T}} \propto T^{-3/2}$, in the limit $T \gg t_0,\omega_0$. This is confirmed by our numerical results for a wide range of the electron-phonon coupling strengths, where all three methods are in agreement; see Fig.~\ref{Fig:mobility}(a) for $1/\sqrt{2} \leq \alpha \leq 2.5$, Fig.~\ref{Fig:mobility}(b) for $0.5\leq \alpha \leq 2$ and Fig.~\ref{Fig:mobility}(c) for $0.5 \leq \alpha \leq 1$.
\pas
While the SCMA gives satisfactory results for high temperatures and intermediate electron-phonon couplings, it deviates from the DMFT at lower temperatures (see, e.g., Fig.~\ref{Fig:mobility}(a) for $\alpha = 2.5$ and Fig.~\ref{Fig:mobility}(b) for $\alpha = 2$) and also for stronger coupling strength (see, e.g., Fig.~\ref{Fig:mobility}(a) for $\alpha > 2.5$ and Fig.~\ref{Fig:mobility}(b) for $\alpha > 2$). At these stronger couplings, the DMFT predicts the non-monotonic mobility, where a region of decreasing mobility with decreasing temperature is ascribed to the hopping transport in phenomenological theories \cite{2003_Fratini, 2016_Fratini}. The strong coupling mobility is better described by the CE than SCMA, although low-temperature results are missing due to our inability to converge
the results with respect to the cutoff $\Lambda$. In Appendix~\ref{AppSec:mobilityMA}, we also give mobility predictions of the MA.

\section{CONCLUSIONS AND OUTLOOK} \label{Sec:discussion}
\poc
In summary, we have presented a comprehensive analysis of the CE method in the context of the Holstein model. The second-order cumulant $C(t)$ is calculated in a broad temperature range for three vibrational frequencies $\omega_0/t_0 = 0.2, 0.5$, and $1$, covering a regime from a weak to strong electron-phonon coupling. We mostly focused on the 1D system in the thermodynamic limit, but some of the results are shown also in 2D and 3D. To avoid numerical instabilities and to reach high numerical precision, we derived a number of analytical expressions and we used the collocation method in calculations of the cumulant, as well as an interpolation scheme for the Fourier transform in corresponding calculations of the spectral functions. The quasiparticle properties, spectral functions, and charge mobility are shown in comparison to the DMFT and SCMA results. The DMFT, which gives close to the exact solution for the Holstein polaron throughout the parameter space \cite{2022_Mitric}, gave a valuable benchmark and facilitated a detailed analysis of the validity of the CE method.
\pas
At weak coupling (roughly corresponding to $m_0/m^* \gtrsim 0.9$) CE, DMFT, and SCMA give very similar spectral functions. Most of the spectral weight for $k=0$ is in the quasiparticle peak, while even a small sideband (satellite) spectral weight is rather well reproduced in all three methods. As the interaction increases, a clear difference in the spectral functions emerges. Nevertheless, the positions of the CE and DMFT quasiparticle and the first satellite peak at low temperatures are in rather good agreement. Furthermore, the overall spectral weight distribution is in a decent agreement even though the satellite peaks are more pronounced in DMFT for stronger electron-phonon coupling. Roughly speaking, there is a decent agreement in 1D up to the interactions corresponding to $m_0/m^* \sim 0.5$. Interestingly, the agreement between the CE and DMFT spectral functions persists also for $k=\pi$, although CE does not capture a tiny quasiparticle peak. In this case, the DMFT spectral weight almost merges to a single broad peak. We note that the difference for $k=\pi$ observed in Ref.~\cite{2022_Robinson_1} is solely due to considering a lattice of finite $N=6$ size. The deviation of CE from the exact solution is most obvious for intermediate momenta where the CE solution merges to a single peak, while the satellite structure is seen in DMFT. At high temperatures, one might suspect that the CE would give the exact spectral functions. However, this is not the case as we showed that the CE gives the exact spectral moments only up to the order $n=4$. We note, that in all these regimes the CE gives slightly better results than the SCMA, while a single-shot MA is adequate only for very weak interactions.
\pas
The spectral functions were used to calculate the charge mobility from the Kubo formula without the vertex corrections. The agreement between DMFT and CE is quite good. This is the case even for stronger electron-phonon coupling where the CE even indicates non-monotonic behavior of $\mu(T)$, with a region of increasing mobility with temperature which is usually assigned to hopping conduction in phenomenological theories. For strong electron-phonon coupling, the CE mobility results are shown only for $T \gtrsim t_0$ since a very small numerical noise at frequencies $\omega \ll E_p$ affects a precise calculation of mobility at lower temperatures. For high temperatures the mobility assumes a universal form: For weak electron-phonon coupling $\mu \propto T^{-2}$, while for stronger coupling $\mu \propto T^{-3/2}$. These high-temperature limits can be obtained also analytically from the CE. 
\pas
{ The CE method can be easily applied to different Hamiltonians, which makes it a particularly attractive method for the calculation of electronic properties beyond the weak-coupling limit in various systems.
In particular, we argue
that it will be most useful in calculations of charge mobility, as has already been done in {\it ab initio} calculations
for SrTiO$_3$ \cite{2019_Zhu} and naphthalene~\cite{2022_Chang}.
While our analysis may suggest that the DMFT appears computationally superior to CE, we note that the numerical efficiency that we achieved with DMFT is restricted to the Holstein model by virtue of the analytic solution for the impurity problem \cite{1997_Ciuchi} and the local Green’s function \cite{2022_Mitric}. For predicting the properties of real materials, the numerical resources within the DMFT are vastly increased and also the issue of nonlocal correlations may emerge,  while the CE remains simple and relatively inexpensive.} Of course, for a definitive answer on the range of validity of CE in connection with {\it ab initio} calculations, one needs to perform a similar analysis for the Fr\"ohlich model and for other models which can be used for realistic description of the electronic spectra and charge transport in real materials. A useful hint in this direction is provided by Ref. \cite{2022_Robinson_2} which shows that the CE, around the bottom of the band, gives
promising results for the spectral function even in the
case when the phonons have a dispersion~\cite{2021_Bonca_Trugman}. Another very interesting question that we leave for further work is a possible contribution of vertex corrections to conductivity. Based on the weak coupling result \cite{1966_Mahan_mobility}, one might assume that their contribution is small for optical phonons, but this remains to be determined in the case of stronger coupling. Our high-temperature results for mobility may also be quite useful when analyzing  a dominant type of electron-phonon coupling in real materials. Still, one needs to be cautious in such analyses since we see that at lower temperatures $\mu(T)$ does not assume a simple universal form.

\section*{Acknowledgments}
The authors acknowledge funding provided by the Institute of Physics Belgrade, 
through the grant by the  { Ministry of Science, Technological Development and Innovation} of the Republic of Serbia. Numerical simulations were performed on 
the PARADOX supercomputing facility at the Scientific Computing Laboratory, 
National Center of Excellence for the Study of Complex Systems, Institute of 
Physics Belgrade.

\clearpage
\appendix

\section{NUMERICAL INTEGRATION SCHEME FOR THE HIGHLY OSCILLATING FUNCTIONS IN THE CE METHOD}\label{Apendix:Numerical_RC}
\poc
We present a numerical integration scheme for the calculation of the cumulant function from Eq.~\eqref{Eq:Cumualnt_wrt_Bessel}. Since $C_{\bf k}(t)$ will be expressed numerically on some $t$ grid $[t_0=0, t_1 \dots t_{G-1}]$, it is much better to divide the integral $\int_{0}^{t}$  from Eq.~\eqref{Eq:Cumualnt_wrt_Bessel} into a sum of integrals of the form $\int_{t_{i-1}}^{t_i}$, where $t_i$ are times from the previously defined $t$ grid. In this manner, we do not integrate over the same interval  multiple times. To shorten the notation, from now on, we denote $a \equiv t_{i-1}$ and $b \equiv t_i$.
\pas
There are two different types of integrals in Eq.~\eqref{Eq:Cumualnt_wrt_Bessel}, and both of them have the following form 
\begin{equation} \label{Eq:general_colocation}
I = \int_a^b  dx \, g(x) e^{i r_1 x} J_0(r_2 x)^n,
\end{equation}
where $g(x)$ is either a linear or a constant function, $r_1 = {\varepsilon_{\bf k} \pm \omega_0}$, and $r_2 = 2t_0$. Numerical integration of Eq.~\eqref{Eq:general_colocation} has already been studied by Levin for arbitrary $r_1$ and $r_2$ and  slowly varying $g(x)$ \cite{1996_Levin}. In the rest of this Appendix, we review this method in the 1D ($n=1$), 2D ($n=2$), and 3D ($n=3$) cases.  
\pas
The main idea is to rewrite the subintegral function as a scalar product of two columns $| \tilde{g}(x) \rangle$ and $| \tilde{J}(x) \rangle$, whose elements are functions
\begin{equation} \label{App:Eq:Scalar_prod_gJ}
I = \int_a^b dx \langle \tilde{g}(x) | \tilde{J}(x) \rangle.
\end{equation}
Column $| \tilde{g}(x) \rangle$ consists exclusively of slowly varying functions, while $| \tilde{J}(x) \rangle$ contains highly oscillating functions, with the property that 
\begin{equation} \label{App:Eq:diff_J}
\frac{ d |\tilde{J} (x) \rangle} { dx} = \hat{A} (x) | \tilde{J}(x) \rangle ,    
\end{equation}
where $\hat{A} (x) $ is a matrix of slowly varying functions. Then, the integral from Eq.~\eqref{App:Eq:Scalar_prod_gJ} can be written as 
\begin{equation}
I = \int_a^b dx \frac{d}{dx} \langle \tilde{f}(x) | \tilde{J}(x) \rangle = 
\langle \tilde{f}(b) | \tilde{J}(b) \rangle - 
\langle \tilde{f}(a) | \tilde{J}(a) \rangle,
\end{equation}
where $| \tilde{f}(x)  \rangle$ satisfies
\begin{equation} \label{App:Eq:Diff_for_f}
\left(
\frac{d}{dx} + \hat{A}^\dagger (x)
\right) | \tilde{f}(x) \rangle = | \tilde{g}(x) \rangle.
\end{equation}
This is then, following Levin \cite{1996_Levin}, solved by formally expanding $
|\tilde{f}(x) \rangle = \sum_{k=1}^{M} u_k(x) [c_k \;\; d_k \;\; \dots]^T    
$ into a basis set of polynomials $u_k(x) = (x-\frac{a+b}{2})^{k-1}$ and determining the unknown polynomial  coefficients $c_k,d_k\dots$ by imposing that Eq.~\eqref{App:Eq:Diff_for_f} is exactly satisfied at $M$ uniformly distributed collocation points $x_j = a + \frac{(j-1)(b-a)}{M-1}$, $j=1\dots M$. The initial problem is thus reduced to a simple linear algebra problem.

\subsection{1D case}
\poc
In the 1D case ($n=1$), columns $| \tilde{g}(x) \rangle$ and $| \tilde{J}(x) \rangle$ assume the following form
\begin{subequations}
\begin{align}
| \tilde{g}(x) \rangle &= [g(x)\;0]^T, \\
| \tilde{J}(x) \rangle &= e^{i r_1 x}[J_0(r_2 x)\;\; J_1(r_2 x)]^T,
\end{align}
\end{subequations}
where $J_0(x)$ and $J_1(x)$ are the Bessel functions of the first kind, of zeroth and first order. The matrix $\hat{A} (x)$, such that  Eq.~\eqref{App:Eq:diff_J} holds, is given by
\begin{equation}
\hat{A} (x) = 
\begin{bmatrix}
i r_1 & -r_2 \\
r_2 & i r_1 - \frac{1}{x}
\end{bmatrix}.
\end{equation}
The unknown coefficients $c_k$ and $d_k$, which determine the column function
\begin{equation}
|\tilde{f}(x) \rangle = \sum_{k=1}^{M} u_k(x) [c_k \;\; d_k]^T,    
\end{equation}
are obtained from the following set of $2M$ linear equations
\begin{equation}
\begin{bmatrix}
\mathcal{C} & \quad & \mathcal{C}^d \\
\quad & \quad & \quad \\
\mathcal{D}^c & \quad &\mathcal{D}
\end{bmatrix}
\begin{bmatrix}
c_1 \\
\vdots \\
c_M \\
d_1 \\
\vdots \\
d_M
\end{bmatrix} = 
\begin{bmatrix}
g(x_1) \\
\vdots \\
g(x_M) \\
0 \\
\vdots \\
0
\end{bmatrix}.
\end{equation}
Here, $\mathcal{C}, \mathcal{C}^d, \mathcal{D}^c, \mathcal{D}$ are $M\times M$ matrices that read as  
\begin{subequations} \label{App:Eq:Cij}
\begin{align} 
\mathcal{C}_{ij} &= u_j'(x_i) - ir_1 u_j(x_i); \quad 
\mathcal{C}^d_{ij} = r_2 u_j(x_i); \\
\mathcal{D}_{ij} &= u_j'(x_i) - \left( ir_1 + \frac{1}{x_i} \right) 
u_j(x_i); \; \mathcal{D}^c_{ij} = -r_2 u_j(x_i).
\end{align}
\end{subequations}
\subsection{2D case}
\poc
In the 2D case, the relevant quantities are given by
\begin{align}
| \tilde{g}(x) \rangle &= [g(x)\;\;0\;\;0]^T, \nonumber \\
| \tilde{J}(x) \rangle &= e^{i r_1 x}[J_0(r_2 x)^2\;\; J_0(r_2 x)J_1(r_2 x) \;\; 
J_1(r_2 x)^2]^T , \nonumber \\
\hat{A} (x) &= 
\begin{bmatrix}
i r_1 & -2r_2 & 0 \\
r_2 & i r_1 - \frac{1}{x} & -r_2 \\
0& 2r_2 & i r_1 - \frac{2}{x}
\end{bmatrix}.
\end{align}
The column $| \tilde{f}(x) \rangle = \sum_{k=1}^M u_k(x) [c_k \;\; d_k \;\; e_k]^T$ is determined by $c_k$, $d_k$ and $e_k$, which are obtained as a solution of the following system of $3M$ linear equations 
%The expansion into the polynomial basis set is analogous $| \tilde{f}(x) \rangle = \sum_{k=1}^M u_k(x) [c_k \;\; d_k \;\; e_k]^T$, while the corresponding system of $3M$ linear equations is determined by
%
\begin{equation}
\begin{bmatrix}
\mathcal{C} & \quad & \mathcal{C}^d &\quad & \mathcal{C}^e\\
\quad & \quad & \quad & \quad & \quad \\
\mathcal{D}^c & \quad &\mathcal{D} &\quad & \mathcal{D}^e \\
\quad & \quad & \quad & \quad & \quad \\
\mathcal{E}^c & \quad &\mathcal{E}^d &\quad & \mathcal{E} \\
\end{bmatrix}
\begin{bmatrix}
c_1 \\
\vdots \\
c_M \\
d_1 \\
\vdots \\
e_1 \\
\vdots 
\end{bmatrix} = 
\begin{bmatrix}
g(x_1) \\
\vdots \\
g(x_M) \\
0 \\
\vdots \\
0 \\
\vdots
\end{bmatrix}.
\end{equation}
Here, $\mathcal{C},\mathcal{C}^d \dots \mathcal{E}$ are $M \times M$ matrices. Elements of $\mathcal{C}_{ij}$ and $\mathcal{C}^d_{ij}$ are the same as in Eq.~\eqref{App:Eq:Cij}, while $\mathcal{C}^e_{ij}=\mathcal{E}^c_{ij}=0$. 
All the other elements are given by:
\begin{align} \label{App:Eq:2d_lineq}
\mathcal{D}_{ij} &= u_j'(x_i) - 
\left( ir_1 + \frac{1}{x_i} \right) u_j(x_i); \nonumber\\
\mathcal{E}_{ij} &= u_j'(x_i) - \left( ir_1 + \frac{2}{x_i} \right) u_j(x_i); \\
\mathcal{D}_{ij}^c &= -2r_2 u_j(x_i);\;
\mathcal{D}_{ij}^e = 2r_2 u_j(x_i);\;
\mathcal{E}_{ij}^d = -r_2 u_j(x_i); \; \nonumber
\end{align}

\subsection{3D case}
\poc
The procedure that was presented so far is actually quite easily generalized to the 3D case as well. Here, the quantities of interest are easily derived and read as
\begin{align} \label{App:Eq:A_3D}
| \tilde{g}(x) \rangle &= [g(x)\;\;0\;\;0\;\;0]^T, \nonumber \\
| \tilde{J}(x)\rangle &= e^{i r_1 x}[J_0(r_2 x)^3\;\; J_0(r_2 x)^2 J_1(r_2 x) \nonumber \\ \;\; 
& \qquad \qquad J_0(r_2 x) J_1(r_2 x)^2 \;\; J_1(r_2 x)^3]^T , \nonumber \\
\hat{A} (x) &= 
\begin{bmatrix}
i r_1 & -3r_2 & 0 & 0\\
r_2 & i r_1 - \frac{1}{x} & -2r_2 & 0 \\
0& 2r_2 & i r_1 - \frac{2}{x} & -r_2 \\
0 & 0 & 3r_2 & ir_1 - \frac{3}{x}
\end{bmatrix}, \nonumber \\
\tilde{f}(x) &= \sum_{k=1}^M u_k(x) [c_k \;\; d_k \;\; e_k\;\;f_k]^T,
\end{align}
where the coefficients $c_k$, $d_k$, $e_k$ and $f_k$ satisfy
\begin{equation}
\begin{bmatrix}
\mathcal{C} & \quad & \mathcal{C}^d &\quad & \mathcal{C}^e &\quad & \mathcal{C}^f\\
\quad & \quad & \quad & \quad & \quad & \quad & \quad \\
\mathcal{D}^c & \quad &\mathcal{D} &\quad & \mathcal{D}^e &\quad & \mathcal{D}^f \\
\quad & \quad & \quad & \quad & \quad & \quad & \quad\\
\mathcal{E}^c & \quad &\mathcal{E}^d &\quad & \mathcal{E}   & \quad & \mathcal{E}^f\\
\quad & \quad & \quad & \quad & \quad & \quad & \quad\\
\mathcal{F}^c & \quad &\mathcal{F}^d &\quad & \mathcal{F}^e   & \quad & \mathcal{F}\\
\end{bmatrix}
\begin{bmatrix}
c_1 \\
\vdots \\
c_M \\
d_1 \\
\vdots \\
e_1 \\
\vdots \\
f_1 \\
\vdots
\end{bmatrix} = 
\begin{bmatrix}
g(x_1) \\
\vdots \\
g(x_M) \\
0 \\
\vdots \\
0 \\
\vdots \\
0 \\
\vdots
\end{bmatrix}.
\end{equation}
Here $\mathcal{C}_{ij}$, $\mathcal{C}^d_{ij}$, $\mathcal{C}^e_{ij}$, $\mathcal{D}_{ij}$, $\mathcal{D}^e_{ij}$, $\mathcal{E}^c_{ij}$ and $\mathcal{E}_{ij}$ are the same as in Eqs.~\eqref{App:Eq:Cij}~and~\eqref{App:Eq:2d_lineq}, while $\mathcal{C}^f_{ij} = \mathcal{F}^c_{ij}=\mathcal{D}^f_{ij} = \mathcal{F}^d_{ij} = 0$. All other elements are given by:
\begin{align} \label{App:Eq:lin_eq_3d}
    \mathcal{E}^d_{ij} &= -2r_2 u_j(x_i);\; \mathcal{E}^f_{ij} = 3r_2 u_j(x_i) ,\nonumber \\
    \mathcal{D}^c_{ij} &= -3r_2 u_j (x_i);\; \mathcal{F}^e_{ij} = -r_2 u_j (x_i),  \\
    \mathcal{F}_{ij} &= u_j'(x_i) - \left( ir_1 + \frac{3}{x_i} \right) u_j(x_i). \nonumber
\end{align}
Thus, our numerical scheme has been completely specified. We note that Eqs.~\eqref{App:Eq:Cij},~\eqref{App:Eq:2d_lineq}~and~\eqref{App:Eq:lin_eq_3d} explicitly demonstrate that our numerical scheme is singular at $x=0$.  This does not pose any problems, as the subintegral function in our initial expression \eqref{Eq:general_colocation} is not highly oscillatory around $x=0$. Therefore, the trapezoid scheme can be applied there.

\section{2D SPECTRAL FUNCTIONS}
\label{AppSec:2DspecF}
\poc
We now examine the CE spectral functions in two dimensions and compare them to the results from DMFT and SCMA. We investigate the Hamiltonian from  Eq.~\eqref{Eq:Holstein_Hamiltonian} on a square lattice and set $\hbar$, $k_B$ and lattice constant to $1$.% and hopping parameters $t_0$ are both set to $1$. 
\pas
In the 2D case, the cumulant function is calculated from Eq.~\eqref{Eq:Cumualnt_wrt_Bessel} by setting $n=2$, and by exploiting the numerical integration scheme from Appendix~\ref{Apendix:Numerical_RC}. The procedure for the implementation of the DMFT and SCMA is the same as explained in Sec.~\ref{Sec:Benchmark_methods}, with the only difference being that Eq.~\eqref{Eq:Analytic_G}   no longer represents the solution for the local Green's function from Eqs.~\eqref{Eq:SCMA_local_Green}~and~\eqref{Eq:DMFT_local_green_int_dos}. 
\begin{figure*}[!t]
    \centering
  \includegraphics[width=7in,trim=0cm 0cm 0cm 0cm]{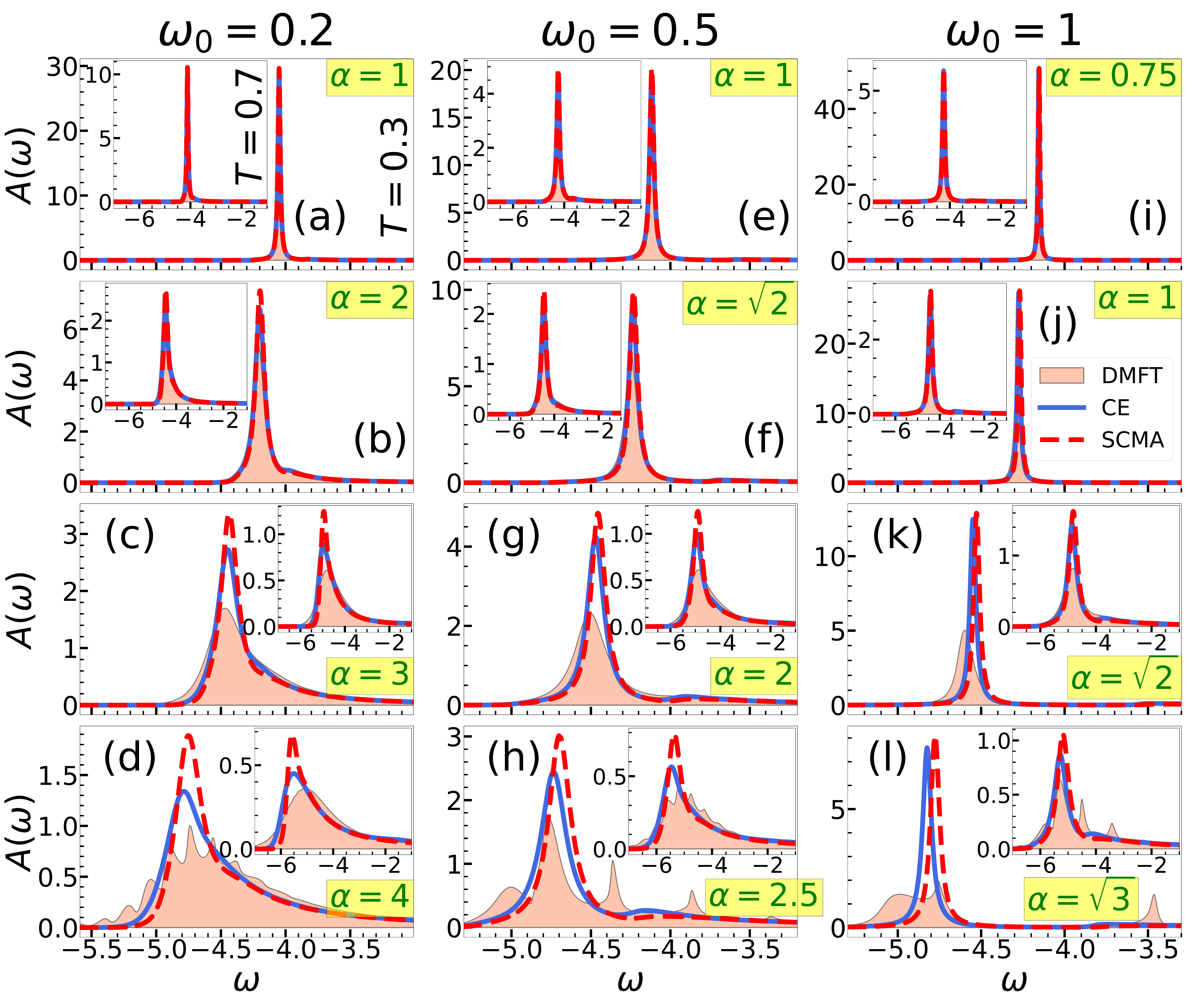} 
 \caption{(a)--(h) Comparison of the CE, DMFT, and SCMA  spectral functions in 2D for $k=0$ and $t_0=1$. The main panels show the results for $T_1=0.3$, while $T_2=0.7$ results are shown in the insets. 
 }
 \label{Fig:Aw_2D}
\end{figure*}
\pas
The local Green's function for the square lattice is obtained as follows. Let us introduce ${B(\omega) \equiv (\omega - \Sigma(\omega))/(2t_0)}$ and rewrite Eq.~\eqref{Eq:DMFT_local_green_int_dos} as 
\begin{equation} \label{Eq:App:D1}
    G(\omega) = 
    - \int_{-\infty}^\infty dx \hat{\rho}(x) \int_{-\infty}^\infty d\varepsilon \frac{e^{ix\varepsilon}}{\varepsilon - 2t_0 B(\omega)}.
\end{equation}
The integral over $\varepsilon$ can be solved using the residue theorem. It is thus important to note that the subintegral function has only a single pole at $\varepsilon_{\mathrm{pole}} = 2t_0 B(\omega)$, that is situated at the upper half-plane, i.e., $\mathrm{Im}B(\omega) >0$ (since $\mathrm{Im}\Sigma (\omega) < 0$). Hence
\begin{equation} \label{Eq:pom_063}
    G(\omega) = -2\pi i \int_{-\infty}^\infty dx \tilde{\rho}(x)e^{2ixt_0 B(\omega)} \theta(x).
\end{equation}
Here $\tilde{\rho}(x)$ is given by Eq.~\eqref{Eq:DOS_Fourier} for $n=2$. Substituting this into Eq.~\eqref{Eq:pom_063} and solving the integral gives
\begin{equation}
    G(\omega) = \frac{K \left(
    \frac{2}{B(\omega)}
    \right)}{B(\omega) \pi t_0},
\end{equation}
where $K(k) \equiv \int_0^{\pi/2} d\theta / \sqrt{1-k^2 \sin^2 \theta}$ is the complete elliptic integral of the first kind.
\pas
Results are presented in Fig.~\ref{Fig:Aw_2D}. We note that  in Figs.~\ref{Fig:Aw_2D}(a)--\ref{Fig:Aw_2D}(d) (Figs.~\ref{Fig:Aw_2D}(i)--(l)) the phonon frequency $\omega_0=0.2$ ($\omega_0=1$) is smaller (larger) than both of the temperatures $T_1=0.3$ and $T_2 = 0.7$ that we are considering. Therefore, we focus on Figs.~\ref{Fig:Aw_2D}(e)--\ref{Fig:Aw_2D}(h) where $T_1 < \omega_0 < T_2$, while other regimes can be analyzed analogously. We see that most of the spectral weight is concentrated in a smaller range of frequencies than in the 1D case; see Figs.~\ref{SubFig:SpecF_w_0.5_k_0}~and~\ref{Fig:Aw_2D}(e)--\ref{Fig:Aw_2D}(h). This is a consequence of the fact that the hopping parameter is always set to unity, while the 2D bandwidth  is twice as large in comparison with the bandwidth in the 1D system. Spectral functions from Figs.~\ref{Fig:Aw_2D}(e)--\ref{Fig:Aw_2D}(g) exhibit 
 qualitatively similar behavior as results for the 1D system in Figs.~\ref{SubFig:SpecF_w_0.5_k_0}(a)--\ref{SubFig:SpecF_w_0.5_k_0}(d). Here, all methods are in agreement and predict that  the quasiparticle peak dominates, while there is only a single tiny satellite structure that is more pronounced at higher temperatures. However, it seems that the satellites are more pronounced in the 1D spectral functions. A much more complicated multi-peak structure is predicted by the DMFT in Fig.~\ref{Fig:mass_renorm}(h), where a large discrepancy can be observed in comparison to the CE and SCMA results. A better agreement is observed for higher temperatures. 
 \pas
 It is interesting to note that while the DMFT frequently gave sharper peaks than other methods in 1D (see Fig.~\ref{SubFig:SpecF_w_0.5_k_0}), here the roles are reversed. This is a consequence of the strong Van Hove singularity at the bottom of the band of a 1D system, which is highly relevant in our case when the concentration of electrons is  very low, while the singularity in the 2D system is weaker and shifted to the center of the band.

\section{A DETAILED STUDY OF THE SPECTRAL FUNCTION FOR \texorpdfstring{$t_0=\omega_0=g=1$}{t₀ = ω₀ = g = 1}~and~\texorpdfstring{$k=\pi$}{k = π}}

\label{App:Compare_Reichman}
\poc
In Sec.~\ref{Sec:SpecF}, we concluded that the CE successfully captures the main features of the spectral functions both at the bottom of the band ($k \approx 0$) and at top of the band ($k \approx \pm \pi$), if the electron-phonon coupling is not too strong. Less promising results were reported in Ref.~\cite{2022_Robinson_1}, where CE was examined on a finite lattice with $N=6$ sites, in the regime $t_0=\omega_0=g=1$ and $k=\pi$, using the finite-temperature Lanczos method (FTLM) \cite{2019_Bonca} as a benchmark. They found that the CE, in addition to the fact that it does not correctly reproduce a quasiparticle peak, predicts that the most prominent feature of the spectrum consists of only a  single broad peak, whereas two distinct peaks are present in the FTLM solution. Here, we show that this discrepancy between the CE and FTLM is significantly reduced in the thermodynamic limit.
%
%The inability of the CE method to capture these two distinct peaks, as reported, is said to be expected even in the thermodynamic limit. While CE results close to the thermodynamic limit were also shown, no reliable benchmark data was available in that case.  
%
\pas
Reference \cite{2022_Robinson_1} emphasized that previous  conclusions are valid only for low-temperature solutions, while CE becomes accurate for $T \ge \omega_0$. This was confirmed by the FTLM, whose spectral functions in this case look like a single broad peak; see Fig.~1(c) from Ref.~\cite{2022_Robinson_1}. However, Fig.~S9 in the Supplemental Material of Ref.~\cite{2022_Mitric}  demonstrates that the spectral function in the thermodynamic limit for $t_0=\omega_0=g=1$, $k=\pi$  consists of  a broad single-peak structure even at $T=0$. This conclusion was reached by carefully examining the finite-size effects using the numerically exact  hierarchical equations of motion method (HEOM). It was established that the system with $N=10$ lattice sites is representative of the thermodynamic limit, although much smaller systems are required for the $k=0$ results.  Furthermore, the same figure shows that two distinct peaks emerge for $N=6$ and $k=\pi$, in accordance with the FTLM results. Hence, CE will provide much better results in the thermodynamic limit than previously expected. We note that for $t_0=\omega_0=g=1$ and finite temperatures one might expect that the required lattice size, representative of the thermodynamic limit, does not exceed $N=10$, as the electron experiences much more scattering compared to the $T=0$ case. This will be crosschecked independently (using the DMFT) in the rest of this Appendix for finite $T$, which satisfies the $T<\omega_0$ condition. 
In that case, we analyze the overall performance of the CE.
\pas
\begin{figure}[!t]
  \includegraphics[width=3.5in,trim=0cm 0cm 0cm 0cm]{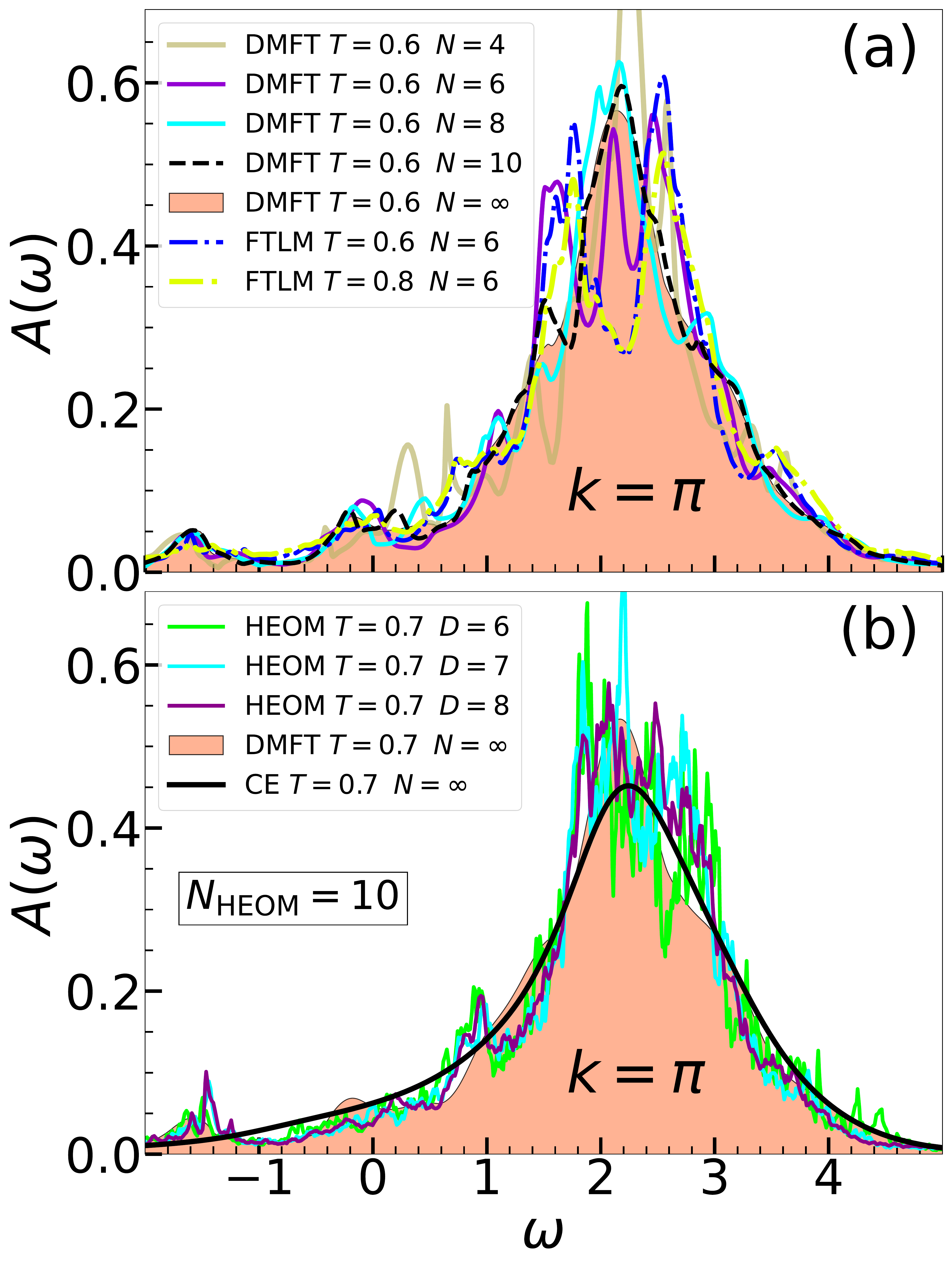} 
 \caption{CE, DMFT, FTLM, and HEOM spectral functions for $t_0=\omega_0=g=1$. (a) Analysis of the finite-size effects. (b) Inspecting the convergence of HEOM data with respect to hierarchy depth $D$. 
 }
 \label{Fig:Compare_Reichman}
\end{figure}
In  Fig.~\ref{Fig:Compare_Reichman}(a) we show the FTLM data, (originally from Ref. \cite{2019_Bonca}) used in Ref. \cite{2022_Robinson_1}, and compare them to the DMFT applied on a system of finite lattice size. We exploit the fact that the corresponding spectral functions (although certainly not as accurate in comparison with the exact solution) provide a rough estimate of how large  $\latsize$ should be to faithfully represent the thermodynamic limit; see Sec.~IV from the Supplemental Material of Ref. \cite{2022_Mitric} for more details. In accordance with the FTLM results, we see that the DMFT spectral function for $\latsize=6$ also predicts distinct peaks around $\omega\approx 1.5$ and $\omega\approx 2.5 $, although there is an additional peak around $\omega \approx 2$. Nevertheless, these results change drastically with increasing $\latsize$ and practically converge for $\latsize=10$. This is the same $N$ as predicted by HEOM at $T=0$. Therefore, the presented FTLM results are not representative of the thermodynamic limit. Additionally, Fig.~\ref{Fig:Compare_Reichman}(a) also shows that FTLM results for $T=0.6$ and $T=0.8$ are quite similar. Hence, our further analysis will be conducted for $T=0.7$ case.
\pas
In  Fig.~\ref{Fig:Compare_Reichman}(b), we present HEOM results for $N=10$ and compare them to CE and DMFT. We note that HEOM has one additional parameter, the so-called hierarchy depth $D$. For details we refer the reader to Ref. \cite{2022_Jankovic}, but we only briefly mention that the numerically exact results are formally obtained in the limit $D\to\infty$. In practice, we always check whether the results converge with respect to $D$, which cannot be increased indefinitely, as finite computer memory presents a limiting factor. We see that the HEOM results have practically converged for $\latsize=10$ and $D=8$. Here, the HEOM solution does not possess the two-peak structure predicted by the FTLM on a smaller lattice size ($\latsize=6$). It actually gives only a single, broad peak around $\omega \approx 2$, which is correctly reproduced by both the CE and the DMFT. Although the CE misses the quasiparticle peak around $\omega \approx -1.5$,  we conclude that CE gives much more accurate results for the thermodynamic limit than for a finite system. 

\section{MOBILITY RESULTS FROM THE ONE-SHOT MIGDAL APPROXIMATION}
\label{AppSec:mobilityMA}
\begin{figure}[!t]
  \includegraphics[width=3.5in,trim=0cm 0cm 0cm 0cm]{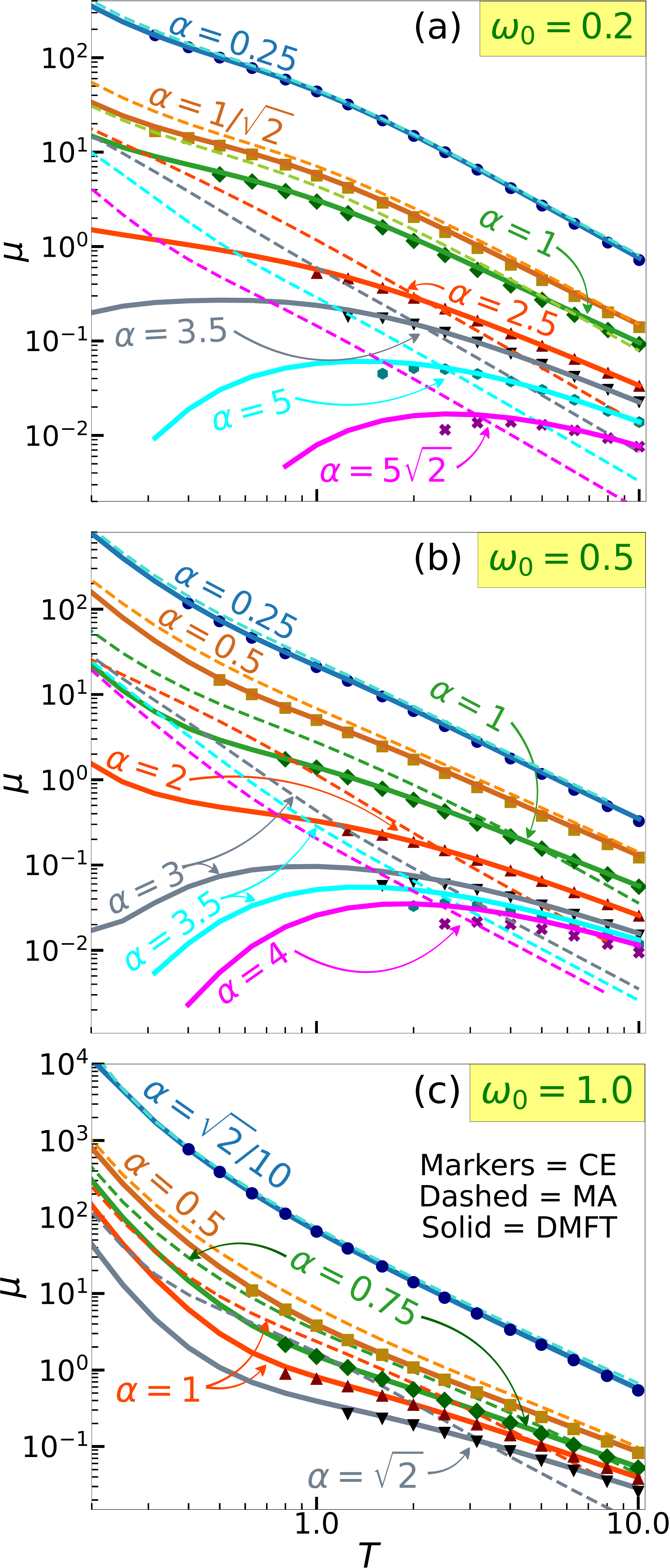} 
 \caption{Temperature dependence of the mobility within CE, DMFT, and MA. Here $t_0=1$.
 }
 \label{Fig:mobility_MA}
\end{figure}

\poc
In Sec.~\ref{Sec:Mobility_diff_methods}, we presented and analyzed the mobility predictions from the CE, DMFT, and SCMA methods. Here, we supplement that study with the data from the one-shot MA (i.e., SCMA without self-consistency). The results are shown in Fig.~\ref{Fig:mobility_MA}. Since the mobility results have already been thoroughly analyzed in Sec.~\ref{Sec:Mobility_diff_methods}, we will here give only brief comments about the performance of the MA. Figure~\ref{Fig:mobility_MA}~(a) shows that MA is practically useless for $\alpha \gtrsim 2.5$. Here, the results are not even qualitatively correct, regardless of the temperature. Even for $\alpha = 1$, the results are still not satisfactory: the predictions for $T<4$ ($T>9$) overestimate (underestimate) the DMFT benchmark.  MA proves to be reliable only for very weak interactions $\alpha \lesssim 1/\sqrt{2}$. Here, the results are better for higher temperatures. This is expected as the MA takes into account only the lowest-order Feynman diagram, while the relevance of higher-order diagrams decreases as the temperature is increased. Similar analysis can be repeated for other phonon frequencies in Figs.~\ref{Fig:mobility_MA}(b)~and~~\ref{Fig:mobility_MA}(c). 

\clearpage
\newpage

%\bibliographystyle{apsrev4-1}
%\bibliographystyle{apsrev4-2}
%\bibliography{refs.bib}


%%%%%%%%%%%%%%%%%%%%%%%%%%%%%%%%%%%%%%%%%%%%%%%%%%%%%%%%%%%%%%%%%%%%%%%%%%%%%%%%%%%
%%%%%%%%%%%%%%%%%%%%%%%%%%%%%%%%%%%%%%%%%%%%%%%%%%%%%%%%%%%%%%%%%%%%%%%%%%%%%%%%%%%
%%%%%%%%%%%%%%%%%%%%%%%%%%%%%%%%%%%%%%%%%%%%%%%%%%%%%%%%%%%%%%%%%%%%%%%%%%%%%%%%%%%
%%%%%%%%%%%%%%%%%%%%%%%%%%%%%%%%%%%%%%%%%%%%%%%%%%%%%%%%%%%%%%%%%%%%%%%%%%%%%%%%%%%
%%%%%%%%%%%%%%%%%%%%%%%%%%%%%%%%%%%%%%%%%%%%%%%%%%%%%%%%%%%%%%%%%%%%%%%%%%%%%%%%%%%
%%%%%%%%%%%%%%%%%%%%%%%%%%%%%%%%%%%%%%%%%%%%%%%%%%%%%%%%%%%%%%%%%%%%%%%%%%%%%%%%%%%
%%%%%%%%%%%%%%%%%%%%%%%%%%%%%%%%%%%%%%%%%%%%%%%%%%%%%%%%%%%%%%%%%%%%%%%%%%%%%%%%%%%
%%%%%%%%%%%%%%%%%%%%%%%%%%%%%%% SUPPLEMENT %%%%%%%%%%%%%%%%%%%%%%%%%%%%%%%%%%%%%%%%
%%%%%%%%%%%%%%%%%%%%%%%%%%%%%%%%%%%%%%%%%%%%%%%%%%%%%%%%%%%%%%%%%%%%%%%%%%%%%%%%%%%
%%%%%%%%%%%%%%%%%%%%%%%%%%%%%%%%%%%%%%%%%%%%%%%%%%%%%%%%%%%%%%%%%%%%%%%%%%%%%%%%%%%
%%%%%%%%%%%%%%%%%%%%%%%%%%%%%%%%%%%%%%%%%%%%%%%%%%%%%%%%%%%%%%%%%%%%%%%%%%%%%%%%%%%
%%%%%%%%%%%%%%%%%%%%%%%%%%%%%%%%%%%%%%%%%%%%%%%%%%%%%%%%%%%%%%%%%%%%%%%%%%%%%%%%%%%
%%%%%%%%%%%%%%%%%%%%%%%%%%%%%%%%%%%%%%%%%%%%%%%%%%%%%%%%%%%%%%%%%%%%%%%%%%%%%%%%%%%
%%%%%%%%%%%%%%%%%%%%%%%%%%%%%%%%%%%%%%%%%%%%%%%%%%%%%%%%%%%%%%%%%%%%%%%%%%%%%%%%%%%

\clearpage
\pagebreak
\newpage

\onecolumngrid
\begin{center}
  \textbf{\large Supplemental Material: Cumulant expansion in the Holstein model: Spectral functions and mobility}\\[.2cm]
  Petar Mitri\'c, Veljko Jankovi\'c, Nenad Vukmirovi\'c, and Darko Tanaskovi\'c \\[.1cm]
  {\itshape Institute of Physics Belgrade,
University of Belgrade, Pregrevica 118, 11080 Belgrade, Serbia}
  \\[1cm]
\end{center}
\twocolumngrid

\setcounter{equation}{0}
\setcounter{figure}{0}
\setcounter{table}{0}
\setcounter{section}{0}
\setcounter{subsection}{0}% and the subsection counter
\renewcommand{\thesection}{\Roman{section}} 
\renewcommand{\thesubsection}{\thesection.\Roman{subsection}}

\renewcommand{\appendixname}{$\,$}
%\titlelabel{\thetitle\ $\! \!$. $\,\,$}

%\setcounter{page}{1}
\makeatletter
\renewcommand{\theequation}{S\arabic{equation}}
\renewcommand{\thefigure}{S\arabic{figure}}
\renewcommand{\bibnumfmt}[1]{[S#1]}
\renewcommand{\citenumfont}[1]{S#1}
\renewcommand{\thetable}{S\arabic{table}}

Here we supplement the main text by giving an alternative derivation of the cumulant function in Sec.~\ref{Supp:Sec:alternative_CE_derivation}, additional spectral functions and heat maps in Sec.~\ref{Supp:Sec:SpecF}, and a comparison of the 1D ground-state energy using DMFT, CE, SCMA, and MA in Sec.~\ref{Supp:Sec:QP_prop}. 

\section{Alternative derivation of the cumulant function in the CE method}
\label{Supp:Sec:alternative_CE_derivation}
\poc
In Sec.~XI of the Supplemental Material in Ref. \cite{Supp_2022_Mitric}, we showed that the Green's function, if there is only a single electron in the band,  can be written as
\begin{equation} \label{Eq:GreenF_single_el}
    G_{\bf k}(t) = -i \theta(t) \langle c_{\bf k}(t) c_{\bf k}^\dagger \rangle_{T,0},
\end{equation}
where:
\begin{subequations}
\begin{align}
    c_{\bf k}(t) &= e^{iHt} c_{\bf k} e^{-iHt}, \\
    H &= H_{\mathrm{el}} + H_{\mathrm{ph}} + H_{\mathrm{el-ph}}, \\
    H_{\mathrm{el}} &= -t_0 \sum_{\langle ij \rangle} \left( c_i^\dagger c_j + \mathrm{H.c.} \right) = \sum_{\bf k} \varepsilon_{\bf k} c_{\bf k}^\dagger c_{\bf k}, \\
    H_{\mathrm{ph}} &= \omega_0 \sum_i a_i^\dagger a_i = \omega_0 \sum_{\bf k} a_{\bf k}^\dagger a_{\bf k}, \\
    H_{\mathrm{el-ph}} &= -g \sum_i c_i^\dagger c_i \left( a_i^\dagger + a_i \right) \nonumber \label{Eq:Hamiltonian_el_ph}\\ &=  
    -\frac{g}{\sqrt{N}} \sum_{\bf k, \bf q} c_{\bf k+q}^\dagger c_{\bf k} \left( a_{\bf q} + a^\dagger_{\bf -q} \right).
\end{align}
\end{subequations}
Here, $N$ is the number of sites (we take $N\to\infty$ in order to get the thermodynamic limit), while $\langle \dots \rangle_{T,0}$ denotes the thermal average over the states with no electrons and 
 arbitrary number of phonons
\begin{equation}
    \langle x \rangle_{T,0} = \frac{\sum_{\phset} \langle 0, \phstate | e^{-H_{\mathrm{ph}}/T} x | 0, \phstate \rangle}{\sum_{\phset} \langle 0, \phstate | e^{-H_{\mathrm{ph}}/T} | 0, \phstate \rangle}.
\end{equation}
For the rest of this section, an arbitrary state with ${n_p}$ phonons and no electrons (since such state is not unique) will be denoted by $|0,{\phstate} \rangle$, while $\sum_{\phset}$ represents the sum over all possible phonon configurations. We also introduce $|{\bf k}, \phstate \rangle \equiv  c_{\bf k}^\dagger |0, \phstate \rangle$ and $Z_{\mathrm{ph}} = \sum_{\phset} \langle 0,{\phstate} | e^{-H_{\mathrm{ph}}/T} | 0, {\phstate} \rangle$.
\pas
Using the fact that $|0,{\phstate} \rangle$ is an eigenstate of both the full and the phononic Hamiltonian $H |0,{\phstate} \rangle = H_{\mathrm{ph}}|0,{\phstate} \rangle = n_p \omega_0 |0,\phstate \rangle$, we see how Eq.~\eqref{Eq:GreenF_single_el} can be written in a more explicit form 
\begin{equation} \label{Eq:derivation_cumualnt_intermediate_1}
    G_{\bf k}(t) = \frac{-i\theta(t)}{Z_{\mathrm{ph}}} \sum_{\phset} e^{i\omega_0 n_p t} e^{-n_p \omega_0/T}
    \langle 0, \phstate | c_{\bf k} e^{-iHt} c_{\bf k}^\dagger | 0, \phstate \rangle.
\end{equation}
The term  $e^{-iHt}$ can be read off from
\begin{equation} \label{Eq:evolution_operator_dirac_pic}
    e^{iH_{\mathrm{el}}t} e^{iH_{\mathrm{ph}}t} e^{-i H t} = T_t \exp \left[ -i \int_0^t dt_1 H_{\mathrm{el-ph}}^{(I)}(t_1) \right],
\end{equation}
which represents two different, but equivalent, forms for the evolution operator in the Dirac picture. Here, $H_{\mathrm{el-ph}}^{(I)}$ is the electron-phonon interaction part of the Hamiltonian in the Dirac picture and $T_t$ is the time-ordering operator.  For the purely phononic part $e^{-iH_{\mathrm{ph}}t}$ we use  $\langle 0,\phstate | e^{-iH_{\mathrm{ph}}t} = e^{-i\omega_0 n_p t} \langle 0,\phstate |$, while purely electronic part $e^{-iH_{\mathrm{el}}t}$ is dealt with analogously   $\langle 0,\phstate | c_{\bf k} e^{-i H_{\mathrm{el}} t} = e^{-i\varepsilon_{\bf k}t} \langle 0,\phstate | c_{\bf k}$. Hence, Eq.~\eqref{Eq:derivation_cumualnt_intermediate_1} becomes
\begin{widetext}
\begin{subequations}
\begin{align} 
    G_{\bf k}(t) &= -\frac{i\theta(t)}{Z_{\mathrm{ph}}} e^{-i\varepsilon_{\bf k}t} \sum_{\phset} e^{-n_p \omega_0 /T} \left\langle 0,\phstate \middle|
    c_{\bf k} T_t \exp  \left[ -i \int_0^t dt_1 H_{\mathrm{el-ph}}^{(I)} (t_1) \right] c_{\bf k}^\dagger \middle| 0,\phstate \right\rangle \\
    &\equiv -i\theta(t) e^{-i\varepsilon_{\bf k}t} \left\langle T_t e^{-i \int_0^t dt_1 H_{\mathrm{el-ph}}^{(I)}(t_1)} \right\rangle_{T,{\bf k}}. \label{Eq:Green_kubo}
\end{align}
\end{subequations}
\end{widetext}
The expressions of the form~\eqref{Eq:Green_kubo} have been extensively studied in the past. As shown in Eq.~(6.10) of Kubo's cumulant paper \cite{Supp_1962_Kubo}, the expectation value with the time-ordering can be written as
\begin{widetext}
\begin{equation} \label{Eq:kubo_cumulant_formula}
    \left\langle T_t e^{-i \int_0^t dt_1 H_{\mathrm{el-ph}}^{(I)}(t_1)} \right\rangle_{T,{\bf k}} = 
    \exp  \left\langle T_t e^{-i \int_0^t dt_1 H_{\mathrm{el-ph}}^{(I)}(t_1)} -1 \right\rangle_{T,{\bf k},c} 
    \equiv e^{C_{\bf k}(t)},
\end{equation}
\end{widetext}
where we defined the cumulant function $C_{\bf k}(t)$. The notation $\langle \dots \rangle_c$ denotes the so-called cumulant average. For our present purposes, we only need to know how the first two cumulant averages are defined:
\begin{subequations}
\begin{align}
    \langle X_1 \rangle_c &= \langle X_1 \rangle \label{Eq:cumulant_vs_moment_1}\\ 
    \langle X_1 X_2 \rangle_c &= \langle X_1 X_2 \rangle - \langle X_1 \rangle \langle X_2 \rangle. \label{Eq:cumulant_vs_moment_2}
\end{align}
\end{subequations}
In general, the cumulant average is defined using the ordinary average, by formally expanding the following expression  in the Taylor series with respect to $\xi_i$ and equating, order by order, the terms on the left- and the right-hand side
\begin{equation}
    \left \langle \exp \sum_j \xi_j X_j \right\rangle = \exp \left\langle \left( \exp \sum_j \xi_j X_j \right) - 1  \right\rangle_c.
\end{equation}
The $-1$ term on the right-hand side is motivated by the fact that the expectation value of the unity operator is equal to $1$. While our paper  focuses on the cumulant of the second-order, there is actually an analytic formula that relates the cumulant average of any order with the ordinary average \cite{Supp_1957_Meeron}. 
\pas
Let us now go back to Eq.~\eqref{Eq:Green_kubo} and use Eq.~\eqref{Eq:kubo_cumulant_formula} to obtain
\begin{equation}
    G_{\bf k}(t) = -i \theta(t) e^{-i\varepsilon_{\bf k}t} e^{C_{\bf k}(t)},
\end{equation}
where 
\begin{equation}
    C_{\bf k}(t) = \sum_{j=1}^\infty \left\langle T_t \frac{(-i)^j}{j!} \int_0^t \prod_{m=1}^j dt_m 
    H_{\mathrm{el-ph}}^{(I)} (t_m) \right\rangle_{T,{\bf k}, c}.
\end{equation}
So far, everything was exact. The approximation, that we now introduce, consists of keeping only the first two terms in the previous equation ($j=1$ and $j=2$ terms) while neglecting everything else. This is known as the second-order cumulant expansion. In the $j=1$ term, the cumulant average coincides with the ordinary average (see Eq.~\eqref{Eq:cumulant_vs_moment_1}), and hence vanishes due to Wick's theorem. As a consequence, the cumulant average can be simply replaced by the ordinary average in the case of $j=2$ term as well; see Eq.~\eqref{Eq:cumulant_vs_moment_2}. Therefore, the second-order cumulant function reads as
\begin{widetext}
\begin{equation} \label{Eq:soc_deriv_1}
    C_{\bf k}(t) = -\frac{1}{2} \int_0^t dt_1 \int_0^t dt_2 \left \langle 
    T_t c_{\bf k} H_{\mathrm{el-ph}}^{(I)}(t_1) H_{\mathrm{el-ph}}^{(I)}(t_2) c_{\bf k}^\dagger 
    \right \rangle_{T,0}.
\end{equation}
\end{widetext}
For a straightforward application of Wick's theorem, it is customary to rewrite electron creation  and annihilation operators in the Dirac picture. In order not to change the already existing time ordering in Eq.~\eqref{Eq:soc_deriv_1}, the annihilation operator is expressed in the final time $c_{\bf k} = e^{i\varepsilon_{\bf k}t} c_{\bf k}^{(I)}(t)$, while the creation operator is expressed in the initial time $c_{\bf k}^\dagger = c_{\bf k}^{\dagger (I)}(0)$. If we also use the explicit form of $H_{\mathrm{el-ph}}^{(I)}(t)$ from Eq.~\eqref{Eq:Hamiltonian_el_ph}, the Eq.~\eqref{Eq:soc_deriv_1} becomes
\begin{widetext}
\begin{equation} \label{Eq:cumulant_before_wick}
    C_{\bf k}(t) = -\frac{g^2}{2N} e^{i\varepsilon_{\bf k} t} \int_0^t dt_1 \int_0^t dt_2 
    \left\langle 
    T_t c_{\bf k}^{(I)}(t) \sum_{{\bf k_1, q_1}} c_{\bf k_1 + q_1}^{\dagger (I)}(t_1) 
    c_{\bf k_1}^{(I)}(t_1)
    A^{(I)}_{\bf q_1}(t_1)
    \sum_{\bf k_2, q_2} c_{\bf k_2 + q_2}^{\dagger (I)} c_{\bf k_2}^{(I)}(t_2) A_{\bf q_2}^{(I)}(t_2) c_{\bf k}^{\dagger (I)}(0)
    \right\rangle_{T,0},
\end{equation}
\end{widetext}
where we introduced the shorthand notation for the phonon part $A_{\bf q} = a_{\bf q} + a_{\bf -q}^\dagger$. Eq.~\eqref{Eq:cumulant_before_wick} is now straightforwardly evaluated using Wick's theorem. Contraction between the phonon degrees of freedom gives \cite{Supp_2000_Mahan}
\begin{equation} 
    \left\langle T_t A_{\bf q_1}^{(I)}(t_1) A_{\bf q_2}^{(I)}(t_2) \right\rangle = 
    \delta_{\bf q_1, -q_2} i D(t_1 - t_2),
\end{equation}
where $iD(t_1 - t_2) = (n_{ph}+1) e^{-i\omega_0 |t_1 - t_2|}  + n_{ph} e^{i\omega_0 |t_1 - t_2|}$ is the phonon propagator, while $n_{ph} = 1/(e^{\omega_0/T} - 1)$ is the Bose factor. Since we are working in the limit of vanishing electron density (single electron in a band), the contraction between the electron creation and annihilation operators does not have a hole part, and hence reads as
\begin{equation}
    \left\langle T_t c_{\bf k}^{(I)}(t_1) c_{\bf q}^{\dagger (I)}(t_2) \right\rangle = 
    \delta_{\bf k,q} \; e^{-i\varepsilon_{\bf k}|t_1 - t_2|} \theta(t_1 - t_2).
\end{equation}
Taking all of this into account, Eq.~\eqref{Eq:cumulant_before_wick} simplifies 
\begin{equation} \label{Eq:soc_deriv_2}
    C_{\bf k}(t) = -\frac{g^2}{2N} \sum_{\bf q} \int_0^t dt_1 \int_0^t dt_2 e^{i(\varepsilon_{\bf k} - \varepsilon_{\bf q})|t_2 - t_1|} iD(t_2-t_1).
\end{equation}
We can get rid of the absolute value by noticing that the contributions for $t_2>t_1$ and for $t_2<t_1$ are equal. It is thus sufficient to restrict ourselves to $t_2>t_1$ and multiply everything by $2$. Also, the expression can be further simplified if we use 
$$e^{i(\varepsilon_{\bf k}-\varepsilon_{\bf q} \pm \omega_0)(t_2-t_1)} = \int_{-\infty}^\infty d\omega e^{-i\omega (t_2 - t_1)} \delta(\omega + \varepsilon_{\bf k} - \varepsilon_{\bf q} \pm \omega_0).$$
Then, the whole $\bf q$ dependence is inside the Dirac delta function, which in combination with the summation over $\bf q$ gives 
\begin{equation}
    \sum_{\bf q}  \delta(\omega + \varepsilon_{\bf k} - \varepsilon_{\bf q} \pm \omega_0) = N \rho(\omega + \varepsilon_{\bf k} \pm \omega_0),
\end{equation}
where $\rho$ is the density of states. It is now straightforward to show that Eq.~\eqref{Eq:soc_deriv_2} reduces to
\begin{align}
    C_{\bf k}(t) &= 
    g^2 \int_{-\infty}^\infty d\omega \frac{e^{-i\omega t} + i\omega t - 1}{\omega^2} \nonumber \\
    &\times
    \left[
    (n_{ph}+1) \rho(\omega + \varepsilon_{\bf k} - \omega_0) +
    n_{ph} \rho(\omega + \varepsilon_{\bf k} + \omega_0)
    \right].
\end{align}
This expression can be rewritten in terms of the Migdal self-energy (see Eq.~(14) from the main text) as follows
\begin{equation} \label{Eq:Cumulant_appendix_final}
    C_{\bf k}(t) = \frac{1}{\pi} \int_{-\infty}^\infty d\omega 
    \frac{|\mathrm{Im}\Sigma^{\mathrm{MA}} (\omega + \varepsilon_{\bf k})|}{\omega^2} 
    (e^{-i\omega t} + i\omega t - 1).
\end{equation}
Hence, we gave an alternative derivation of the cumulant function $C_{\bf k}(t)$, where the self-energy in the Migdal approximation emerges more explicitly than in Eq.~(7) of the main text.
\pas
We note that the cumulant expansion method that we have now presented is 
 analogous to the linked cluster expansion for the thermodynamic potential $F$ in statistical mechanics. This is a consequence of the same mathematical form of $C_{\bf k} (t) = \ln \left( G_{\bf k}(t)/G_{{\bf k}, 0}(t) \right)$ and $F = \ln (Z/Z_0)$, where $Z$ and $Z_0$ are the partition function of the full and noninteracting theories.

\vfill
\clearpage
\newpage

\section{Spectral Functions} \label{Supp:Sec:SpecF}
\begin{figure}[!b]
  \includegraphics[width=3.2in,trim=0cm 0cm 0cm 0cm]{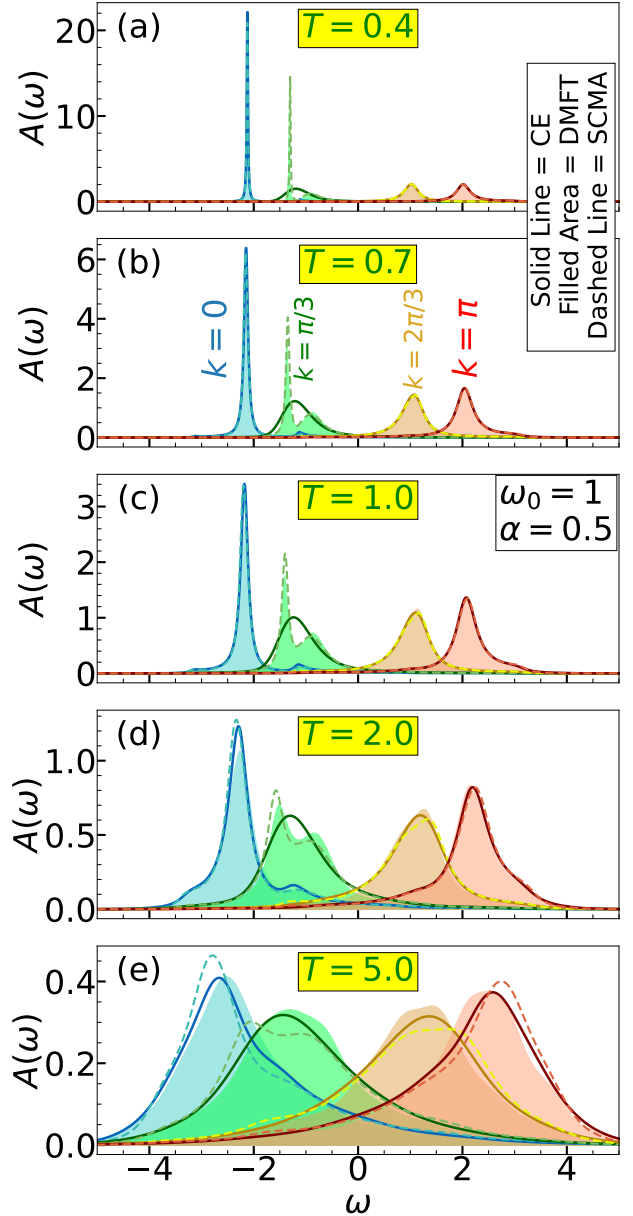} 
 \caption{(a)--(e) Comparison of CE, DMFT, and SCMA spectral functions in the weak coupling regime, for a wide range of temperatures. Here $t_0=\omega_0=1$ and $\alpha=0.5$.
 }
 \label{AppFig:SpecF_w=1_weak}
\end{figure}

\begin{figure*}
\setcounter{subfigure}{0}
%\captionsetup[subfigure]{format=hang,singlelinecheck=false,justification=RaggedRight}
\captionsetup[subfigure]{format=hang,singlelinecheck=false,justification=RaggedRight, labelsep=space }
\renewcommand\thesubfigure{\roman{subfigure}}
\begin{center}
\begin{minipage}[t]{\linewidth}
\centering
\subfloat[ \label{AppFig:SpecF_w=1_k=0}  (a)--(h) Spectral functions for $\omega_0=1$ and $k=0$. In the left panels $T=0.4$, while $T=1$ in the right panels. Insets show the integrated spectral weights $I(\omega)=\int_{-\infty}^\infty A(\omega) d\omega$.]{ \includegraphics[width=0.48\linewidth]{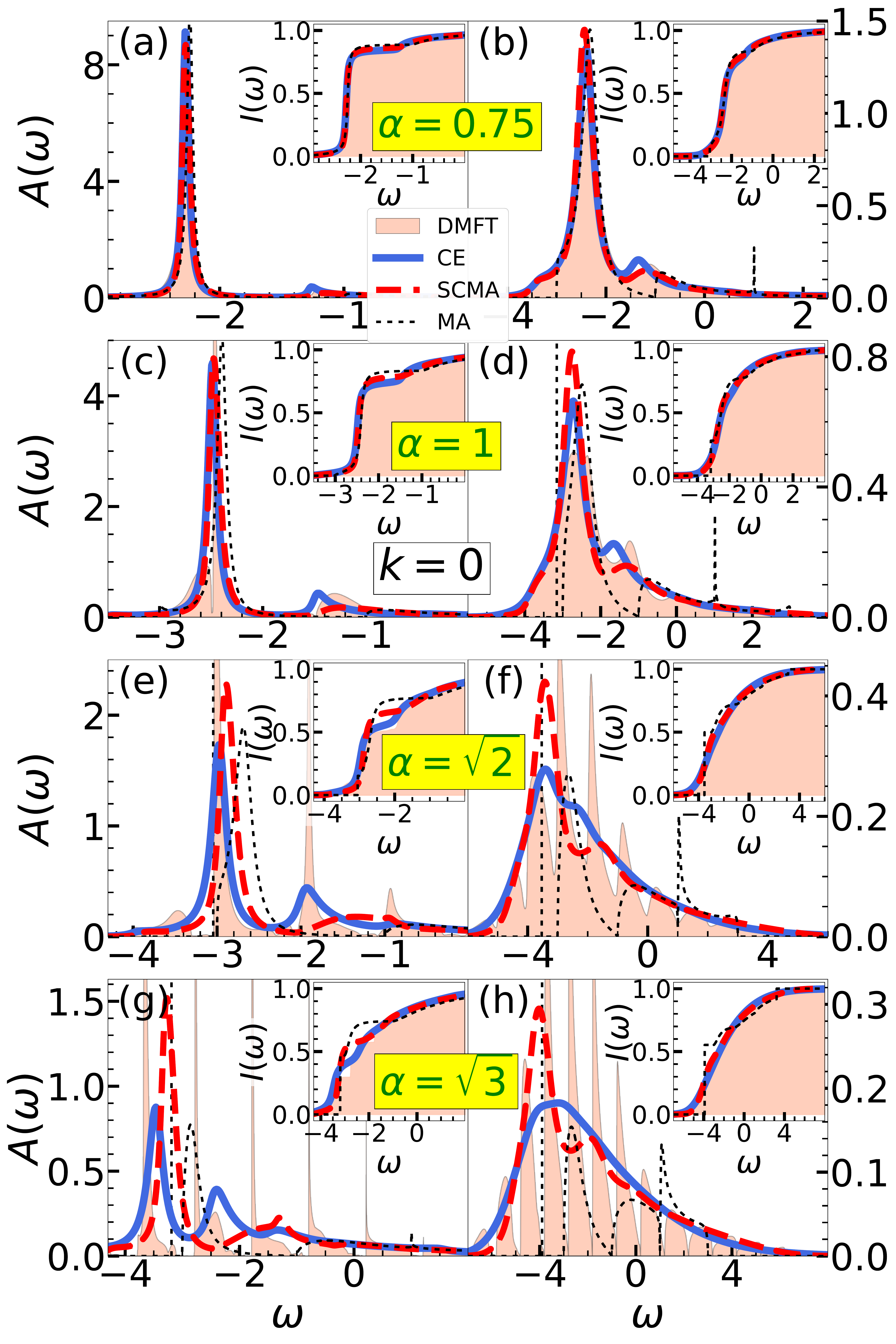}}
%\quad
\;\;\;
\subfloat[ \label{AppFig:SpecF_w=1_k=pi} (a)--(h) Spectral functions for $\omega_0=1$ and $k=\pi$. In the left panels $T=0.4$, while $T=1$ in the right panels. Insets show the integrated spectral weights $I(\omega)=\int_{-\infty}^\infty A(\omega) d\omega$.]{\includegraphics[width=0.48\linewidth]{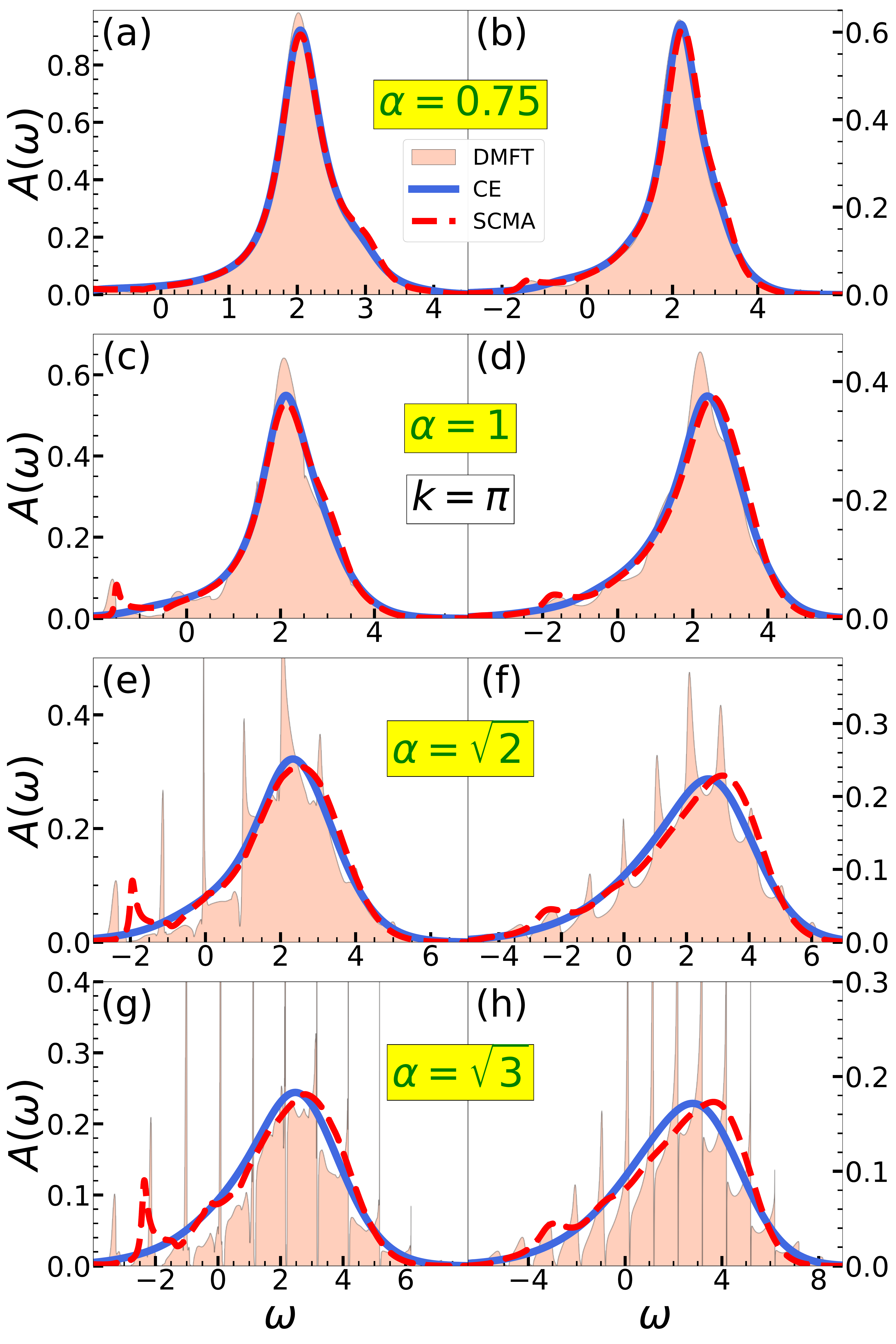}}
\end{minipage}
\end{center}
\hfill
\begin{center}
\begin{minipage}[b]{\linewidth}
\subfloat[ \label{AppFig:SpecF_w=1_k=0,pi_highT} \centering Spectral functions at higher 
temperatures for $\omega_0=1$ and $k=0,\pi$.]{\includegraphics[width=\linewidth]{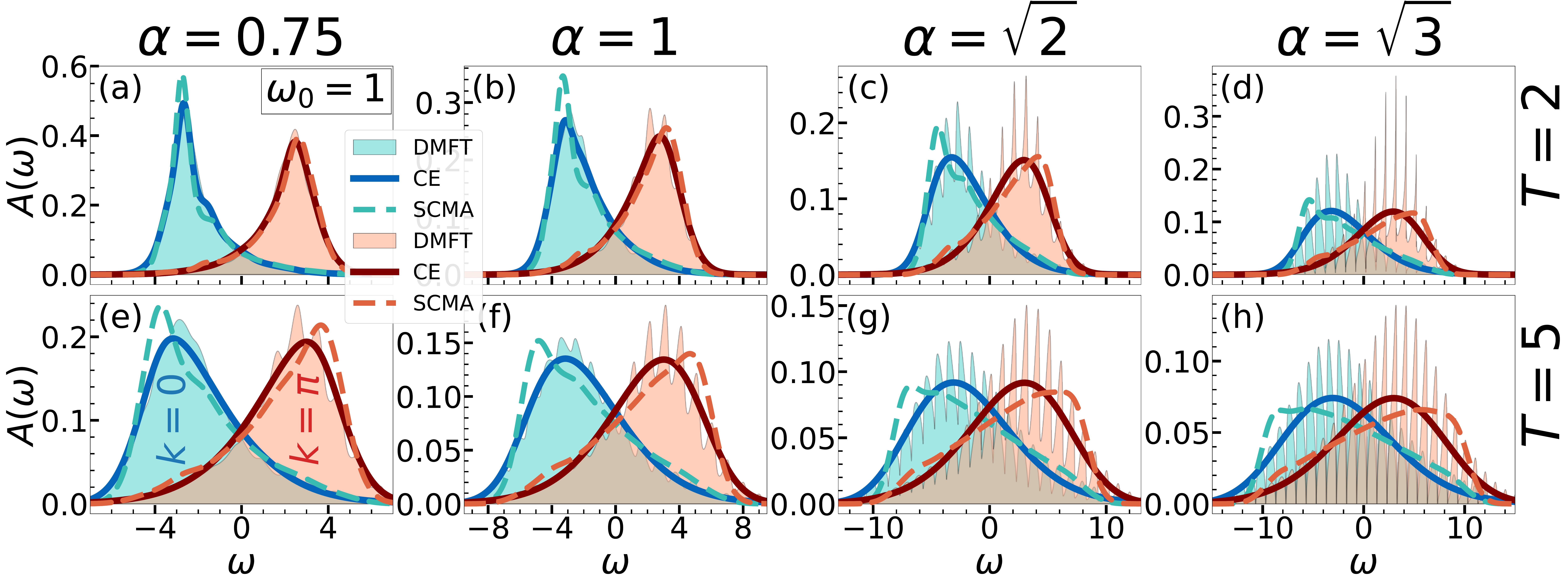}}
\end{minipage}
\end{center}

\caption[Impact of ...]{Comparison of the CE, DMFT, SCMA, and MA spectral functions in 1D  for $t_0=\omega_0 = 1$.}
\label{AppFig:SpecF_w=1_k=0,pi}
\end{figure*}

In Sec.~III of the main text, we presented  spectral functions $A(\omega)$ and heat maps for $\omega_0=0.5$. Here, we present a large number of results for  $\omega_0=1$, $\omega_0=0.2$, as well as some additional results for $\omega_0=0.5$ that are organized as follows:
\begin{enumerate}
    \item Results for $\omega_0=1$:
    \begin{itemize}
        \item Fig.~\ref{AppFig:SpecF_w=1_weak}: $A(\omega)$ in the weak coupling regime for a wide range of temperatures and momenta.
        \item Fig.~\ref{AppFig:SpecF_w=1_k=0,pi}:         $A(\omega)$ in the weak, intermediate and strong electron-phonon coupling regimes for $k=0$ and $k=\pi$:
        \begin{itemize}
            \item Fig.~\ref{AppFig:SpecF_w=1_k=0}: $k=0$ at $T=0.4$ and $T=1$.
            \item Fig.~\ref{AppFig:SpecF_w=1_k=pi}: $k=\pi$ at $T=0.4$ and $T=1$.
            \item Fig.~\ref{AppFig:SpecF_w=1_k=0,pi_highT}: $k=0,\pi$ at $T=2$ and $T=5$.
        \end{itemize}
        \item Fig.~\ref{AppFig:SpecF_w=1_k=pi3,2pi3}:         $A(\omega)$ in the weak, intermediate and strong electron-phonon coupling regimes for $k=\pi/3$ and $k=2\pi/3$:
        \begin{itemize}
            \item Fig.~\ref{AppFig:SpecF_w=1_k=pi3,2pi3_lowT}: $T=0.4$.
            \item Fig.~\ref{AppFig:SpecF_w=1_k=pi3,2pi3_moderateT}: $T=1$.
            \item Fig.~\ref{AppFig:SpecF_w=1_k=pi3,2pi3_highT}: $T=2$ and $T=5$.
        \end{itemize}
        \item Fig.~\ref{AppFig:HeatPlots_w=1}: Heat maps 
        \begin{itemize}
            \item Fig.~\ref{AppFig:HeatPlots_w=1_T=0.4}:  $T=0.4$.
            \item Fig.~\ref{AppFig:HeatPlots_w=1_T=1}:  $T=1$.
        \end{itemize}
    \end{itemize}
    \item Results for $\omega_0=0.5$:
    \begin{itemize}
        \item Fig.~\ref{AppFig:SpecF_w=1_k=pi3,2pi3}: $A(\omega)$ in the weak, intermediate, and strong electron-phonon coupling regimes for $k=\pi/3$ and $k=2\pi/3$:
        \begin{itemize}
            \item Fig.~\ref{AppFig:SpecF_w=0.5_k=pi3,2pi3_T=0.3}: $T=0.3$.
            \item Fig.~\ref{AppFig:SpecF_w=0.5_k=pi3,2pi3_T=0.7}: $T=0.7$.
            \item Fig.~\ref{AppFig:SpecF_w=0.5_k=pi3,2pi3_highT}: $T=2$ and $T=5$.
        \end{itemize}
    \end{itemize}
    \item Results for $\omega_0=0.2$:
    \begin{itemize}
        \item Fig.~\ref{AppFig:SpecF_w=0.2_k=0,pi}: $A(\omega)$ in the weak, intermediate and strong coupling regimes for $k=0$ and $k=\pi$:
        \begin{itemize}
            \item Fig.~\ref{SAppFig:SpecF_w=0.2_k=0}: $k=0$ at $T=0.3$ and $T=0.7$.
            \item Fig.~\ref{AppFig:SpecF_w=0.2_k=pi}: $k=\pi$ at $T=0.3$ and $T=0.7$.
            \item Fig.~\ref{AppFig:SpecF_w=0.2_k=0,pi_highT}: $k=0,\pi$ at $T=2$ and $T=5$.
        \end{itemize}
        \item Fig.~\ref{AppFig:SpecF_w=0.2_k=pi3,2pi3}: $A(\omega)$ in the weak, intermediate and strong coupling regimes for $k=\pi/3$ and $k=2\pi/3$:
        \begin{itemize}
            \item Fig.~\ref{AppFig:SpecF_w=0.2_k=pi3,2pi3_T=0.3}: $T=0.3$.
            \item Fig.~\ref{AppFig:SpecF_w=0.2_k=pi3,2pi3_T=0.7}: $T=0.7$.
            \item Fig.~\ref{AppFig:SpecF_w=0.2_k=pi3,2pi3_highT}: $T=2$ and $T=5$.
        \end{itemize}
        \item Fig.~\ref{AppFig:HeatPlots_w=0.2}: Heat maps:
        \begin{itemize}
            \item Fig.~\ref{AppFig:HeatPlots_w=0.2_T=0.4}: $T=0.3$.
            \item Fig.~\ref{AppFig:HeatPlots_w=0.2_T=1}: $T=0.7$.
        \end{itemize}
    \end{itemize}
\end{enumerate}

%
%\clearpage
\newpage

\onecolumngrid
\twocolumngrid

%%%%%%%%%%%%%%%%%%%%%%%%%%%%%%%%%%%%%%%%%%%%%%%%%%%%%%
%%%%%%%%%%%%%%%%%%%%%%%%%%%%%%%%%%%%%%%%%%%%%%%%%%%%%
%%%%%%%%%%%%%%%%%%%%%%%%%%%%%%%%%%%%%%%%%%%%%%%%%%%%%%%

\onecolumngrid
\begin{figure*}
\setcounter{subfigure}{0}
%\captionsetup[subfigure]{format=hang,singlelinecheck=false,justification=RaggedRight}
\captionsetup[subfigure]{format=hang,singlelinecheck=false,justification=RaggedRight, labelsep=space }
\renewcommand\thesubfigure{\roman{subfigure}}
\begin{center}
\begin{minipage}[t]{\linewidth}
\centering
\subfloat[ \label{AppFig:SpecF_w=1_k=pi3,2pi3_lowT} (a)--(h) Spectral functions for $\omega_0=1$ and  $T=0.4$.  In the left panels $k=\pi/3$, while $k=2\pi/3$ in the right panels.]{\includegraphics[width=0.48\linewidth]{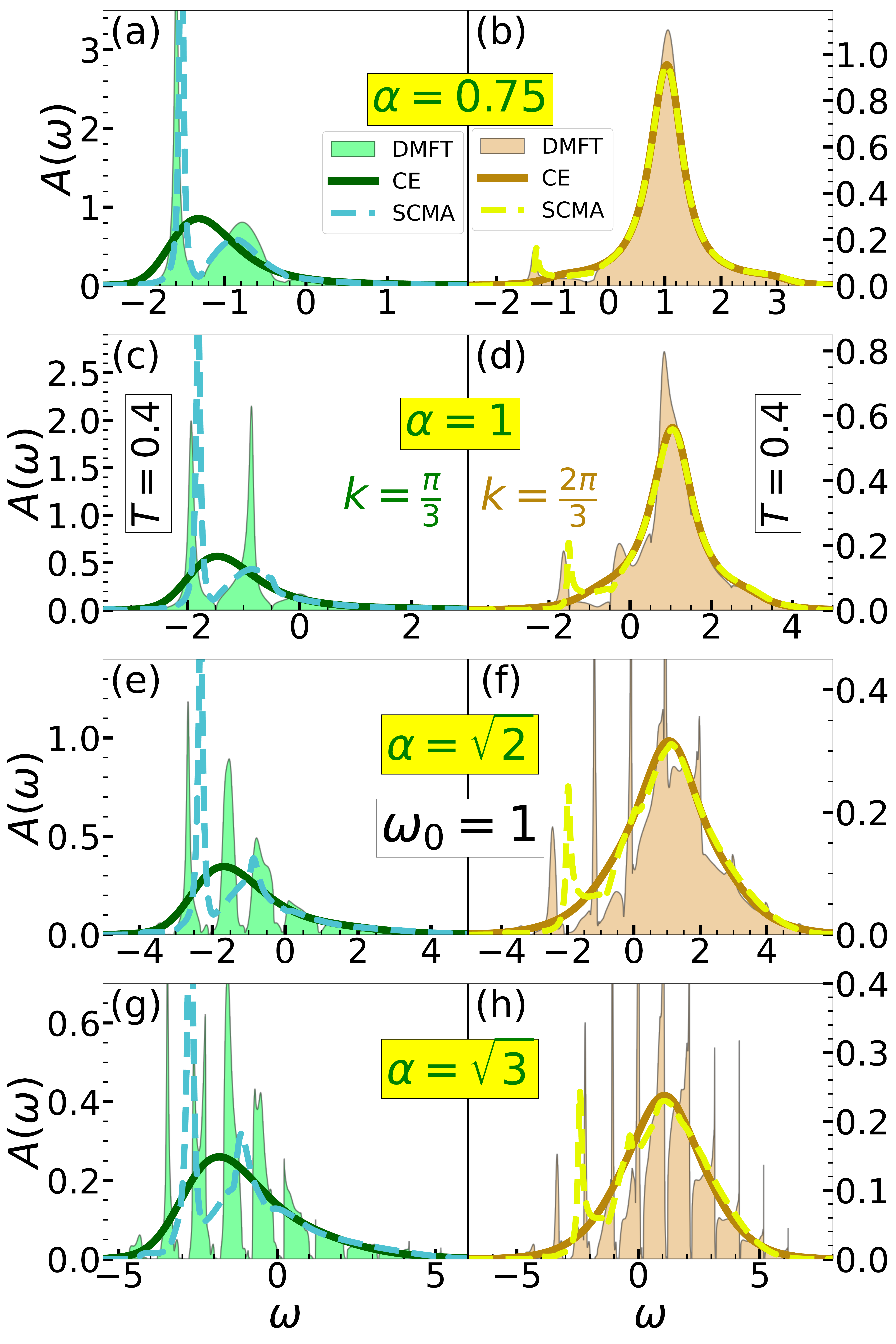}}
%\quad
\;\;\;
\subfloat[ \label{AppFig:SpecF_w=1_k=pi3,2pi3_moderateT} (a)--(h) Spectral functions for $\omega_0=1$ and $T = 1$. In the left panels $k=\pi/3$, while $k=2\pi/3$ in the right panels. ]{\includegraphics[width=0.48\linewidth]{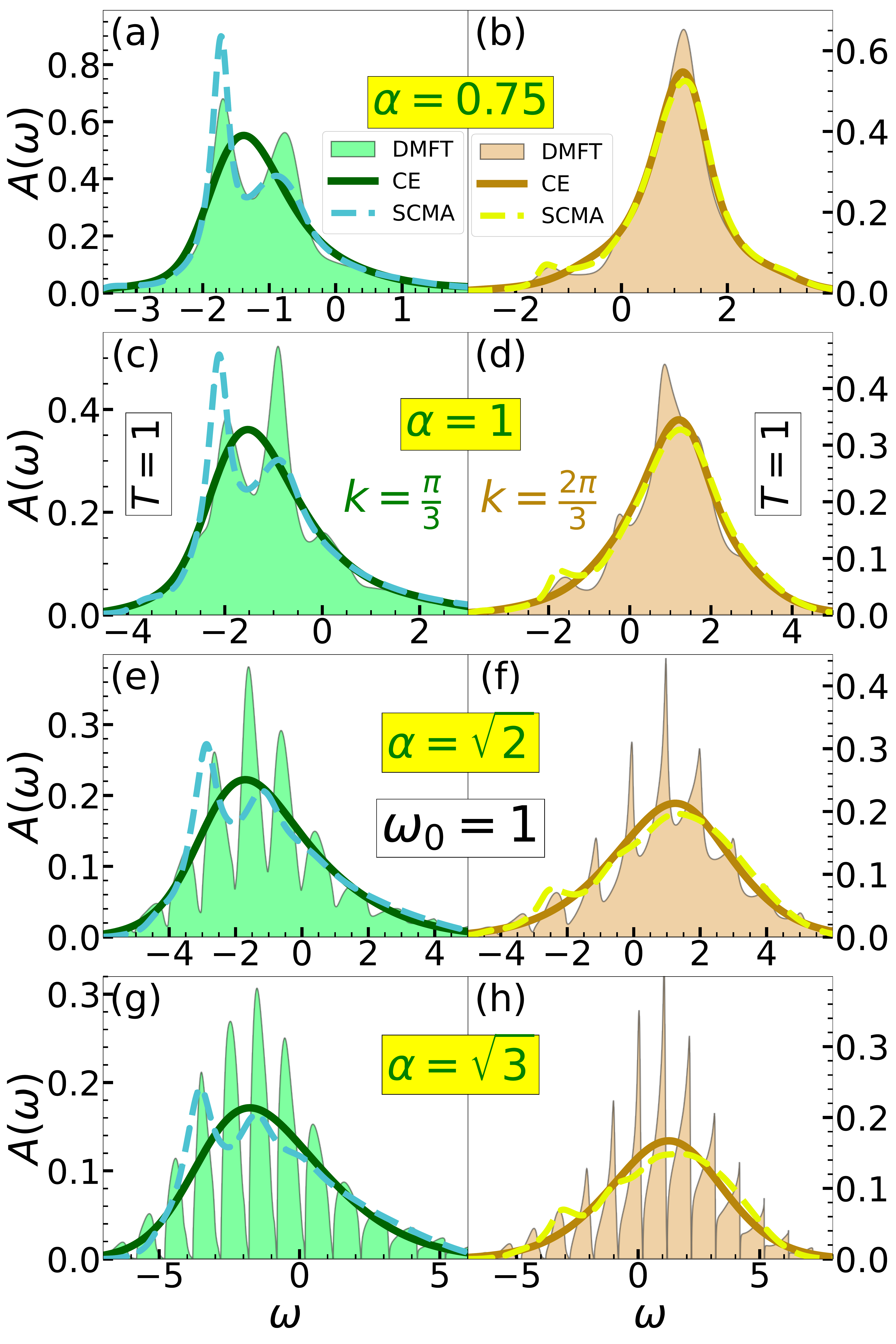}}
\end{minipage}
\end{center}
\hfill
\begin{center}
\begin{minipage}[b]{\linewidth}
\subfloat[ \label{AppFig:SpecF_w=1_k=pi3,2pi3_highT} \centering Spectral functions at higher 
temperatures for $\omega_0=1$ and $k=\pi/3, 2\pi /3$.]{\includegraphics[width=\linewidth]{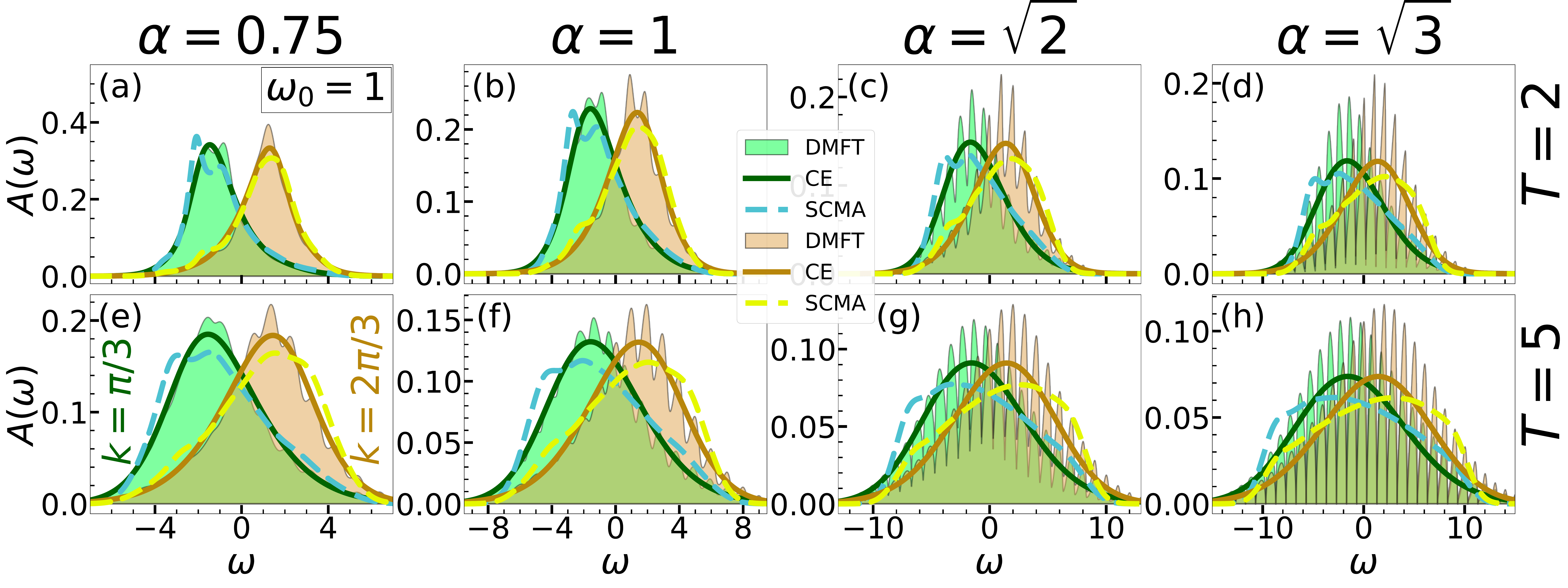}}
\end{minipage}
\end{center}

\caption[Impact of ...]{Comparison of the CE, DMFT, and SCMA   spectral functions in 1D  for $t_0 = \omega_0 = 1$ and $k=\pi/3, 2\pi /3 $.}
\label{AppFig:SpecF_w=1_k=pi3,2pi3}
\end{figure*}
\twocolumngrid

%%%%%%%%%%%%%%%%%%%%%%%%%%%%%%%%%%%%%%%%%%%%%%%%%%
%%%%%%%%%%%%%%%%%%%%%%%%%%%%%%%%%%%%%%%%%%%%%%%%%%
%%%%%%%%%%%%%%%%%%%%%%%%%%%%%%%%%%%%%%%%%%%%%%%%%%

\onecolumngrid
\begin{figure*}
\setcounter{subfigure}{0}
\captionsetup[subfigure]{format=hang,singlelinecheck=false,justification=RaggedRight, labelsep=quad }
\renewcommand\thesubfigure{\roman{subfigure}}
\begin{center}
\begin{minipage}[t]{\linewidth}
\centering
\subfloat[ \label{AppFig:HeatPlots_w=1_T=0.4} (a)--(h) Heat maps for $T=0.4$. In the left panels, we present CE results, while the DMFT benchmark is presented in
the right panels. All plots use the same color coding.]{\includegraphics[width=0.48\linewidth]{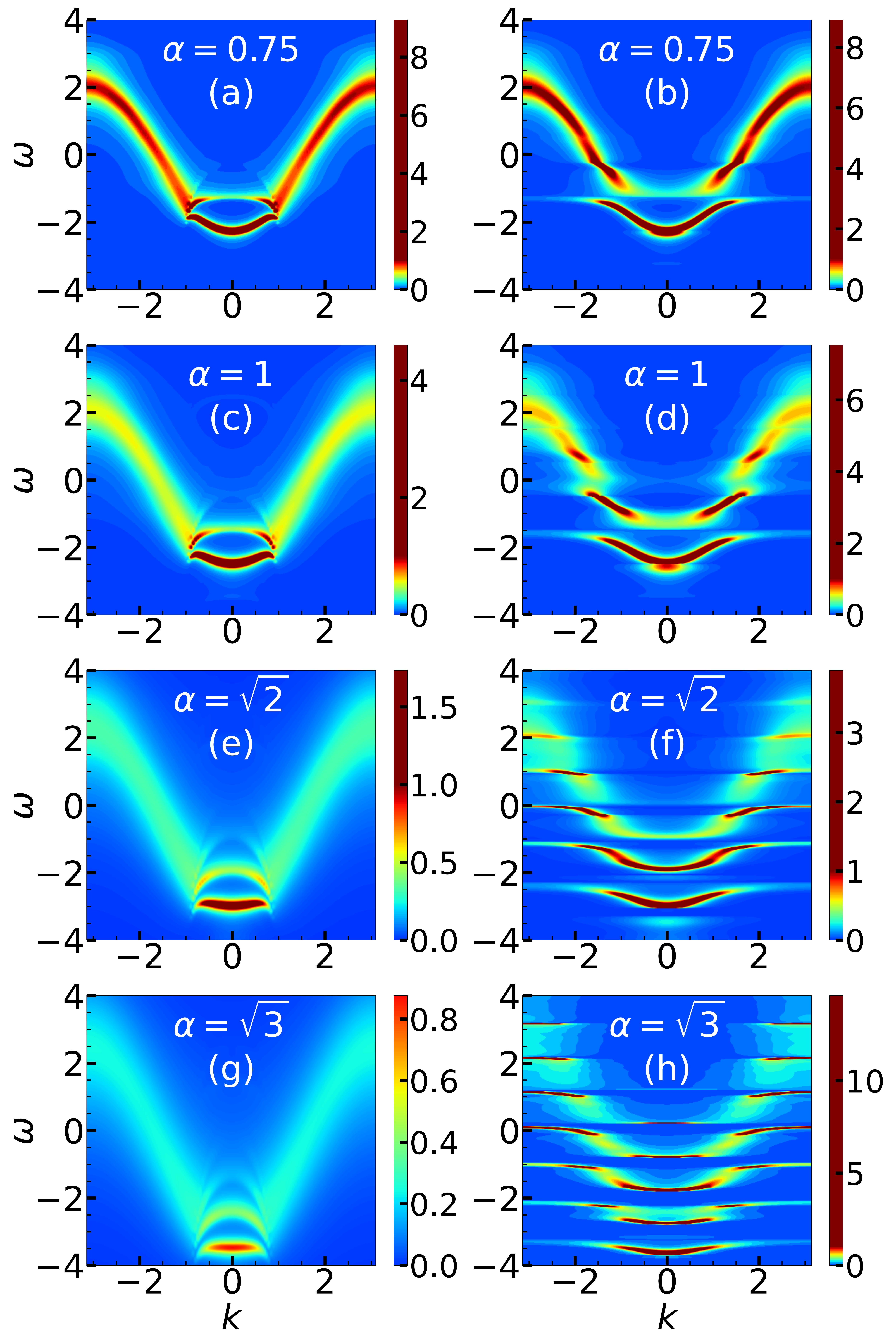}}
%\quad
\;\;\;
\subfloat[ \label{AppFig:HeatPlots_w=1_T=1} (a)--(h) Heat maps for $T=1$.  In the left panels, we present CE results, while the DMFT benchmark is presented in
the right panels. All plots use the same color coding.]{\includegraphics[width=0.48\linewidth]{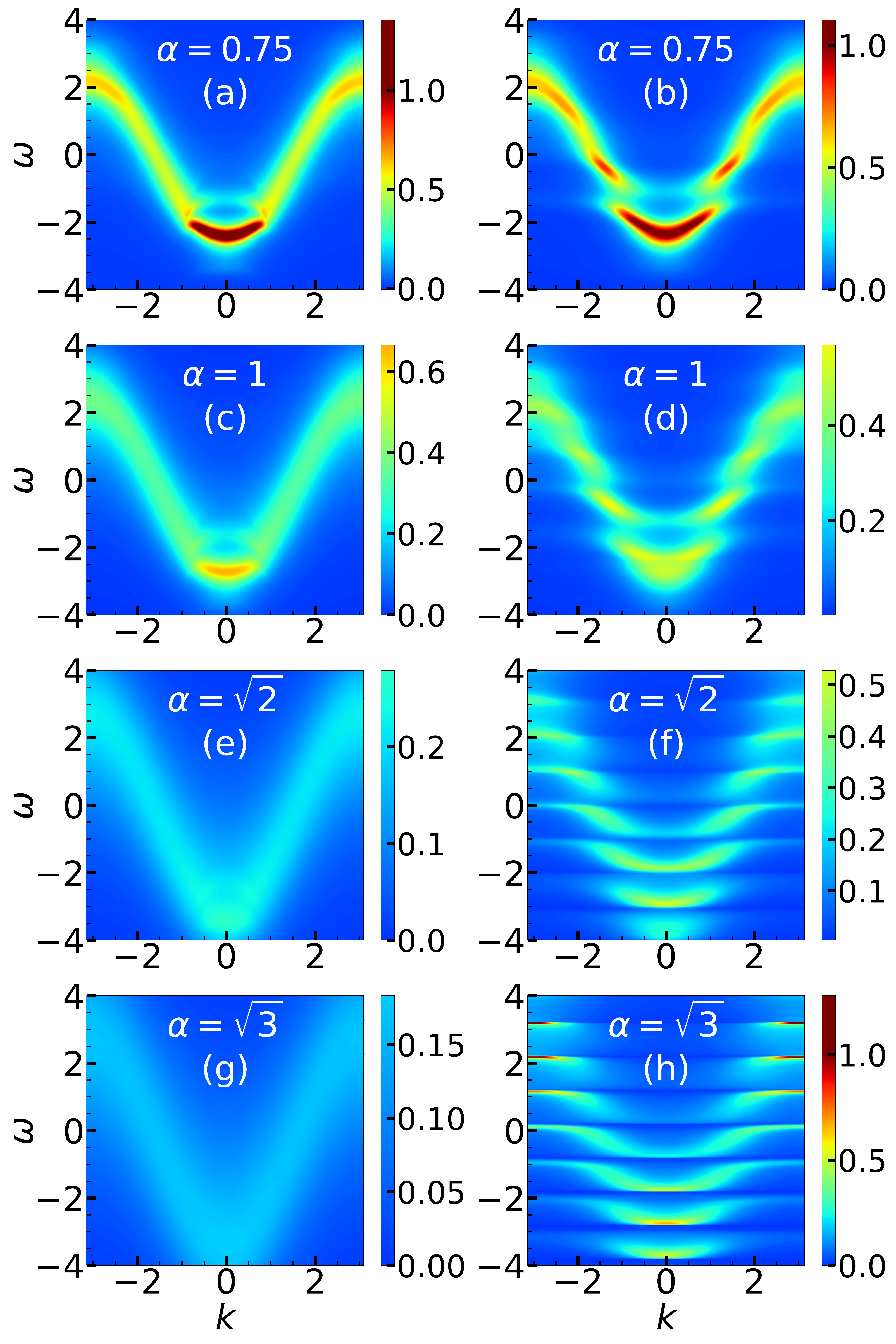}}
\end{minipage}
\end{center}
% \hfill
% \begin{center}
% \begin{minipage}[b]{\linewidth}
% \subfloat[ \label{SubFig:SpecF_w_1_highT} Spectral functions at higher 
% temperatures.]{\includegraphics[width=\linewidth]{./Pics/HighT_w=1.0_other_k.pdf}}
% \end{minipage}
% \end{center}

\caption[Impact of ...]{Comparison of the CE and  DMFT heat maps for $t_0=\omega_0 = 1$.}
\label{AppFig:HeatPlots_w=1}
\end{figure*}
\twocolumngrid
%%%%%%%%%%%%%%%%%%%%%%%%%%%%%%%%%%%%%%%%%%%%%%%%%%%%%%%
%%%%%%%%%%%%%%%%%%%%%%%%%%%%%%%%%%%%%%%%%%%%%%%%%%%%%%%%
%%%%%%%%%%%%%%%%%%%%%%%%%%%%%%%%%%%%%%%%%%%%%%%%%%%%%

\onecolumngrid
\begin{figure*}
\setcounter{subfigure}{0}
\captionsetup[subfigure]{format=hang,singlelinecheck=false,justification=RaggedRight, labelsep=space }
\renewcommand\thesubfigure{\roman{subfigure}}
\begin{center}
\begin{minipage}[t]{\linewidth}
\centering
\subfloat[ \label{AppFig:SpecF_w=0.5_k=pi3,2pi3_T=0.3} (a)--(h) Spectral functions for $\omega_0=0.5$ and $T=0.3$. In the left panels $k=\pi / 3$, while $k=2\pi /3 $ in the right panels. ]{\includegraphics[width=0.48\linewidth]{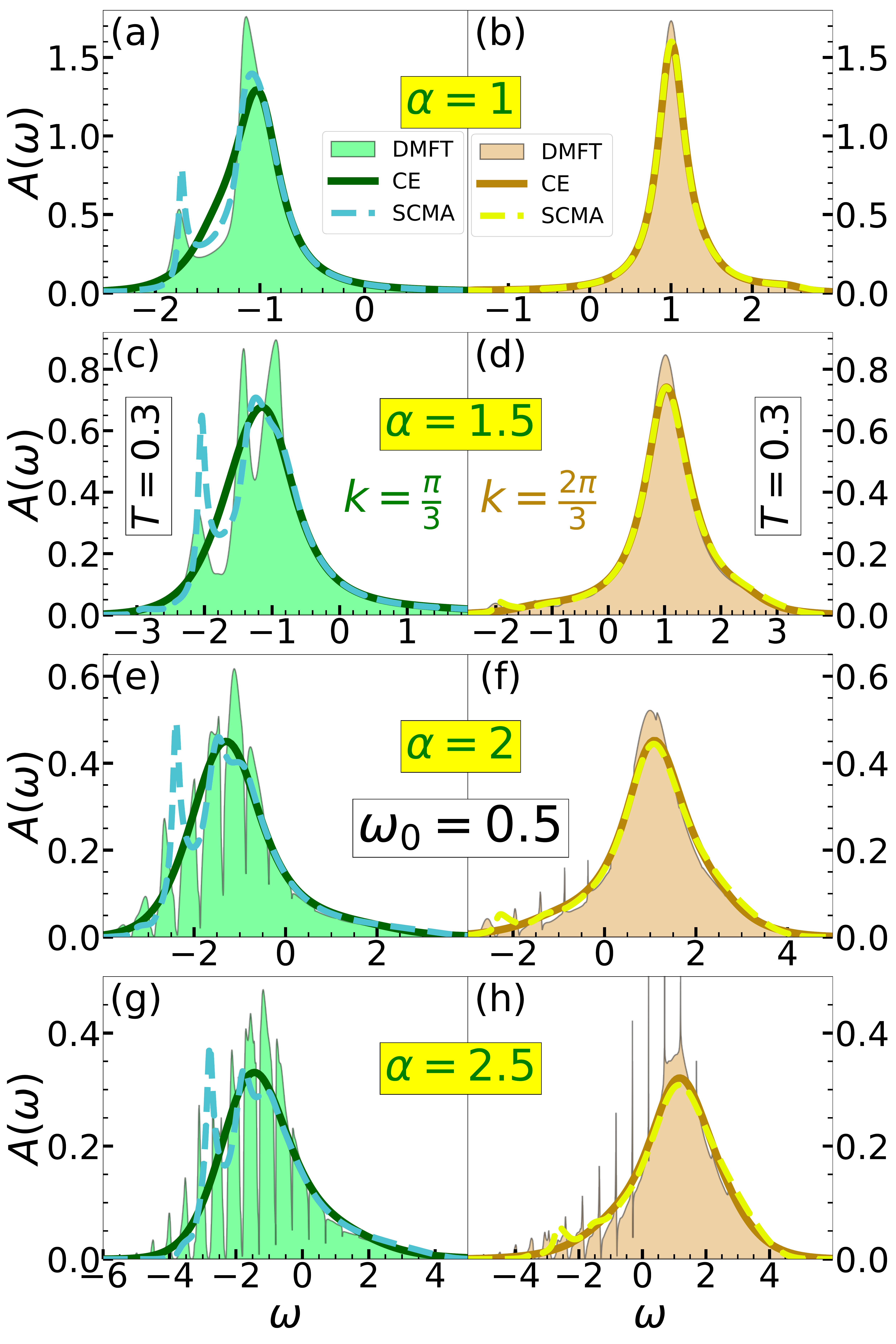}}
%\quad
\;\;\;
\subfloat[ \label{AppFig:SpecF_w=0.5_k=pi3,2pi3_T=0.7} (a)--(h) Spectral functions for $\omega_0=0.5$ and $T=0.7$. In the left panels $k=\pi / 3$, while $k=2\pi /3 $ in the right panels. ]{\includegraphics[width=0.48\linewidth]{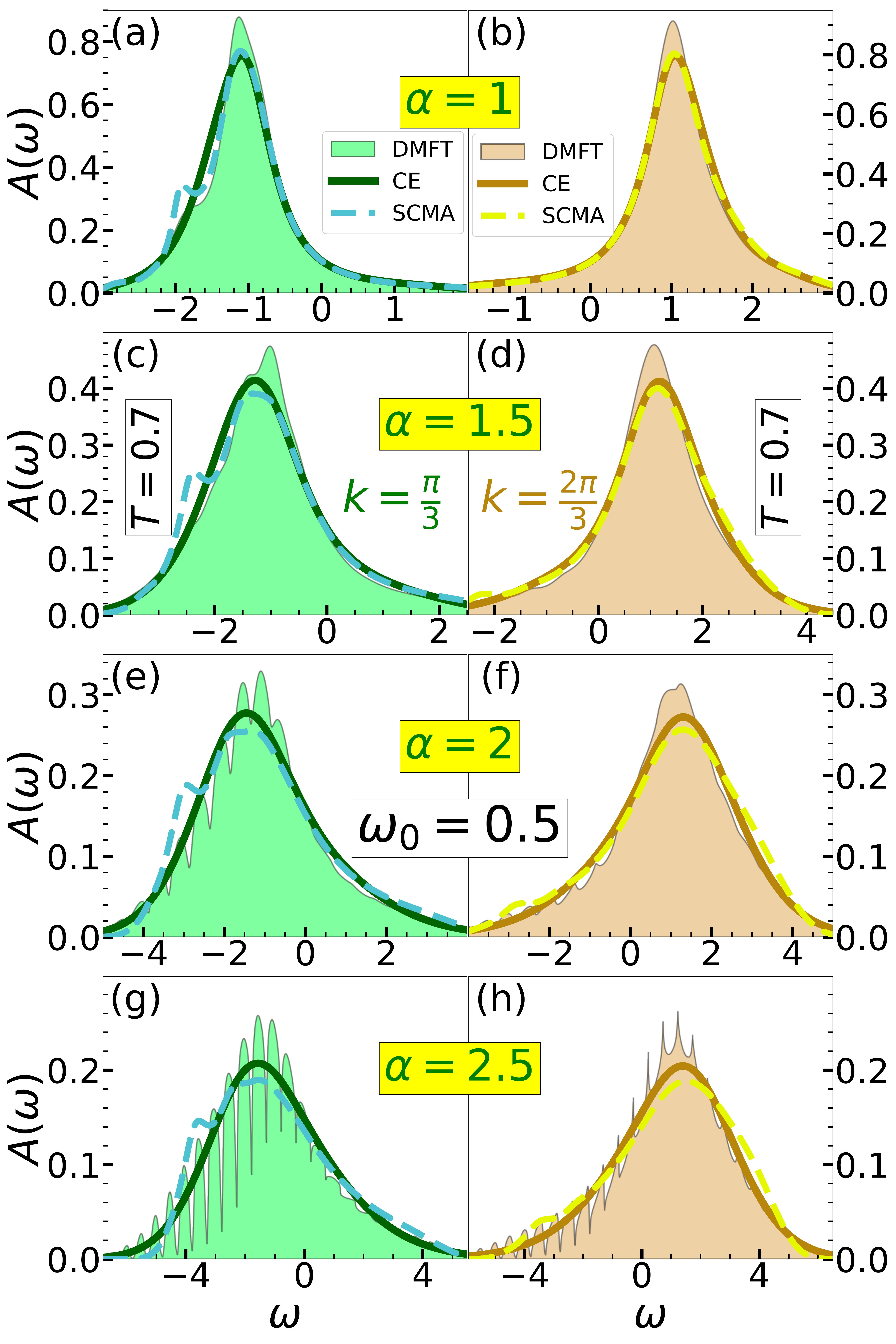}}
\end{minipage}
\end{center}
\hfill
\begin{center}
\begin{minipage}[b]{\linewidth}
\subfloat[ \label{AppFig:SpecF_w=0.5_k=pi3,2pi3_highT} \centering Spectral functions at higher 
temperatures for $\omega_0=0.5$ and $k=\pi/3, 2\pi /3$.]{\includegraphics[width=\linewidth]{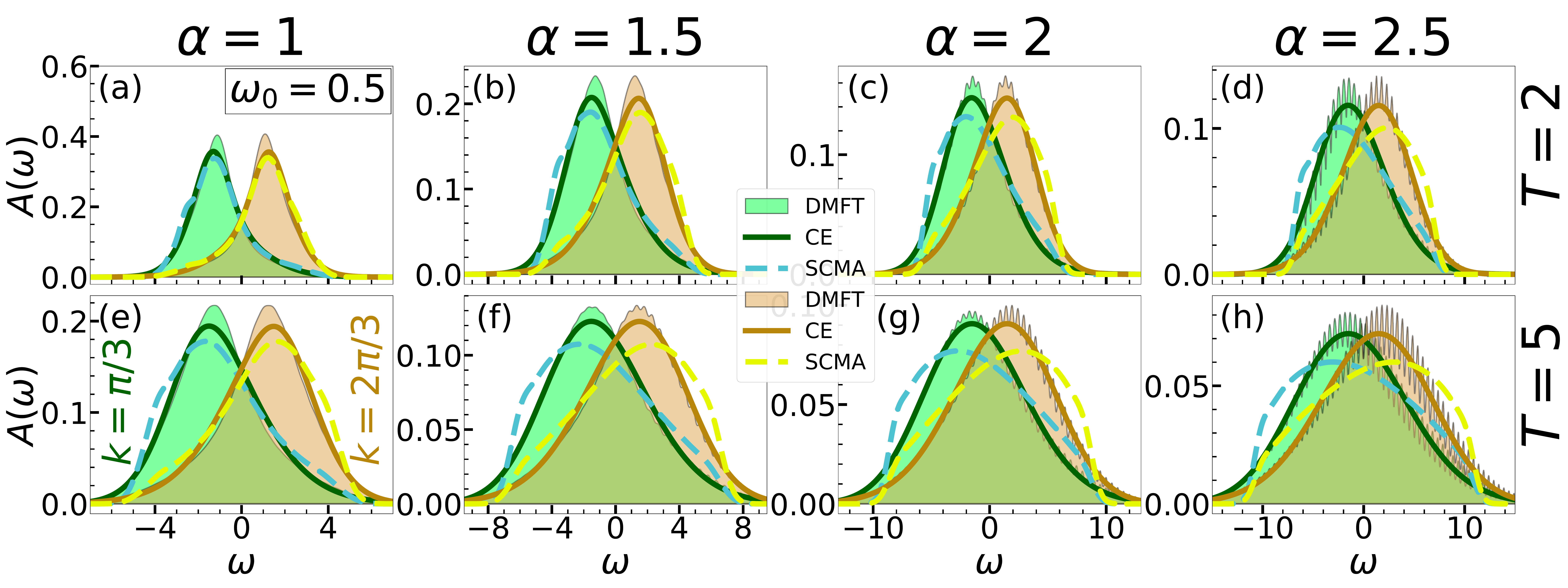}}
\end{minipage}
\end{center}

\caption[Impact of ...]{Comparison of the CE, DMFT, and SCMA   spectral functions in 1D  for $t_0=1$, $\omega_0 = 0.5$ and $k=\pi/3, 2\pi / 3$.}
\label{AppFig:SpecF_w=0.5_k=pi3,2pi3}
\end{figure*}
%\twocolumngrid

%%%%%%%%%%%%%%%%%%%%%%%%%%%%%%%%%%%%%%%%%%%%%%%%%%%%%%%
%%%%%%%%%%%%%%%%%%%%%%%%%%%%%%%%%%%%%%%%%%%%%%%%%%%%%%%%
%%%%%%%%%%%%%%%%%%%%%%%%%%%%%%%%%%%%%%%%%%%%%%%%%%%%%

%\onecolumngrid
\begin{figure*}
\setcounter{subfigure}{0}
\captionsetup[subfigure]{format=hang,singlelinecheck=false,justification=RaggedRight, labelsep=space }
\renewcommand\thesubfigure{\roman{subfigure}}
\begin{center}
\begin{minipage}[t]{\linewidth}
\centering
\subfloat[ \label{SAppFig:SpecF_w=0.2_k=0} (a)--(h) Spectral functions for $\omega_0=0.2$ and $k=0$. In the left panels $T=0.3$, while $T=0.7$ in the right panels. Insets show the integrated spectral weight $I(\omega) = \int_{-\infty}^\infty A(\omega) d\omega$.]{\includegraphics[width=0.48\linewidth]{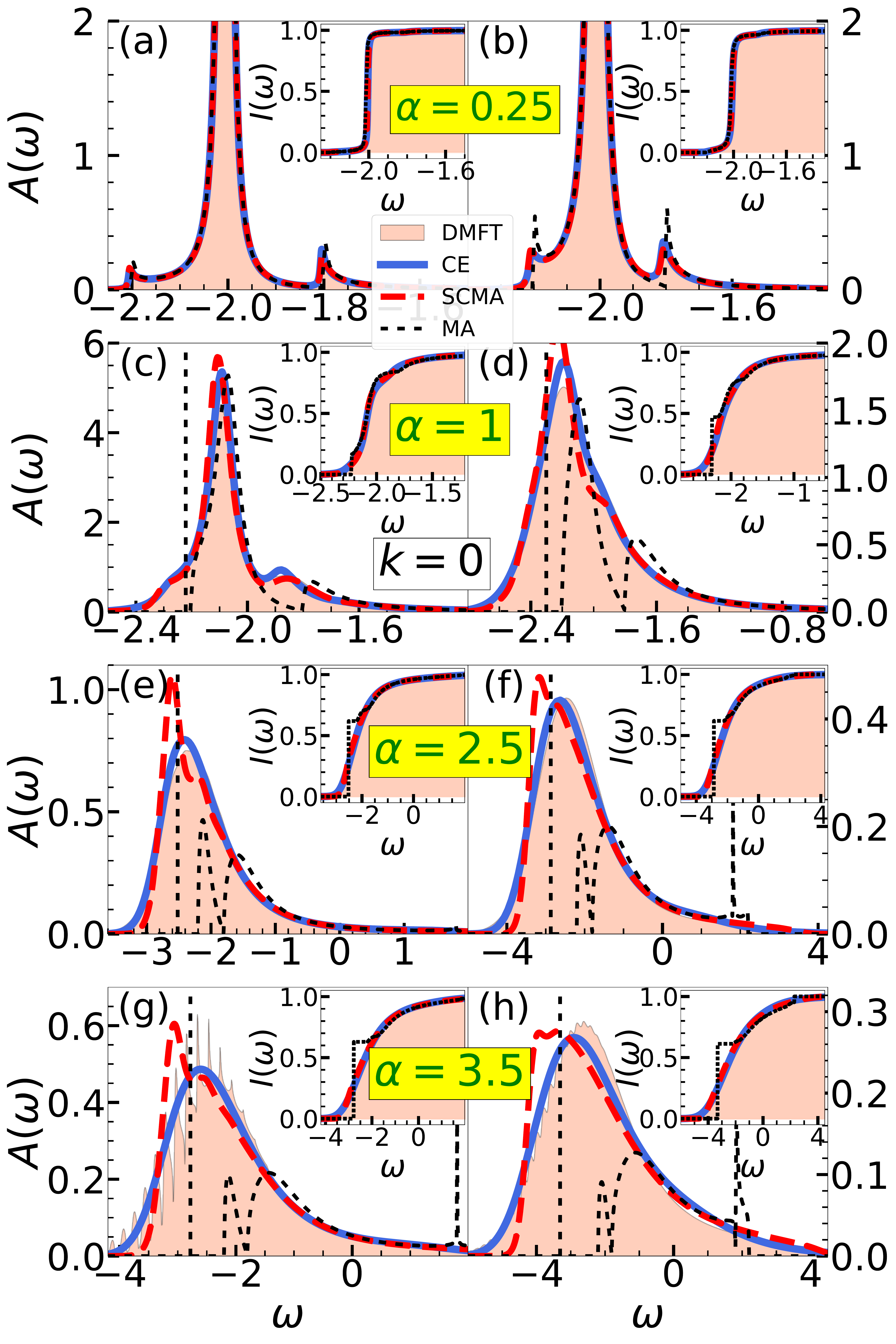}}
%\quad
\;\;\;
\subfloat[ \label{AppFig:SpecF_w=0.2_k=pi} (a)--(h) Spectral functions for $\omega_0=0.2$ and $k=\pi$. In the left panels $T=0.3$, while $T=0.7$ in the right panels. Insets show the integrated spectral weight $I(\omega) = \int_{-\infty}^\infty A(\omega) d\omega$.]{\includegraphics[width=0.48\linewidth]{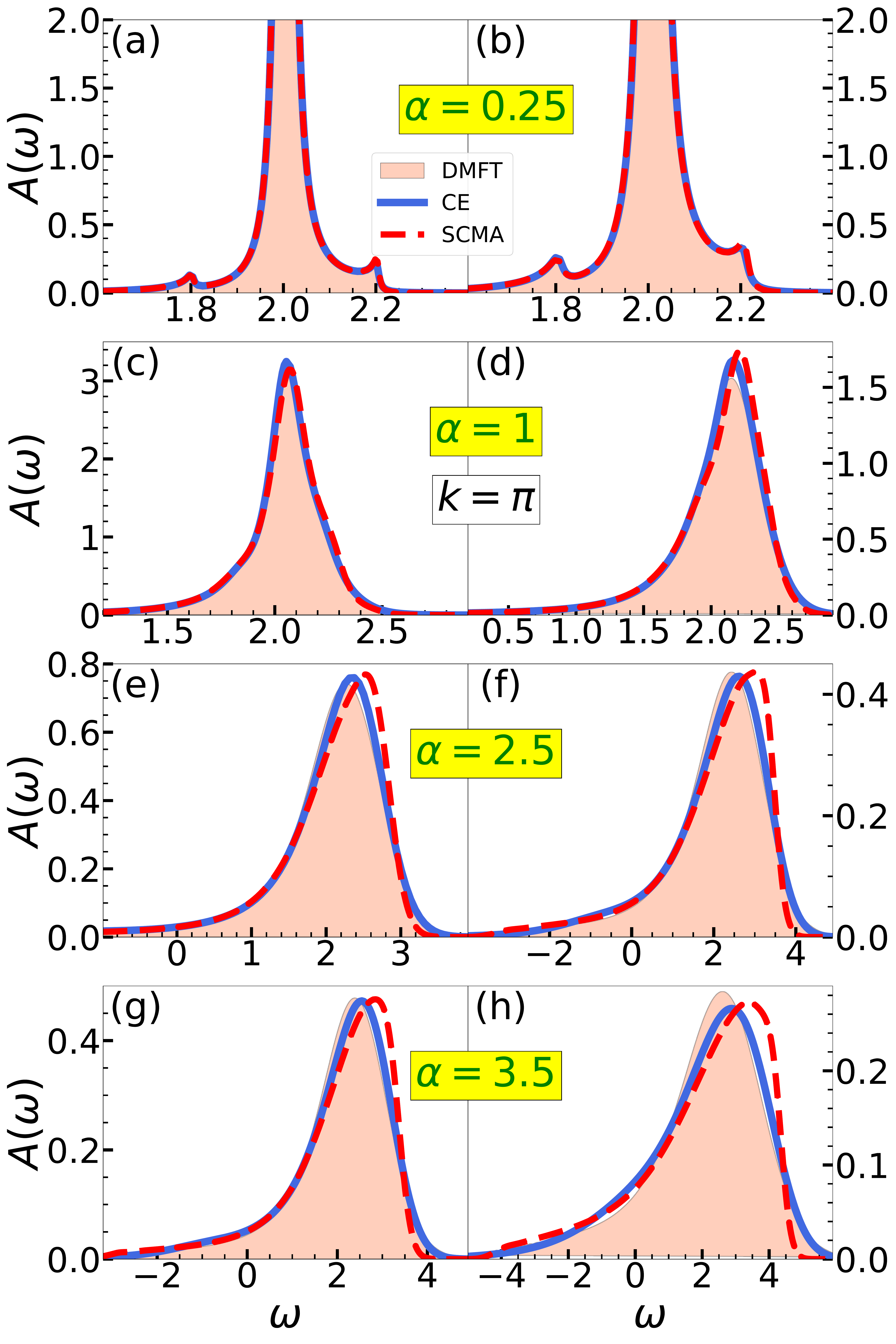}}
\end{minipage}
\end{center}
\hfill
\begin{center}
\begin{minipage}[b]{\linewidth}
\subfloat[ \label{AppFig:SpecF_w=0.2_k=0,pi_highT} \centering  Spectral functions at higher 
temperatures for $\omega_0=0.2$ and $k=0,\pi$.]{\includegraphics[width=\linewidth]{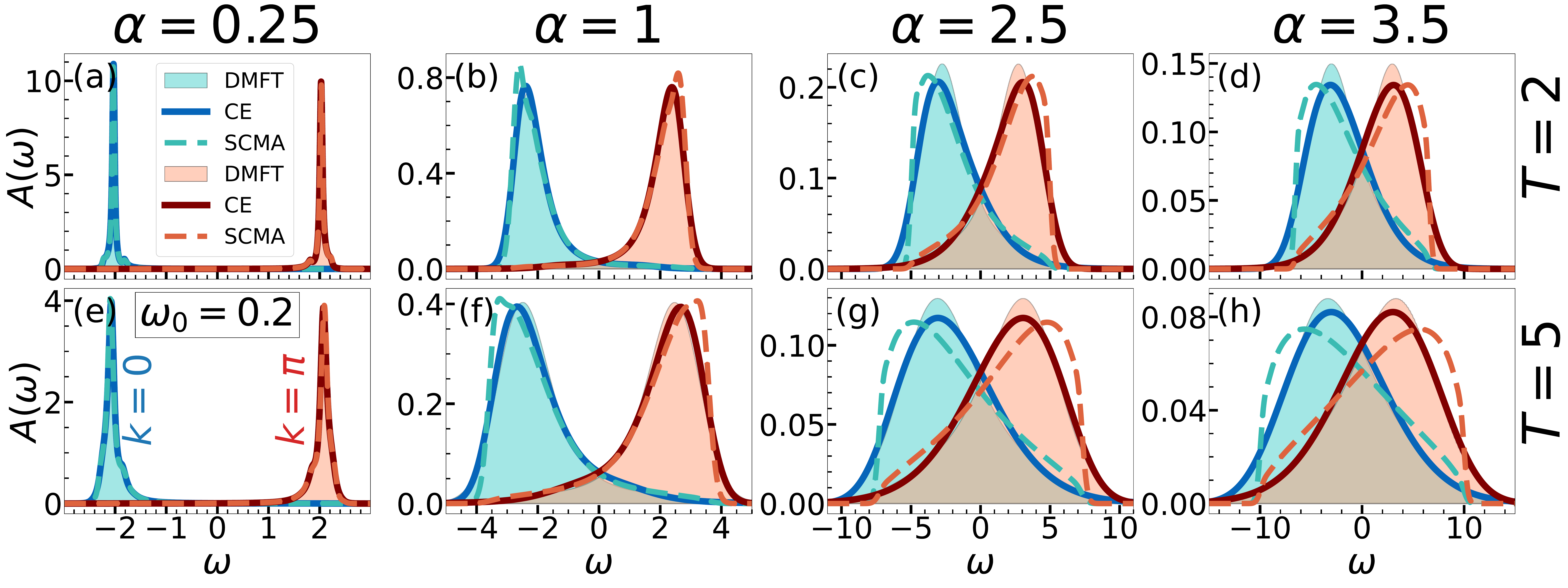}}
\end{minipage}
\end{center}

\caption[Impact of ...]{Comparison of the CE, DMFT, SCMA, and MA spectral functions in 1D  for $t_0=1$, $\omega_0 = 0.2$, and $k=0,\pi$.}
\label{AppFig:SpecF_w=0.2_k=0,pi}
\end{figure*}
%\twocolumngrid

%%%%%%%%%%%%%%%%%%%%%%%%%%%%%%%%%%%%%%%%%%%%%%%%%%%%%%%
%%%%%%%%%%%%%%%%%%%%%%%%%%%%%%%%%%%%%%%%%%%%%%%%%%%%%%%%
%%%%%%%%%%%%%%%%%%%%%%%%%%%%%%%%%%%%%%%%%%%%%%%%%%%%%

%\onecolumngrid
\begin{figure*}
\setcounter{subfigure}{0}
\renewcommand\thesubfigure{\roman{subfigure}}
\captionsetup[subfigure]{format=hang,singlelinecheck=false,justification=RaggedRight, labelsep=space }
\begin{center}
\begin{minipage}[t]{\linewidth}
\centering
\subfloat[ \label{AppFig:SpecF_w=0.2_k=pi3,2pi3_T=0.3} (a)--(h) Spectral functions for $\omega_0=0.2$ and  $T=0.3$. In the left panels $k=\pi /3$, while $k=2\pi / 3$ in the right panels.]{\includegraphics[width=0.48\linewidth]{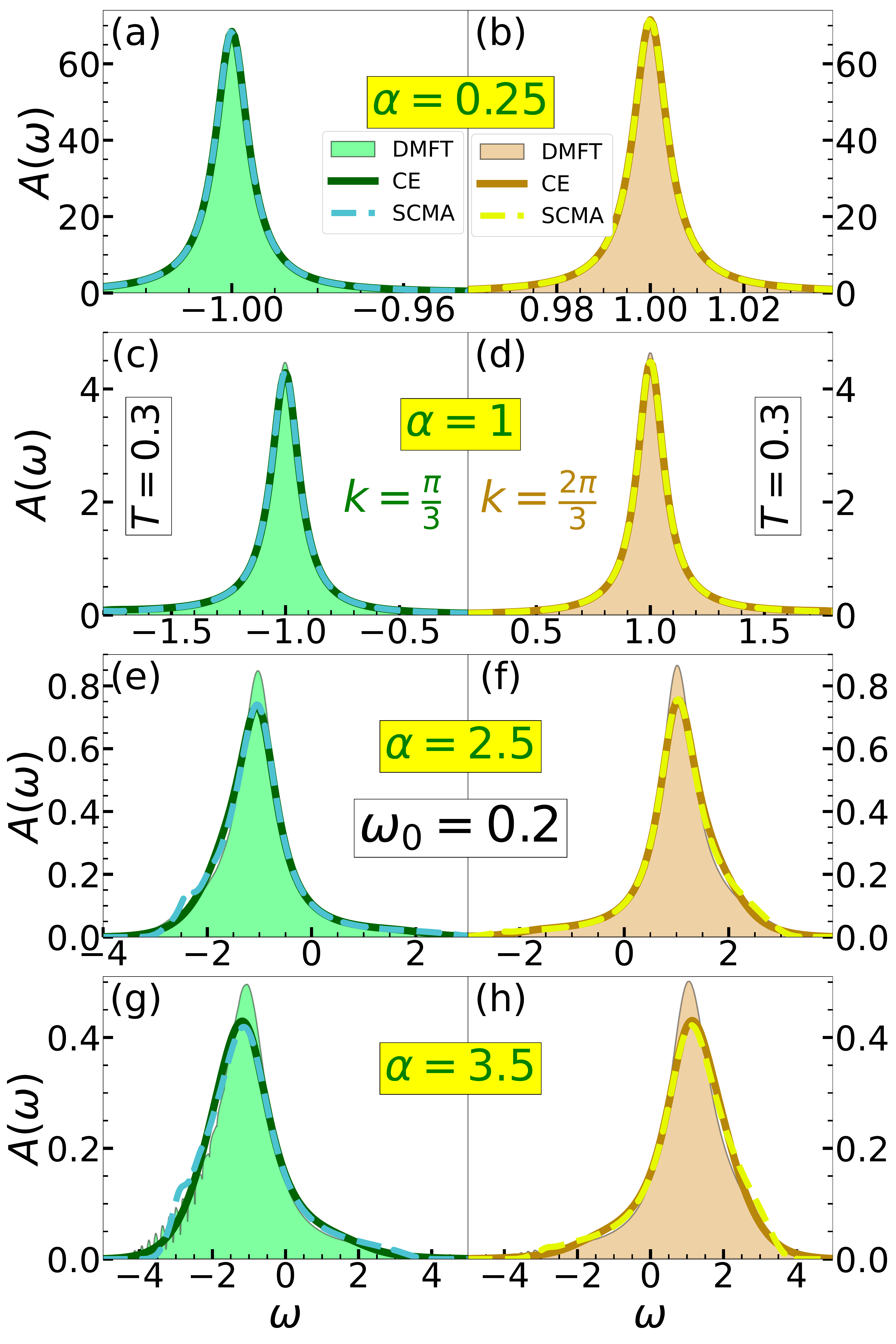}}
%\quad
\;\;\;
\subfloat[ \label{AppFig:SpecF_w=0.2_k=pi3,2pi3_T=0.7} (a)--(h) Spectral functions for $\omega_0=0.2$ and  $T=0.7$. In the left panels $k=\pi /3$, while $k=2\pi / 3$ in the right panels.]{\includegraphics[width=0.48\linewidth]{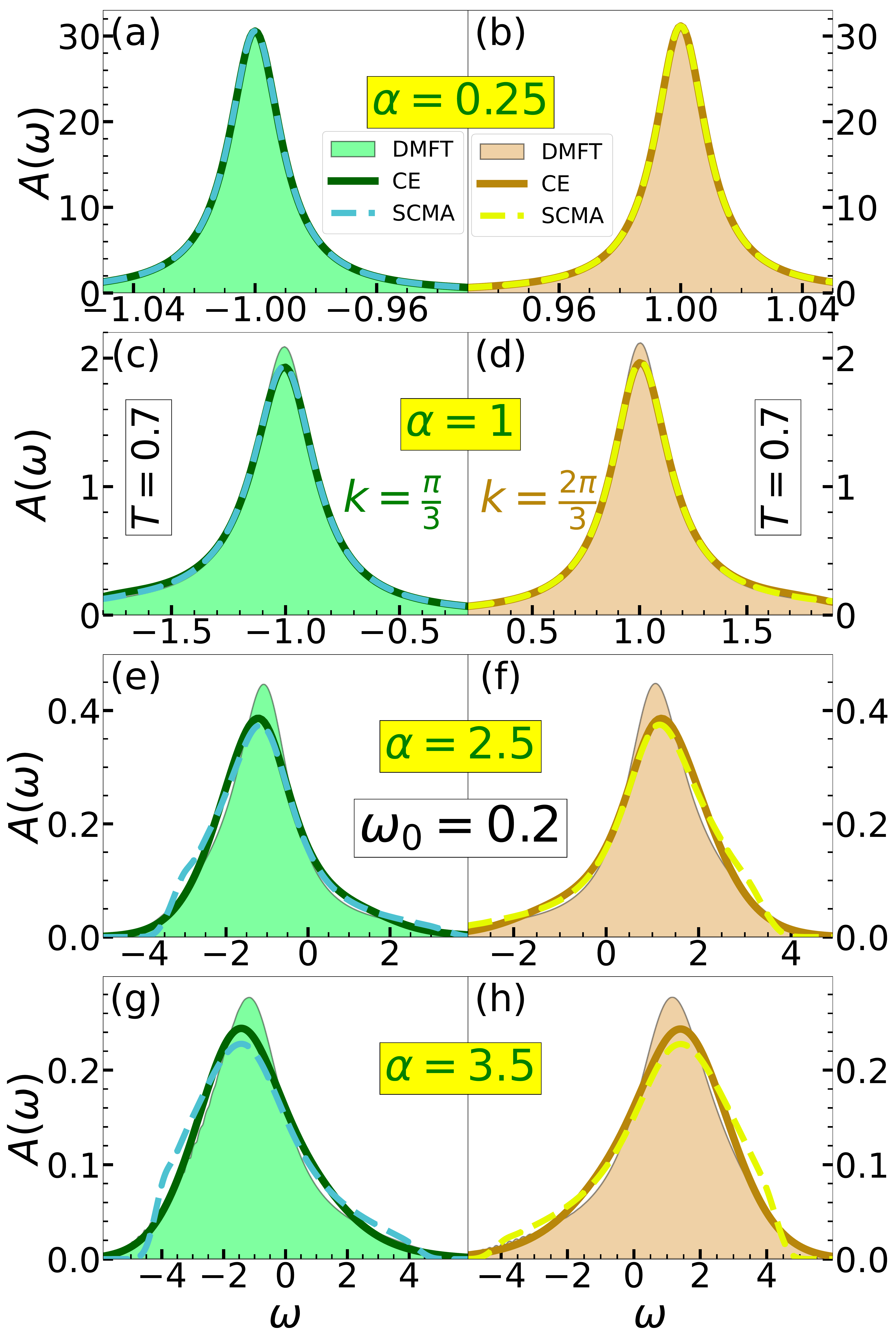}}
\end{minipage}
\end{center}
\hfill
\begin{center}
\begin{minipage}[b]{\linewidth}
\subfloat[ \label{AppFig:SpecF_w=0.2_k=pi3,2pi3_highT} \centering Spectral functions at higher 
temperatures for $\omega_0=0.2$ and $k=\pi / 3, 2\pi / 3$.]{\includegraphics[width=\linewidth]{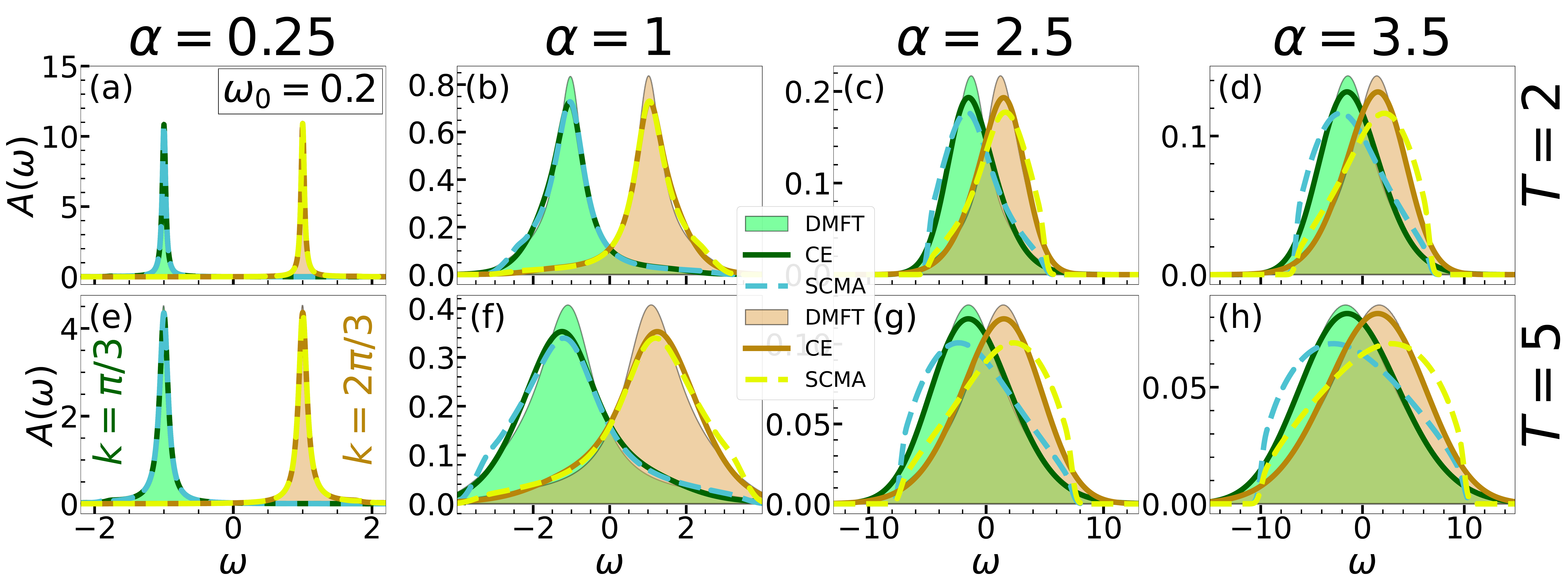}}
\end{minipage}
\end{center}

\caption[Impact of ...]{Comparison of the CE, DMFT, and SCMA   spectral functions in 1D  for $t_0=1$,  $\omega_0 = 0.2$ and $k=\pi/3, 2\pi /3$.}
\label{AppFig:SpecF_w=0.2_k=pi3,2pi3}
\end{figure*}
%\twocolumngrid
%%%%%%%%%%%%%%%%%%%%%%%%%%%%%%%%%%%%%%%%%%%%%%%%%%%%%%%%%%%%%%%%%%%%%5
%%%%%%%%%%%%%%%%%%%%%%%%%%%%%%%%%%%%%%%%%%%%%%%%%%%%%%%%%%%%%%%%%%%%%%
%%%%%%%%%%%%%%%%%%%%%%%%%%%%%%%%%%%%%%%%%%%%%%%%%%%%%%%%%%%%%%%%%%%%%%
\begin{figure*}
\setcounter{subfigure}{0}
\captionsetup[subfigure]{format=hang,singlelinecheck=false,justification=RaggedRight, labelsep=quad }
\renewcommand\thesubfigure{\roman{subfigure}}
\begin{center}
\begin{minipage}[t]{\linewidth}
\centering
\subfloat[ \label{AppFig:HeatPlots_w=0.2_T=0.4} (a)--(h) Heat maps for $T = 0.3$. In the left panels, we present CE results, while the DMFT benchmark is presented in the right panels. Panels (c)--(h) use the same color coding, while panels (a) and (b) use different color coding.]{\includegraphics[width=0.48\linewidth]{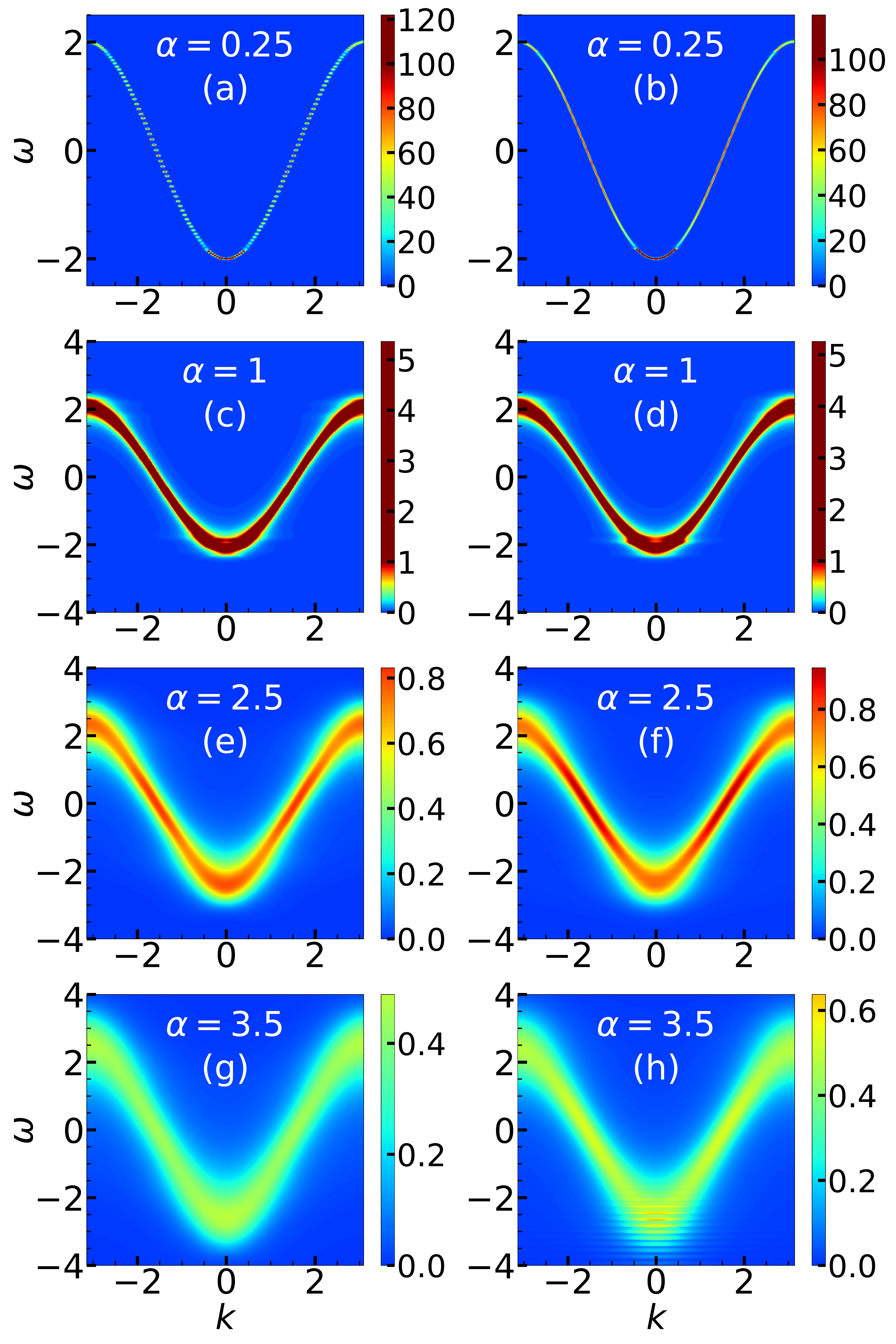}}
%\quad
\;\;\;
\subfloat[ \label{AppFig:HeatPlots_w=0.2_T=1} (a)--(h) Heat maps for $T = 0.7$. In the left panels, we present CE results, while the DMFT benchmark is presented in the right panels. Panels (c)--(h) use the same color coding, while panels (a) and (b) use different color coding.]{\includegraphics[width=0.48\linewidth]{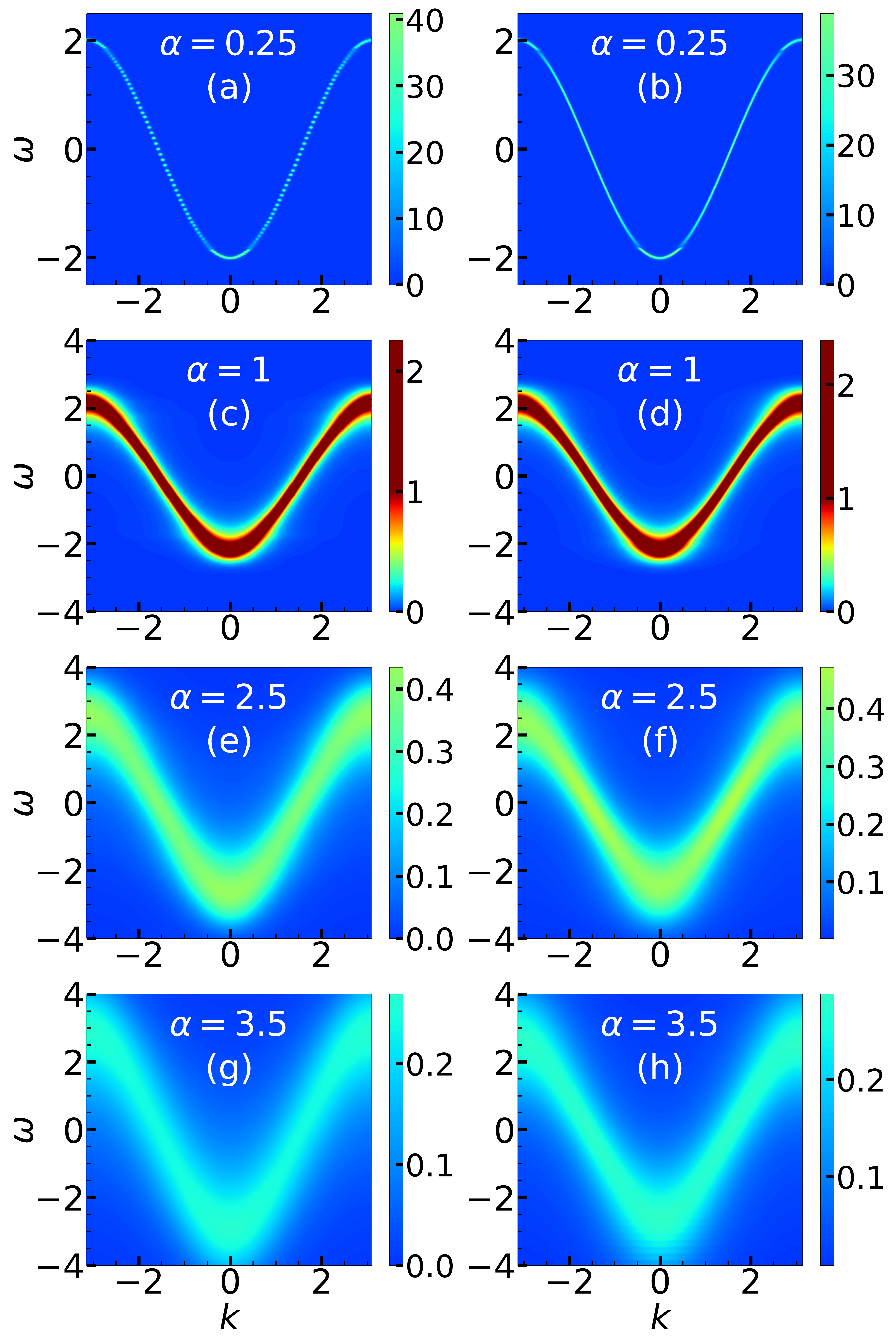}}
\end{minipage}
\end{center}
% \hfill
% \begin{center}
% \begin{minipage}[b]{\linewidth}
% \subfloat[ \label{SubFig:SpecF_w_1_highT} Spectral functions at higher 
% temperatures.]{\includegraphics[width=\linewidth]{./Pics/HighT_w=1.0_other_k.pdf}}
% \end{minipage}
% \end{center}

\caption[Impact of ...]{Comparison of the CE and  DMFT heat maps for $t_0=1$ and $\omega_0 = 0.2$.}
\label{AppFig:HeatPlots_w=0.2}
\end{figure*}

\clearpage
\newpage

\twocolumngrid
\section{Quasiparticle properties} \label{Supp:Sec:QP_prop}
\begin{figure}[!h]
  \includegraphics[width=3.2in,trim=0cm 0cm 0cm 0cm]{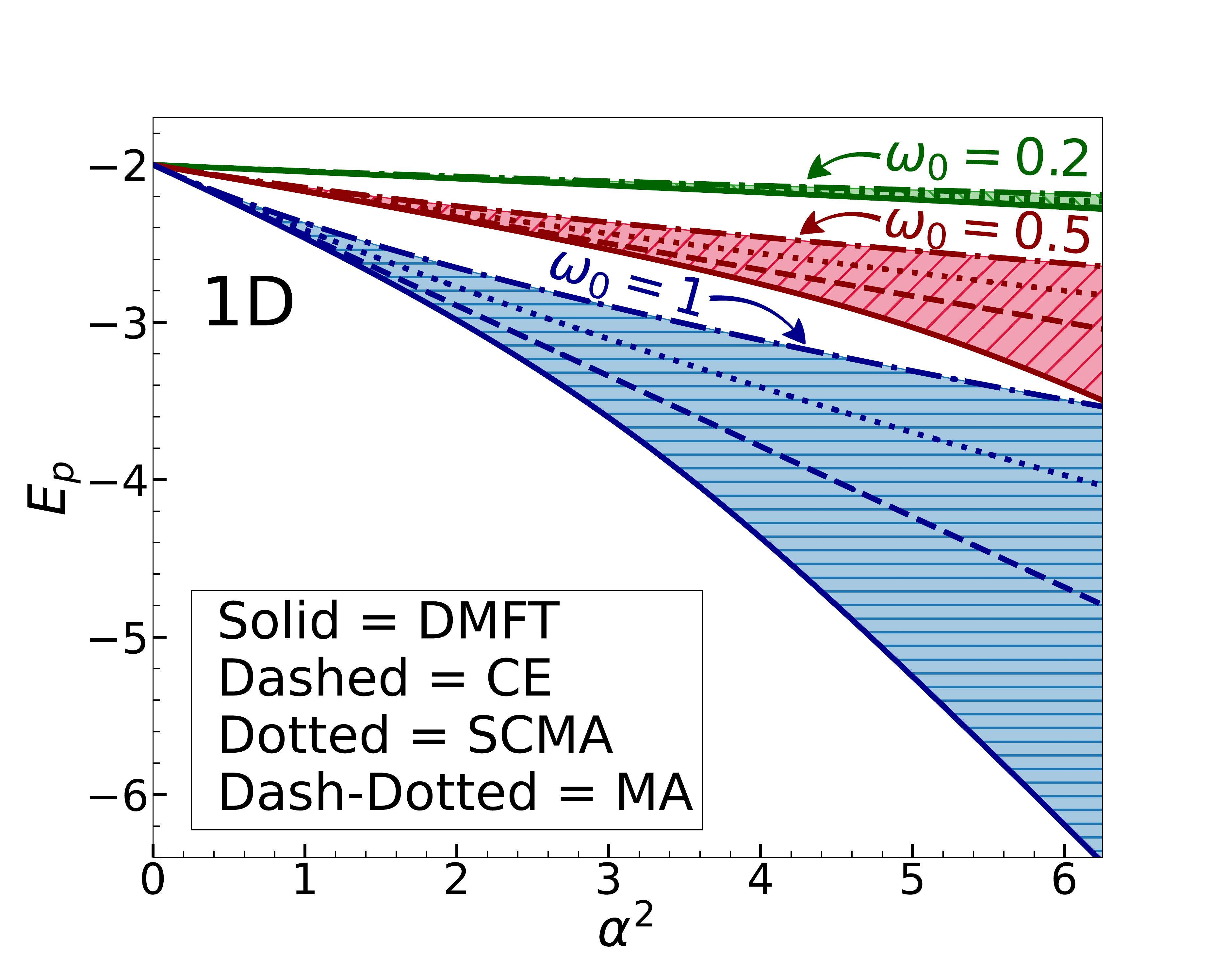} 
 \caption{Ground-state energy within DMFT, CE, SCMA, and MA for the one-dimensional Holstein model with $t_0=1$.
 }
 \label{AppFig:qp}
\end{figure}

In Sec.~IV of the main text, we showed and analyzed the quasiparticle properties of CE, DMFT, and SCMA methods. Here we supplement that study by including the predictions of the  Migdal approximation for the ground-state energy in 1D. The results are shown in Fig.~\ref{AppFig:qp}. We emphasize that  the predictions of the DMFT benchmark  are practically identical to the exact numerical results \cite{Supp_2022_Mitric}. These results readily demonstrate how much improvement to the simplest approximation (MA) is provided by including the self-consistency (SCMA) and by employing the cumulant expansion method (CE).

%
%\clearpage
%\input{Appendix}
%\bibliographystyle{apsrev4-1}
%\bibliography{refs.bib}

\end{document}